\documentclass[11pt,urlcolor=blue, linkcolor=blue]{article}
\pdfoutput=1
\usepackage{cite}

\usepackage{amsmath, amsthm, amssymb}
\usepackage{ifpdf}
\ifpdf
  \usepackage[pdftex]{graphicx}
  \usepackage{epstopdf}
\else
  \usepackage[dvips]{graphicx}
\fi
\parskip=\baselineskip

\usepackage[usenames, dvipsnames]{color}
\usepackage{mathrsfs}
\usepackage{marginnote}

\allowdisplaybreaks[1]

\definecolor{mygray}{gray}{0.6}

\usepackage{upgreek}
\usepackage{bbm}

\newenvironment{myfont}[2][]{\csname#2\endcsname[#1]}{}

\newcommand{\bea}{\begin{eqnarray}}
\newcommand{\eea}{\end{eqnarray}}
\def\be{\begin{equation}}
\def\ee{\end{equation}}

\usepackage{color}
\usepackage[colorlinks,citecolor=blue]{hyperref}
\definecolor{red}{rgb}{1,0,0}
\definecolor{blue}{rgb}{0,0,1}
\definecolor{dblue}{rgb}{0,0,0.4}
\definecolor{green}{rgb}{0,1,0}
\definecolor{black}{rgb}{0,0,0}
\definecolor{white}{rgb}{1,1,1}

\definecolor{brn}{rgb}{.8,.4,.0}
\definecolor{redo}{rgb}{1,.5,.0}
\definecolor{ddgrn}{rgb}{0,0.4,0}
\definecolor{dgrn}{rgb}{0,0.55,0}
\definecolor{dbl}{rgb}{0,0,0.5}

\usepackage[bbgreekl]{mathbbol}
\usepackage{amscd}

\newcommand{\bpm}{\begin{pmatrix}}
\newcommand{\epm}{\end{pmatrix}}
\newcommand{\bmm}{\begin{matrix}}
\newcommand{\emm}{\end{matrix}}

\usepackage{slashed}
\usepackage[makeroom]{cancel}
\usepackage[normalem]{ulem}
\usepackage{soul}
\newcommand{\stkout}[1]{\ifmmode\text{\sout{\ensuremath{#1}}}\else\sout{#1}\fi}

\usepackage{tikz}
\usetikzlibrary{arrows}
\usetikzlibrary{calc}
\pgfmathsetseed{3}
\usetikzlibrary{shapes.arrows}

\newcommand{\normord}[1]{:\mathrel{#1}:}

\numberwithin{equation}{section}

\begin{document}

\begin{titlepage}
\begin{flushright}
\end{flushright}
\vskip .5in

\begin{center}

{\bf\LARGE{The chiral SYK model}}


\vskip 1.5cm
\Large{Biao Lian$^{1}$, S. L. Sondhi$^{2}$, Zhenbin Yang$^{2}$}

\vskip.5cm

 \vskip.2cm
{\small{\textit{$^1$
Princeton Center for Theoretical Science, Princeton University,
Princeton, NJ 08544, USA}\\}}
 \vskip.3cm
{\small{\textit{$^2$Department of Physics, Princeton University,
Princeton, NJ 08544, USA}\\}}


\end{center}
\vskip.5cm
\baselineskip 16pt
\begin{abstract}
We study the generalization of the Sachdev-Ye-Kitaev (SYK) model to a $1+1$ dimensional chiral SYK model of $N$ flavors of right-moving chiral Majorana fermions with all-to-all random 4-fermion interactions. The interactions in this model are exactly marginal, leading to an exact scaling symmetry. We show the Schwinger-Dyson equation of this model in the large $N$ limit is exactly solvable. In addition, we show this model is integrable for small $N\le6$ by bosonization. Surprisingly, the two point function in the large $N$ limit has exactly the same form as that for $N=4$, although the four point functions of the two cases are quite different. The ground state entropy in the large $N$ limit is the same as that of $N$ free chiral Majorana fermions, leading to a zero ground state entropy density. The OTOC of the model in the large $N$ limit exhibits a non-trivial spacetime structure reminscent of that found by Gu and Kitaev \cite{Gu:2018jsv} for generic SYK-like models. Specifically we find a Lyapunov regime inside an asymmetric butterfly cone, which are signatures of quantum chaos, and that the maximal velocity dependent Lyapunov exponent approaches the chaos bound $2\pi/\beta$ as the interaction strength approaches its physical upper bound. Finally, the model is integrable for (at least) $N\le6$ but chaotic in the large $N$ limit, leading us to conjecture that there is a transition from integrability to chaos as $N$ increases past a critical value.
\end{abstract}
\end{titlepage}

\tableofcontents

\section{Introduction}\label{Sec1}

The last few years have seen an enormous amount of activity around the Sachdev-Ye-Kitaev (SYK) model \cite{Sachdev:1992fk, Polchinski:2016xgd,Maldacena:2016hyu,Kitaev:2017awl}. This activity is a consequence of the confluence of several themes. First, the model is partially solvable in a tractable large $N$ limit
which it shares with the class of so-called tensor models \cite{Witten:2016iux,gurau,KPT} and the literature on dynamical mean field theory. This large $N$ limit is distinct from the large $N$ limits of vector and matrix models. Second,
the SYK model exhibits many body quantum chaos or ergodicity as can be seen by computing the four point
out of time order correlator (OTOC) in the same large $N$ limit. This has fed into the current intense
interest in such foundational questions in quantum statistical mechanics. Finally, features of the model suggest that it is the holographic dual to some theory of quantum gravity in $1+1$ dimensions, and the prospect of a solvable model of holography has naturally excited enormous interest.

The SYK Model lives in $0+1$ dimensions---it is essentially a model on a quantum dot---and various authors have considered extensions to higher dimensions. One family of extensions builds higher dimensional models out of
lattices of SYK dots, and this has led to interesting results on transport and on non-Fermi liquid behavior \cite{Gu:2016oyy,Berkooz:2016cvq,Blake:2017,davison2017,Jian:2017unn,chen2017,cai2018,zhangp2018}. A second family of extensions involves nonchiral homogeneous models with interactions in the new large $N$ limit which can be viewed either as arising from random couplings or from tensorial fields \cite{Turiaci:2017zwd,Murugan:2017eto,Berkooz:2017efq,narayan2017,Klebanov:2016xxf,Giombi:2017dtl}. In particular, the random extension to $1+1$ dimensional nonchiral fermions shows marginally irrelevant SYK interactions \cite{Berkooz:2017efq}. In this paper we present a third extension, this time to a homegenous \emph{chiral} system in $1+1$ dimensions, where we study the resulting \emph{chiral SYK model} at large $N$.


Before turning to our model, we briefly recapitulate some aspects of the original Sachdev-Ye-Kitaev (SYK) model \cite{Sachdev:1992fk, Polchinski:2016xgd,Maldacena:2016hyu,Kitaev:2017awl}.
This model contains $N$ Majorana fermions $\psi_i$, which have all-to-all $q$-fermion random interactions, where $q\ge2$ is an even number. The $0+1 \equiv 1$d action of the SYK model is
\begin{equation}
S_{\text{SYK}}=\int dt\left(\frac{i}{2}\sum_{i}\psi_i\partial_t\psi_i -i^{\frac{q}{2}}\sum_{i_1\cdots i_q}J_{i_1\cdots i_q}\psi_{i_1}\cdots\psi_{i_q}\right)\ ,
\end{equation}
where $J_{i_1\cdots i_q}$ are a set of random Gaussian couplings satisfying $\langle J_{i_1\cdots i_q}^2 \rangle=J^2(q-1)!/N^{q-1}$. In the large $N$ limit, the SYK model has a large ground state entropy due to the existence of highly degenerate spin glass like states. At low temperatures, an asymptotic conformal symmetry emerges from the collection of those states. The conformal symmetry is weakly broken and the corresponding soft mode physics is captured by a Schwarzian action. The low temperature dynamics governed by this action has been argued to be maximally chaotic or scrambled. In this context this means that the out-of-time-order correlators (OTOCs)
grow as $1-{1\over N}e^{\lambda (t-t_0)}$ with time $t$ and that the Lyapunov exponent $\lambda$ approaches the so-called ``chaos bound''\footnote{In systems with locally bounded Hilbert spaces there does not appear to be such a Lyapunov regime, hence our qualifier.}  $2\pi/\beta$, where $\beta$ is the inverse of temperature \cite{Shenker:2013pqa,Shenker:2014cwa,Maldacena:2015waa}.
It turns out that the Schwarzian action can also be viewed as emergent from the description of Near Extremal Black Holes when we consider gravitational dynamics near the AdS$_2$ throat
\cite{Sachdev:2010prl,Almheiri:2014cka,Jensen:2016pah,Maldacena:2016upp,Engelsoy:2016xyb}.
In the gravitational system, the black hole horizon provides a large ground state degeneracy and the conformal symmetry is the asymptotic symmetry of AdS$_2$.
With the Schwarzian action, the SYK model provides a concrete example of holographic dual system for near extremal black holes, although it is not clear whether the full model has an exact gravitational dual.
Many striking conclusions about near extremal black holes have been drawn from this duality,  such as the construction of a traversable wormhole \cite{Maldacena:2017axo,Maldacena:2018lmt,kim2019} and black hole level statistics \cite{Cotler:2016fpe,Saad:2018bqo,Saad:2019lba}.

Turning now to our paper, we present a generalization of the SYK model to $N$ flavors of $1+1$ dimensional 
chiral Majorana fermions with all-to-all interactions (see action (\ref{Eq-action})), which we call the \emph{chiral SYK model}. The interactions are uniform in the spacetime and random with respect to fermion flavors.
Since fermions in $1+1$ dimensions have a scaling dimension $1/2$, any $q>4$ fermion interactions will be irrelevant, so we only focus on the $q=4$ fermion interactions. From the perspective of Hibert space dimension, the model we study here is the minimal generalization of the SYK model into higher spacetime dimensions. In condensed matter physics, chiral fermions can exist on the edge of $2+1$ dimensional gapped chiral topological states, and can interact with each other on the edge. For instance, chiral complex fermions occur on the edge of integer quantum Hall states \cite{laughlin1981,halperin1982} or Chern insulators \cite{haldane1988,qi2006}, each of which is equivalent to two chiral Majorana fermions. Odd number of chiral Majorana fermions can be realized on the edge of non-Abelian fractional quantum Hall states \cite{moore1991}, $p+ip$ chiral topological superconductors \cite{read2000,qi2011} or chiral spin systems \cite{kitaev2006}.

Unlike nonchiral systems, the 4-fermion interactions of $1+1$ dimensional chiral fermions are generically exactly marginal. Besides, such 4-fermion interactions break the Lorentz symmetry explicitly, since the product of 4 chiral fermion fields has conformal spin $2$ instead of zero. For example, in the celebrated chiral Luttinger model \cite{wen1990,chang2003,cano2014} of two complex chiral fermions, the 4-fermion interaction leads to a spin-charge velocity separation independent of energy scale, which is therefore exactly marginal; meanwhile, the Lorentz symmetry is absent since spin and charge excitations have different velocities. In our $1+1$ dimensional chiral SYK model in the large $N$ limit, the 4-fermion interactions are also exactly marginal and break the Lorentz symmetry. We show this explicitly by solving the two point function of the model exactly. Therefore, all the features of our model are energy scale independent.

The exact marginality of 4-fermion interactions in our chiral SYK model leads to a scaling symmetry, which enables us to solve the model exactly. In polar coordinates $re^{i\theta}=\tau+ix$ of imaginary time $\tau$ and spatial position $x$, the scaling symmetry restricts the zero temperature two point function to be proportional to $r^{-1}$, therefore the $1+1$ dimensional Schwinger-Dyson equations in the large $N$ limit reduce to $1$ dimensional equations in the boost angle $\theta$, which are exactly solvable.
The exact two point function at finite temperature $\beta^{-1}$ can also be obtained by observations based on the zero temperature solution. We find that the two point function in real space takes a form of the product of two $0+1$ dimensional SYK propagators in the conformal limit moving at two different velocities $u_{\pm}$ determined by the interaction strength ($u_+>u_->0$).
With the exact solution, we calculate the ground state entropy by directly evaluating the large $N$ effective action (\ref{Eq-FN}), and find it is the same as the ground state entropy of $N$ flavors of free chiral Majorana fermions. Depending on the spatial boundary condition, the ground state entropy could be either $0$ (anti-periodic boundary condition) or $N\log 2$ (periodic boundary condition, which has $N$ zero modes). In any case, the ground state entropy density vanishes, so there are no spin-glass like states in the model.
We also investigate the behavior of the OTOC using the retarded kernel method in \cite{Maldacena:2016hyu,Kitaev:2017awl}, and identify a tilted butterfly cone where the OTOC grows exponentially along a space-time trajectory with fixed velocity.
The butterfly cone exactly coincides with the causality cone, which has edges at velocities $u_+$ and $u_-$ (Figure \ref{exponent}(e)). The butterfly cone can be further divided into three regions which behave differently: for the region with small velocity $x/t$ near the left butterfly cone edge, the $1/N$ piece of the OTOC behaves as $\exp\left[{2\pi \over \beta}(u_-^{-1}x-t)\right]$.
For the region in the middle of butterfly cone, the $1/N$ piece of the OTOC grows as $\exp\left[\overline{\lambda} t-{\#(t-v_c x)^2\over t}\right]$, where $0<\overline{\lambda}<{2\pi \over \beta}$, and $u_-<v_c<u_+$. When the interaction strength approaches its allowed upper bound (where $u_-\rightarrow0$), the Lyapunov exponent $\overline{\lambda}$ approaches the chaos bound $2\pi/\beta$. For the large velocity region near the right butterfly cone edge, the $1/N$ piece of OTOC grows as $\exp\left[{2\pi \over \beta}(t-u_+^{-1}x)\right]$, where the exponent in the $t$ direction saturates the chaos bound $2\pi/\beta$. A similar saturation of the chaos bound near the butterfly edge was observed in \cite{Gu:2016oyy}, and recently explained on generic grounds by \cite{Gu:2018jsv} using the ladder identity. However, we note that this is not the case near the left butterfly edge of our model, where the exponent in the $t$ direction is $-2\pi/\beta$, i.e., opposite to the chaos bound value.
It is interesting that our model has both vanishing ground state entropy density and maximal Lyapunov exponent when the interaction strength approaches its upper bound, which is not a feature expected from black holes.

We also study the model at small finite $N$. In particular, in the cases $N=4,5$ and $6$, the model is seen to be integrable by bosonization. The $N=4$ model is nothing but the chiral Luttinger model with two complex fermions, which has a spin-charge separation.
Surprisingly, we find that the $N=4$ fermion two point function is exactly of the same form as the large $N$ two point function.
We conjecture that this coincidence may be due to the fact that the model has an SO($N$) symmetry for both $N=4$ and in the large $N$ limit (upon averaging over the random couplings). Together with the scaling symmetry, the SO($N$) symmetry may already pin down the form of two point function.
Nevertheless, the four point function for $N=4$ is completely different from that in the large $N$ limit.  This is because the $N=4$ case is integrable while the large $N$ case is chaotic. As a result, the four point function for $N=4$ only has power-law or exponential decay (up to a background constant from conserved charges), and is insensitive to time orderings.
For $N=5$, upon a flavor basis rotation, the model decouples into an $N=4$ model plus a free chiral Majorana fermion.
For $N=6$, by a flavor basis rotation and bosonization, we can rewrite the model as three free chiral bosons with three distinct velocities. For finite $N\ge7$, the model becomes much more complicated and probably unsolvable. We conjecture that the model has a transition from integrablity to chaos when $N$ increases, and this transition probably happens between $N=6$ and $N=7$. This would also be the transition from non-thermalization to thermalization among the $N$ chiral Majorana flavors. In chiral topological condensed matter systems, thermalization of chiral states on the edge could affect the thermal measurements in experiments \cite{banerjee2018,simon2018,feldman2018,ma2019}.

While the bosonization of the $0+1$ dimensional SYK model has been attempted \cite{murugan2018}, we note that our chiral SYK model for any $N$ can always be written straightforwardly in the bosonized representation. For $N\le6$, it is advantageous to solve the model in the bosonized representation; while in the large $N$ limit, the fermion representation is more convenient.

The paper is organized as follows:
In Sec. \ref{Sec-model}, we introduce the $1+1$ dimensional chiral SYK model for general number of chiral Majorana fermion flavors $N$, and find the energy-momentum tensor of the model.
In Sec. \ref{Sec-N=4}, we solve the model for $N=4$ via bosonization. We then calculate the thermal quantities and the four point function for $N=4$, and show the OTOC in this case only has exponential decay.
In Sec. \ref{Sec-SYK}, we investigate the chiral SYK model in the large $N$ limit by solving the Schwinger-Dyson equations exactly. Based on the exact two point function, we discuss its thermal properties and ground state entropy, and examine the large $N$ OTOC for signatures of chaos and operator spreading.
In section \ref{Sec-56}, we give the exact solutions for $N=5$ and $N=6$, and discuss what might happen at $N\ge7$ and its consequence for thermalization. Finally, we briefly summarize our results in Sec. \ref{Sec-con}.

\section{The 1+1 dimensional chiral SYK model}\label{Sec-model}
In this section, we describe the $1+1$ dimensional chiral SYK model of chiral Majorana fermions which we will study in this paper.
We will introduce the Lagrangian and the energy momentum tensor of the model. In particular, the model requires a proper choice of point splitting regularization for the operator product expansion (OPE) of the chiral Majorana fermion fields, which we will specify.

\subsection{The Lagrangian}
In $1+1$ dimensions, a free chiral Majorana fermion field $\psi$ has a scaling dimension $1/2$, so a $q$-fermion interaction term has a scaling dimension $q/2$, which is marginal if $q=4$, and irrelevant if $q>4$. This was considered as a major obstacle in the generalization of the SYK model to $1+1$ dimensions, since it is important to have the interaction relevant in the original $0+1$ dimensional SYK model. However, it is still worthwhile to examine the effect of the marginal random $4$-fermion interaction in $1+1$ dimensions, and whether some features of the SYK model are preserved.

In this paper, we study the direct generalization of the $q=4$ SYK model to $N$ flavors of $1+1$ dimensional right-moving chiral Majorana fermions $\psi_i$ ($1\le i\le N$), which has the action
\begin{equation}\label{Eq-action}
S=\int dtdx\mathcal{L}=\int dt dx \left(\frac{i}{2}\sum_{i=1}^N\psi_i(\partial_t+\partial_x)\psi_i +\sum_{1\le i<j<k<l\le N}J_{ijkl}\psi_i\psi_j\psi_k\psi_l\right)\ ,
\end{equation}
where $\psi_i$ satisfies the anticommutation relation $\{\psi_i(t,x),\psi_j(t,x')\}=\delta_{ij}\delta(x-x')$, and $\mathcal{L}$ denotes the Lagrangian density. The couplings $J_{ijkl}$ are real, and antisymmetric with respect to any two indices. In the large $N$ case, we assume $J_{ijkl}$ obey the random Gaussian distribution with
\begin{equation}\label{Eq-J2}
\begin{split}
&\langle J_{ijkl}\rangle=0\ ,\qquad \\
&\langle J_{ijkl}J_{i'j'k'l'}\rangle=\frac{3!J^2}{(N-1)(N-2)(N-3)} \delta_{ii'}\delta_{jj'}\delta_{kk'}\delta_{ll'} \quad (i<j<k<l,\ i'<j'<k'<l')\ ,
\end{split}
\end{equation}
where the coupling strength $J\ge0$ is kept constant when we take the large $N$ limit. Note that here $J_{ijkl}$ is only random with respect to Majorana fermion flavor indices, while is uniform in the entire spacetime. This is fundamentally different from the chiral interactions studied in \cite{Berkooz:2016cvq} which are spatially disordered. As we shall show, the spacetime uniform 4-fermion interaction $J_{ijkl}$ we study here is exactly marginal, thus is important at all energy scales.

The chiral 4-fermion interactions $J_{ijkl}$ explicitly break the Lorentz symmetry, and thus also break the conformal symmetry. This is because the chiral Majorana fermion $\psi_i$ carries a conformal spin $1/2$, thus the 4-fermion interaction term has a nonzero conformal spin $2$ and is not Lorentz invariant. The absence of Lorentz symmetry is, however, not a problem in condensed matter systems. For example, the $N$ chiral Majorana fermions can live on the edge of $N$ copies of the $2+1$d $p+ip$ chiral topological superconductor, which has no intrinsic Lorentz symmetry.

The Euler-Lagrange equation of motion of the action (\ref{Eq-action}) can be easily found to be
\begin{equation}\label{Eq-EOM}
i(\partial_t+\partial_x)\psi_i+\sum_{1\le j<k<l\le N}J_{ijkl}\psi_j\psi_k\psi_l=0\ .
\end{equation}
Throughout this paper, we will work in both real time $t$ and imaginary time $\tau=it$, depending on which one is more convenient.

\subsection{Point splitting regularization and the energy momentum tensor}\label{Sec-PS}
In quantum field theories (QFTs), the product of quantum fields at the same spacetime coordinate needs to be regularized carefully. The regularization is usually done by finding the operator product expansion (OPE) of two fields at split spacetime points $(t,x)$ and $(t',x')$, taking the limit $t'\rightarrow t$ and $x'\rightarrow x$, and keeping the normal ordered part of the limit. In QFTs without Lorentz invariance, the regularization may depend on the direction of point splitting, so one needs to specify the point splitting direction.

In this paper, we specify the point splitting to be in the $x$ direction whenever we regularize the product of field operators at the same spacetime point. Namely, for instance, we define the product of two chiral Majorana fermion fields as $\psi_i(t,x)\psi_j(t,x)=\lim_{x'\rightarrow x}\normord{\psi_i(t,x)\psi_j(t,x')}$, where $\normord{\mathcal{O}}$ stands for the normal ordering of operator $\mathcal{O}$. This point splitting choice is conventional in condensed matter systems, where there is generically an ultraviolet (UV) cutoff in the $x$ direction (lattice constant, etc) playing the role of the point splitting. Such a point splitting in the $x$ direction leads to the commutation relation on a constant time slice
\begin{equation}\label{Eq-commu2}
\begin{split}
&[(\psi_i\psi_j)(t,x),(\psi_k\psi_l)(t,x')] \\ = &\frac{i}{2\pi}(\delta_{ik}\delta_{jl}-\delta_{il}\delta_{jk})\partial_x\delta(x-x') -\delta(x-x')(\delta_{ik}\psi_j\psi_l-\delta_{il}\psi_j\psi_k+\delta_{jl}\psi_i\psi_k -\delta_{jk}\psi_i\psi_l)\ ,
\end{split}
\end{equation}
which can be derived from the OPE of free chiral Majorana fields $\psi_i$ (Appendix \ref{App-OPE}).
Note that $i\psi_i\psi_j$ ($1\le i<j\le N$) form the representation of the current operators of SO$(N)_1$, thus Eq. (\ref{Eq-commu2}) is nothing but the commutation relation of the SO$(N)_1$ Kac-Moody algebra. For the model (\ref{Eq-action}) to be self-consistently defined, we require this commutation relation on a constant time slice to persist when the chiral Majorana fermions $\psi_i$ are no longer free ($J>0$).

The Hamiltonian of the model can be obtained via the Legendre transformation as
\begin{equation}\label{Eq-H}
H=\int dx\mathcal{H}\ ,\qquad \mathcal{H}=-\frac{i}{2}\sum_{i=1}^N\psi_i\partial_x\psi_i -\sum_{1\le i<j<k<l\le N}J_{ijkl}\psi_i\psi_j\psi_k\psi_l\ ,
\end{equation}
where $\mathcal{H}$ is the energy density. Note that the kinetic term can also be written into the Sugawara form, $-\frac{i}{2}\sum_{i}\psi_i\partial_x\psi_i= -\frac{\pi}{N-1} \sum_{i<j}\normord{\psi_i\psi_j}\normord{\psi_i\psi_j}$ (Appendix \ref{App-OPE}). Therefore, the kinetic term and the 4-fermion interaction term $\psi_i\psi_j\psi_k\psi_l$ are of the same 4-fermion form (with scaling dimension $2$ and conformal spin $2$). This suggests that the interaction couplings $J_{ijkl}$ are exactly marginal, which we will verify by calculations in the rest of our paper. This is in contrast to the nonchiral interaction in the Thirring model, which is either marginally irrelevant (for large $N$) \cite{Berkooz:2016cvq} or marginally relevant \cite{gross1974}.

We can also derive the energy momentum tensor $T^{\mu}_{\ \nu}$ of the model. By definition, $T^{0}_{\ 0}$ has the physical meaning of the energy density, so we have $T^{0}_{\ 0}=\mathcal{H}$ as given by Eq. (\ref{Eq-H}). The energy current $T^{x}_{\ 0}$ can be derived from the energy conservation law $\partial_tT^{0}_{\ 0}+\partial_xT^{x}_{\ 0}=i[H,\mathcal{H}]+\partial_xT^{x}_{\ 0}=0$. Making use of the equation of motion (\ref{Eq-EOM}), the commutation relation (\ref{Eq-commu2}) and the correlation (\ref{Eq-J2}), we can obtain the energy current averaged over all random couplings $J_{ijkl}$ (Appendix \ref{App-OPE})
\begin{equation}\label{Eq-current}
T^{x}_{\ 0}=\frac{i}{2}\sum_{i=1}^N\psi_i\left(\partial_t-\frac{J^2}{4\pi^2}\partial_x\right)\psi_i\ .
\end{equation}
In a similar way, we can derive the momentum density and the momentum current components of $T^{\mu}_{\ \nu}$ as (Appendix \ref{App-OPE})
\begin{equation}
T^{0}_{\ x}=-\frac{i}{2}\sum_{i=1}^N\psi_i\partial_x\psi_i\ ,\qquad T^{x}_{\ x}= \frac{i}{4}\sum_{i=1}^N\psi_i(\partial_t-\partial_x)\psi_i\ .
\end{equation}
Due to the absence of Lorentz symmetry at $J>0$, the energy momentum tensor $T^{\mu}_{\ \nu}$ is not a symmetric tensor. We note that the expression of $T^{\mu}_{\ \nu}$ depends on the choice of the point splitting direction; here our results are for point splitting along the $x$ direction.

In a gapped $2+1$ dimensional condensed matter system with a chiral edge theory on its $1+1$ dimensional edge, the energy current $j_\mathcal{E}=\langle T^{x}_{\ 0}\rangle_\beta$ as a function of temperature $\beta^{-1}$ gives a thermal Hall conductance $\kappa_{xy}=\partial j_\mathcal{E}/\partial(\beta^{-1})$. For scaling invariant edge theories, the energy current is of the form $j_\mathcal{E}=\pi c/12\beta^2$, and yields a thermal Hall conductance $\kappa_{xy}=\pi c/6\beta$, where $c$ is a constant. For conformal field theories (CFTs), $c=c_R-c_L$ is the right-moving central charge $c_R$ minus the left-moving central charge $c_L$ \cite{kane1997}. In our case, the chiral model (\ref{Eq-action}) is a CFT of $N$ free chiral Majorana fermions when $J=0$, which has $c_R=N/2$ and $c_L=0$. Therefore, we conclude that our model has $\kappa_{xy}=N\pi/12\beta$ when $J=0$. When $J>0$, the model is no longer a CFT but is still scaling invariant, and we need to calculate the coefficient $c$ using Eq. (\ref{Eq-current}).

\section{Exact solution for $N=4$ via bosonization}\label{Sec-N=4}
The $1+1$ dimensional chiral SYK model (\ref{Eq-action}) with $N=4$ flavors of chiral Majorana fermions is integrable via bosonization, which is equivalent to the celebrated chiral Luttinger model with spin-charge separation in condensed matter physics. $N=4$ is also the minimal number of flavors which allows a 4-fermion interaction. In this section, we review this exact solution for $N=4$, and examine the two point function and OTOC of $\psi_i$, as a representative of non-chaotic integrable models. Furthermore, in Sec. \ref{Sec-56} we shall show the model is also integrable for $N=5$ and $N=6$, but is probably no longer integrable for all $N\ge7$.

Since there is only one coupling parameter $J_{1234}$ for $N=4$, it is meaningless to talk about the distribution of $J_{ijkl}$ as defined by Eq. (\ref{Eq-J2}). Instead, we simply set $J_{1234}=J\ge0$ as a constant. Note that this definition of $J$ is consistent with that defined in Eq. (\ref{Eq-J2}) for $N=4$.

It is convenient to define two chiral complex fermion fields $c_\uparrow=(\psi_1+i\psi_2)/\sqrt{2}$ and $c_\downarrow=(\psi_3+i\psi_4)\sqrt{2}$, distinguished by a spin index $\uparrow$ or $\downarrow$. The action (\ref{Eq-action}) for $N=4$ can then be rewritten as
\begin{equation}\label{Eq-complex}
S=\int dt dx \left(i\sum_{\sigma=\uparrow,\downarrow}c_\sigma^\dag(\partial_t+\partial_x)c_\sigma -J n_\uparrow n_\downarrow\right)\ ,
\end{equation}
where $n_\sigma=c^\dag_\sigma c_\sigma$ is the spin $\sigma$ fermion density. In this form, Eq. (\ref{Eq-complex}) is simply the chiral Luttinger model, which can be exactly solved by bosonization and exhibits a spin-charge separation physics. Physically, it could describe the edge of a Chern number $\nu=2$ quantum Hall state with a short-range repulsive interaction.

We comment that the $N=4$ action (\ref{Eq-complex}) has an SO(4) symmetry upon rotations among the four chiral Majorana fields $\psi_i$ \cite{yang1989,yang1990}. In contrast, there is generically no SO($N$) symmetry for $N\ge5$ due to the presence of couplings $J_{ijkl}$ which are non-symmetric under SO($N$) rotations. Therefore, the $N=4$ model is more symmetric.

\subsection{Bosonization of the action}\label{Sec-boson-action}
Here we briefly review the bosonization procedure of the model action (\ref{Eq-complex}). The bosonization is done by redefining the fermion operators in terms of vertex operators
\begin{equation}\label{Eq-boson}
c_{\sigma}(t,x)=\eta_\sigma e^{i\phi_\sigma(t,x)}\ ,\qquad c_{\sigma}^\dag(t,x)=\eta^\dag_\sigma e^{-i\phi_\sigma(t,x)}\quad (\sigma=\uparrow,\downarrow)\ ,
\end{equation}
where $\phi_\sigma(t,x)$ are two scalar boson fields satisfying the commutation relation
\begin{equation}
[\phi_\sigma(t,x),\phi_{\sigma'}(t,x')]=i\pi\delta_{\sigma\sigma'}\text{sgn}(x-x')\ ,
\end{equation}
with $\text{sgn}(x)$ denoting the sign of $x$. The coefficients $\eta_\sigma$ and $\eta_\sigma^\dag$ are the Klein factors \cite{haldane1981,heidenreich1980,delft1998}, which recover the anti-commutation relation between different flavors of fermion fields, and do not appear in fermion bilinears.
By imposing a point splitting regularization in the $x$ direction, one can derive the following normal ordered identities in an OPE expansion (Appendix \ref{App-OPE}):
\begin{equation}
n_\sigma=c^\dag_\sigma c_\sigma=\frac{\partial_x\phi_\sigma}{2\pi}\ ,\qquad -ic^\dag_\sigma \partial_x c_\sigma=\frac{(\partial_x\phi_\sigma)^2}{4\pi}\ .
\end{equation}
The action (\ref{Eq-complex}) can then be bosonized into
\begin{equation}
S=-\int dtdx\left(\frac{1}{4\pi}\sum_{\sigma=\uparrow,\downarrow}\partial_x\phi_\sigma(\partial_t\phi_\sigma +\partial_x\phi_\sigma)+ \frac{J}{4\pi^2}\partial_x\phi_\uparrow\partial_x\phi_\downarrow \right)\ .
\end{equation}
Note that the action $S$ only contains bilinears of $\phi_\sigma$, so the boson fields $\phi_\sigma$ are free. By redefining two new boson fields $\phi_\pm=(\phi_\uparrow\pm\phi_\downarrow)/\sqrt{2}$, we can rewrite the action as
\begin{equation}\label{Eq-bS}
S=-\int dtdx\left(\sum_{\alpha=\pm} \frac{1}{4\pi}\partial_x\phi_\alpha(\partial_t\phi_\alpha +u_\alpha\partial_x\phi_\alpha)\right)\ ,
\end{equation}
where $\phi_+$ and $\phi_-$ are two independent chiral boson fields decoupled from each other. The velocities $u_+$ and $u_-$ of the two chiral boson fields are given by
\begin{equation}\label{Eq-upm}
u_\pm=1\pm\frac{J}{2\pi}\ ,
\end{equation}
respectively. From this solution, we see that the Lorentz symmetry is indeed explicitly broken when $J>0$, since there are two different ``speeds of light" $u_+$ and $u_-$. Besides, we also see that the interaction $J$ is exactly marginal as we expected, which simply modifies the boson velocities $u_\pm$ in an energy scale independent way.

\begin{figure}[htbp]
\begin{center}
\includegraphics[width=6in]{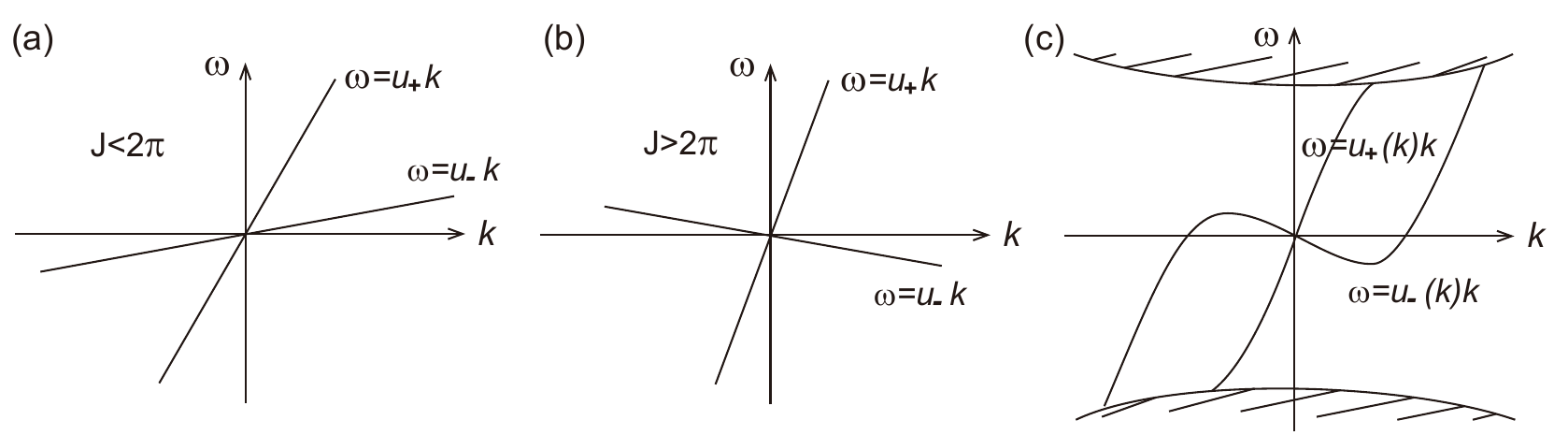}
\end{center}
\caption{(a) Illustration of the dispersions $\omega=u_\pm k$ of the two free bosons $\psi_\pm$ for $0<J<2\pi$. (b) The dispersions at $J>2\pi$, which has an incorrect total chirality and is unphysical. (c) Illustration of the physical understanding of the $J>2\pi$ case when the chiral model (\ref{Eq-bS}) describes the edge of a bulk gapped condensed matter system. The shaded area denotes the bulk states. For $J>2\pi$, the chirality of $\phi_-$ is flipped at small $k$, but has to be reversed back at large $k$ by UV nonlinearities, so that the total chirality is preserved.}
\label{largeK}
\end{figure}

Eq. (\ref{Eq-upm}) also indicates a physical bound $0\le J<2\pi$ for the interaction strength $J$. This ensures that both velocities $u_\pm$ are positive, and thus the total chirality of the model is kept invariant, as shown in Fig. \ref{largeK}(a). If $J>2\pi$, one would have the chirality of the boson field $\phi_-$ flipped as shown in Fig. \ref{largeK}(b), which is unphysical. This is because the action (\ref{Eq-complex}) is derived based on the normal ordering assumption that all the $k<0$ ($k>0$) fermion states are occupied (empty); while dispersions as shown in Fig. \ref{largeK}(b) would have both $k<0$ and $k>0$ occupied states and invalidate the normal ordering assumption.

We may still try to understand what happens when $J>2\pi$ if the $1+1$ dimensional chiral SYK model describes the edge theory of a bulk gapped $2+1$ dimensional condensed matter system. Physically, as long as the bulk gap does not close, the chirality of the edge should be preserved, as protected by the (Bogoliubov-de Gennes) Chern number of the ground state in the bulk. In fact, to understand the dispersion of $J>2\pi$, one has to take into account the irrelevant nonlinearities at high energies in a physical condensed matter system. For instance, a condensed matter system always has a spatial UV cutoff $\ell_0$ (lattice constant, etc), so one would expect the delta function interaction $Jn_\uparrow n_\downarrow$ to be smeared into a finite range $\ell_0$ interaction $\bar{J}(x-x')n_\uparrow(x)n_\uparrow(x')$, and $J=\int dx\bar{J}(x)$ only describes the interaction at the low energy limit. The model is then no longer scaling invariant. This will modify the velocities of the two boson fields $\phi_\pm$ into $u_\pm(k)=1\pm\tilde{J}(k)/2\pi$, where $k$ is the momentum, and $\tilde{J}(k)$ is the Fourier transform of $J(x)$. Due to the cutoff $\ell_0$, one has $\tilde{J}(k)\rightarrow J$ when $|k|\ll\ell_0^{-1}$, and $\tilde{J}(k)\rightarrow 0$ when $|k|\gtrsim\ell_0^{-1}$. Therefore, when $J>2\pi$, the dispersions of the edge chiral bosons up to very high energies will be as shown in Fig. \ref{largeK}(c), where the shaded area stands for the bulk states. In particular, the velocity of $\phi_-$ is negative near $k=0$, but goes back to positive when $|k|\gtrsim\ell_0^{-1}$. Therefore, the ground state of $J>2\pi$ is severely changed from that of $J<2\pi$, and the low energy theory becomes three right-moving chiral bosons and one left-moving chiral boson, which keeps the total chirality preserved but is no longer purely chiral.

In this paper, we shall only discuss interaction strengths $J$ inside the physical bound $0\le J<2\pi$, so that we do not need to worry about any UV nonlinearities, and the model is purely chiral and exactly scaling invariant. At zero temperature, the bosonic two point function (without time ordering) of $\phi_\alpha$ from the action (\ref{Eq-bS}) can be derived as
\begin{equation}
\langle \phi_\alpha(t,x)\phi_{\alpha'}(0,0)\rangle =-\delta_{\alpha\alpha'}\log\left[2\pi i(u_\alpha t-x-i0^+)\right]\ ,
\end{equation}
where $\alpha,\alpha'=\pm$, and $0^+$ stands for a real positive infinitesimal number. Since the bosons are free, their two point function at finite temperature $\beta^{-1}$ can be derived by summing over all the zero temperature two point functions at time $t+im\beta$ ($m\in\mathbb{Z}$):
\begin{equation}\label{Eq-b2p}
\begin{split}
\langle \phi_\alpha(t,x)\phi_{\alpha'}(0,0)\rangle_\beta &=\sum_{m=-\infty}^{\infty} \langle \phi_\alpha(t+im\beta,x)\phi_{\alpha'}(0,0)\rangle_{\beta=\infty} \\
&=-\delta_{\alpha\alpha'}\log\left\{2i\beta u_\alpha \sinh\left[\frac{\pi}{\beta}\left(t-u_\alpha^{-1}x-i0^+\right)\right]\right\}\ .
\end{split}
\end{equation}
All the $n$ point functions of boson fields $\phi_\pm$ can then be obtained from the above two point function and the Wick's theorem.

The bosonic action (\ref{Eq-bS}) allows us to calculate the energy momentum tensor $T^{\mu}_{\ \nu}$ and thermal quantities in terms of boson fields $\phi_{\alpha}$. For instance, the energy density and energy current are given by
\begin{equation}
T^0_{\ 0}=\frac{1}{4\pi}\sum_{\alpha=\pm} u_\alpha\partial_x\phi_\alpha\partial_x\phi_\alpha\ ,\qquad T^x_{\ 0}=-\frac{1}{4\pi}\sum_{\alpha=\pm} u_\alpha\partial_t\phi_\alpha\partial_x\phi_\alpha\ ,
\end{equation}
which can be derived from Noether's theorem. With the bosonic two point function in Eq. (\ref{Eq-b2p}), we can calculate the values of $T^0_{\ 0}$ and $T^x_{\ 0}$ at temperature $\beta^{-1}$ by performing a point splitting $\epsilon$ in the $x$ direction, and then take the limit $\epsilon\rightarrow0$. This leads to an energy density
\begin{equation}\label{Eq-bT00}
\mathcal{E}=\langle T^0_{\ 0}\rangle_\beta=-\frac{1}{4\pi}\sum_{\alpha=\pm} u_\alpha\partial_x^2\langle\phi_\alpha(t,x)\phi_\alpha(0,0)\rangle_\beta\Big|_{t\rightarrow 0,x\rightarrow \epsilon}=\frac{-1}{2\pi\epsilon^2}+\frac{\pi}{12\beta^2}(u_+^{-1}+u_-^{-1})+\mathcal{O}(\epsilon^2)\ ,
\end{equation}
and an energy current
\begin{equation}\label{Eq-bTx0}
j_\mathcal{E}=\langle T^x_{\ 0}\rangle_\beta=\frac{1}{4\pi}\sum_{\alpha=\pm} u_\alpha\partial_t\partial_x\langle\phi_\alpha(t,x)\phi_\alpha(0,0)\rangle_\beta\Big|_{t\rightarrow 0,x\rightarrow \epsilon}=-\frac{u_+^{2}+u_-^{2}}{4\pi\epsilon^2}+\frac{\pi}{6\beta^2} +\mathcal{O}(\epsilon^2)\ .
\end{equation}
The $\epsilon^{-2}$ terms in both expressions come from the vacuum expectation values, thus should be subtracted. Therefore, we find the physical energy density and energy current are
\begin{equation}\label{Eq-T4}
\mathcal{E}=\langle T^0_{\ 0}\rangle_\beta=\frac{\pi}{12\beta^2}(u_+^{-1}+u_-^{-1})\ ,\qquad j_\mathcal{E}=\langle T^x_{\ 0}\rangle_\beta=\frac{\pi}{6\beta^2}\ .
\end{equation}
In the bosonized representation here, different choice of point splitting direction in the above calculations will only affect the $\epsilon^{-2}$ terms, and does not affect the physical result of $\mathcal{E}$ and $j_\mathcal{E}$.

Eq. (\ref{Eq-T4}) leads to a thermal Hall conductance $\kappa_{xy}=\partial j_\mathcal{E}/\partial(\beta^{-1})=\pi/3\beta$. This is in agreement with the expectation that a gapped condensed matter system with $N$ flavors of chiral Majorana fermions on the edge has a quantized thermal Hall conductance $\kappa_{xy}=N\pi/12\beta$ \cite{kane1997}, where $N=4$ here. We can also derive the thermal entropy density $\mathcal{S}$ from $\beta^{-1}[\partial \mathcal{S}/\partial(\beta^{-1})]=\partial \mathcal{E}/\partial(\beta^{-1})$, which gives $\mathcal{S}=(\pi/6\beta)(u_+^{-1}+u_-^{-1})$. The zero temperature entropy density of the free bosons should be zero, which is used here.

\subsection{Fermion two point function for $N=4$}
Since we are mainly interested in fermion correlation functions in this paper, it is useful to refermionize the above bosonized picture for $N=4$. First, we derive the average fermion two point function (without time ordering) at temperature $\beta^{-1}$, which is defined as
\begin{equation}
G_\beta(t,x)=\frac{1}{4}\sum_{i=1}^4\langle \psi_i(t,x)\psi_i(0,0)\rangle_\beta=\frac{1}{4}\sum_{\sigma= \uparrow,\downarrow}\langle c_\sigma(t,x)c_\sigma^\dag(0,0)+ c_\sigma^\dag(t,x)c_\sigma(0,0)\rangle_\beta\ .
\end{equation}
By Eq. (\ref{Eq-boson}), we can replace $c_\sigma$ and $c_\sigma^\dag$ by vertex operators $e^{i\phi_\sigma}$ and $e^{-i\phi_\sigma}$, respectively. Then by Wick's theorem, one has $\langle e^{\pm i\phi_\sigma(t,x)}e^{\mp i\phi_\sigma(0,0)}\rangle_\beta=e^{\langle\phi_\sigma(t,x)\phi_\sigma(0,0)\rangle_\beta}$, so the fermion two point function can be derived as
\begin{equation}\label{Eq-G4}
\begin{split}
G_\beta(t,x)&
= \frac{1}{2}\sum_{\sigma= \uparrow,\downarrow}e^{\langle\phi_\sigma(t,x)\phi_\sigma(0,0)\rangle_\beta} =\prod_{\alpha=\pm} e^{\frac{1}{2}\langle\phi_\alpha(t,x)\phi_\alpha(0,0)\rangle_\beta}\\
=& \frac{1}{2i\beta\sqrt{u_+u_-}}\frac{1}{\sqrt{\sinh\left[ \frac{\pi}{\beta}(t-u_+^{-1}x-i0^+)\right] \sinh\left[ \frac{\pi}{\beta}(t-u_-^{-1}x-i0^+)\right]}}\ ,
\end{split}
\end{equation}
where we have used the two point function of boson fields $\phi_\pm=(\phi_\uparrow\pm\phi_\downarrow)/\sqrt{2}$ in Eq. (\ref{Eq-b2p}).

To verify Eq. (\ref{Eq-G4}) is the correct fermion two point function, we can calculate the finite temperature energy density $\mathcal{E}=\langle T^0_{\ 0}\rangle_\beta$ and energy current $j_\mathcal{E}=\langle T^x_{\ 0}\rangle_\beta$ in the fermion basis using $G_\beta(t,x)$ in Eq. (\ref{Eq-G4}). By the expression of $\langle T^0_{\ 0}\rangle_\beta=\mathcal{H}$ in Eq. (\ref{Eq-H}) and the equation of motion (\ref{Eq-EOM}), we can rewrite the energy density as
\begin{equation}
\langle T^0_{\ 0}\rangle_\beta=\langle -\frac{i}{2}\sum_{i=1}^4\psi_i\partial_x\psi_i -J\psi_1\psi_2\psi_3\psi_4 \rangle_\beta= \frac{i}{4}\sum_{i=1}^4\langle\psi_i(\partial_t-\partial_x)\psi_i\rangle_\beta\ .
\end{equation}
This expression involves products of fermion fields at the same spacetime point, which therefore requires a point-splitting regularization, which is chosen to be the $x$ direction in our paper. By performing a point splitting $\epsilon$ in the $x$ direction, the expression becomes a derivative of the two point function $G_\beta$, and we find the energy density given by
\begin{equation}\label{Eq-E4}
\mathcal{E}=\langle T^0_{\ 0}\rangle_\beta=-i(\partial_t-\partial_x) G_\beta(t,x)\Big|_{t\rightarrow 0, x\rightarrow\epsilon}=\frac{1}{\pi\epsilon^2}+\frac{\pi}{12\beta^2}(u_+^{-1}+u_-^{-1}) +\mathcal{O}(\epsilon^2)\ .
\end{equation}
Similarly, the energy current $j_\mathcal{E}=\langle T^x_{\ 0}\rangle$ defined by Eq. (\ref{Eq-current}) can also be rewritten as a derivative of $G_\beta$ via a point splitting $\epsilon$ in the $x$ direction, and can then be computed as
\begin{equation}\label{Eq-jE4}
j_\mathcal{E}=\langle T^x_{\ 0}\rangle_\beta=-2i\left(\partial_t-\frac{J^2}{4\pi^2}\partial_x\right) G_\beta(t,x)\Big|_{t\rightarrow 0, x\rightarrow\epsilon}=\frac{u_+^{-2}+u_-^{-2}}{2\pi\epsilon^2}+\frac{\pi}{6\beta^2} +\mathcal{O}(\epsilon^2)\ ,
\end{equation}
where the definition of $u_\pm$ in Eq. (\ref{Eq-upm}) is used. As one can see, the physical pieces of $\mathcal{E}$ and $j_\mathcal{E}$ proportional to $\beta^{-2}$ are consistent with the results in Eq. (\ref{Eq-T4}) derived in the boson picture, although their vacuum $\epsilon^{-2}$ terms do not agree with those in Eqs. (\ref{Eq-bT00}) and (\ref{Eq-bTx0}). This is because fermions and bosons have different vacuum energies. We note that in the fermionic picture here, it is important to have the point splitting direction in the $x$ direction.

\subsection{Fermion four point function for $N=4$}

We can further calculate any fermionic $n$ point functions for $N=4$ by rewriting them in terms of vertex operators of $\phi_\pm$. Here we calculate the averaged four point function
\begin{equation}
\mathcal{F}_4(t_1,x_1,\cdots,t_4,x_4)\equiv \frac{1}{4^2}\sum_{i,j=1}^4\langle \psi_j(t_1,x_1)\psi_i(t_2,x_2) \psi_j(t_3,x_3) \psi_i(t_4,x_4) \rangle_\beta
\end{equation}
for $N=4$. One can rewrite the fermion fields $\psi_i$ as vertex operators $(e^{i\phi_\sigma}\pm e^{-i\phi_\sigma})/\sqrt{2}$ of the boson fields $\phi_{\uparrow,\downarrow}$. By Wick's theorem, the $n$ point correlation of vertex operators $e^{iq_\alpha\phi_\alpha}$ satisfies $\langle \prod_{j=1}^{n}e^{iq_j\phi_{\alpha_j}}\rangle=\delta_{\sum_{j}q_j,0}\exp\left(-\sum_{\langle ij\rangle}q_iq_j\langle \phi_{\alpha_i}\phi_{\alpha_j} \rangle\right)$. With this formula and some rearrangements, we arrive at
\begin{equation}\label{Eq-4pf4}
\mathcal{F}_4(t_1,x_1,\cdots,t_4,x_4)=-\frac{1}{4}G_\beta^{13} G_\beta^{24}\left( \frac{G_\beta^{12}G_\beta^{34}}{G_\beta^{14}G_\beta^{23}} + \frac{G_\beta^{14}G_\beta^{23}}{G_\beta^{12}G_\beta^{34}} + \frac{\widetilde{G}_\beta^{12}\widetilde{G}_\beta^{34}}{\widetilde{G}_\beta^{14}\widetilde{G}_\beta^{23}} + \frac{\widetilde{G}_\beta^{14}\widetilde{G}_\beta^{23}}{\widetilde{G}_\beta^{12}\widetilde{G}_\beta^{34}}  \right)\ ,
\end{equation}
where we have defined the abbreviated notation $G_\beta^{ij}=G_\beta(t_i-t_j,x_i-x_j)$ for the two point function given by Eq. (\ref{Eq-G4}), while the function $\widetilde{G}_\beta^{ij}=\widetilde{G}_\beta(t_i-t_j,x_i-x_j)$ is defined by
\begin{equation}
\widetilde{G}_\beta(t,x)=\frac{e^{\frac{1}{2}\langle\phi_+(t,x)\phi_+(0,0)\rangle_\beta}} {e^{\frac{1}{2}\langle\phi_-(t,x)\phi_-(0,0)\rangle_\beta}}=\sqrt{\frac{u_-\sinh\left[ \frac{\pi}{\beta}(t-u_-^{-1}x-i0^+)\right]}{u_+\sinh\left[ \frac{\pi}{\beta}(t-u_+^{-1}x-i0^+)\right]}}\ .
\end{equation}
One can easily check that Eq. (\ref{Eq-4pf4}) reduces to the free fermion four point function when $J=0$, in which case $\widetilde{G}_\beta(t,x)\equiv1$. If we set $(t_2-i\frac{\beta}{2},x_2)=(t_4,x_4)=(0,0)$ and $(t_1-i\frac{\beta}{2},x_1)=(t_3,x_3)=(t +i\frac{\beta}{4},x)$, Eq. (\ref{Eq-4pf4}) becomes the definition of the regularized equal thermal circle separation OTOC (see Eq. (\ref{Eq-4pfInf})), which can be simplified as
\begin{equation}
\mathcal{F}_4(t,x)=\frac{1}{2\beta^2u_+u_-}\tanh\left[\frac{2\pi}{\beta}(t-u_+^{-1}x)\right] \tanh\left[\frac{2\pi}{\beta}(t-u_-^{-1}x)\right]\ .
\end{equation}

\begin{figure}[htbp]
\begin{center}
\includegraphics[width=6.5in]{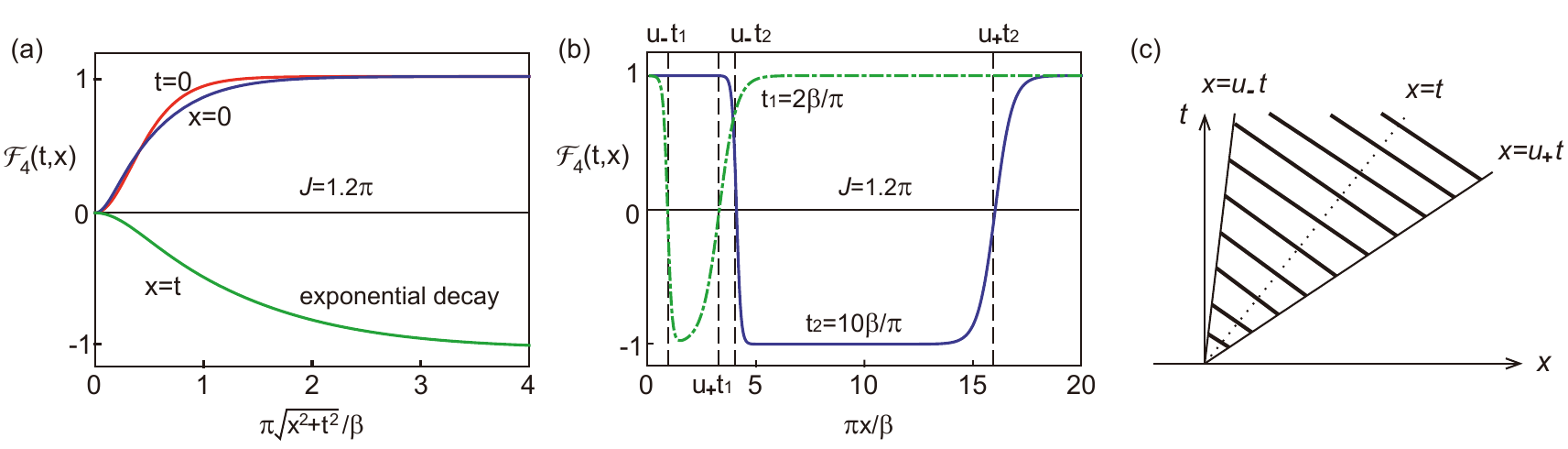}
\end{center}
\caption{(a) The normalized OTOC $\mathcal{F}_4(t,x)$ for $N=4$ along the $t$ direction ($x=0$), $x$ direction ($t=0$) and the $x=t$ line, where we have set $J=1.2\pi$. (b) The snapshot of the normalized OTOC $\mathcal{F}_4(t,x)$ as a function of $x$ at fixed time $t_1=2\beta/\pi$ (green dashed-dotted line) and $t_2=10\beta/\pi$ (blue solid line), respectively. The vertical dashed lines denote the positions $x/t=u_-$ and $x/t=u_+$ at time $t_1=10\beta/\pi$ and $t_2=10\beta/\pi$, respectively. (c) At late time $t$, The normalized OTOC $\mathcal{F}_4(t,x)$ approaches $-1$ in the shaded region $u_-t<x<u_+t$, and approaches $1$ outside the region. This shaded region is also the causal cone of the retarded Greens function for $N=4$.}
\label{4PC}
\end{figure}

Fig. \ref{4PC}(a) shows normalized $\mathcal{F}_4(t,x)$ along the $x=0$ direction, the $t=0$ direction and the $t=x$ line when $J=1.2\pi$. Fig. \ref{4PC}(b) shows the snapshot of normalized $\mathcal{F}_4(t,x)$ as a function of $x$ at fixed time $t=2\beta/\pi$ (green dashed-dotted line) and $t=10\beta/\pi$ (blue solid line), respectively. At late time $t$, one can see that $\mathcal{F}_4(t,x)$ decays to $-1$ in the region $u_-t<x<u_+t$ as denoted by the shaded area in Fig. \ref{4PC}(c), while decays to $1$ in the regions $x<u_-t$ and $x>u_+t$. Along a direction $x=vt$ with velocity away from $u_-$ and $u_+$, the OTOC $\mathcal{F}_4(t,x)$ always decays exponentially in $t$ towards $1$ or $-1$ for $t\gtrsim\beta$. This can be seen more explicitly by noting that $\mathcal{F}_4(t,x)$ can be expanded in nonnegative powers of the exponentially decaying function $e^{-2\pi t/\beta}$.
For integrable models, the OTOCs have no fundamental difference from the TOCs.

We note that the shaded region $u_-t<x<u_+t$ in Fig. \ref{4PC}(c) is exactly the causal cone of the model, which can be seen from the retarded fermion Green's function
\begin{equation}
G_R(t,x)= \frac{1}{\beta\sqrt{u_+u_-}}\frac{\Theta(t-u_+^{-1}x)\Theta(u_-^{-1}x-t)}{\sqrt{\sinh\left[ \frac{\pi}{\beta}(t-u_+^{-1}x)\right] \sinh\left[ \frac{\pi}{\beta}(u_-^{-1}x-t)\right]}}\ ,
\end{equation}
obtained by analytical continuation of the two point function (\ref{Eq-G4}). Therefore, the behavior of the OTOC function $\mathcal{F}_4(t,x)$ is intuitively understandable. Outside the causal cone in the large $t$ or large $x$ limit, the two $\psi_i$ fields do not communicate with the two $\psi_j$ fields, so $\mathcal{F}_4(t,x)$ tends to two decoupled two point functions $-G^{13}_\beta G^{24}_\beta$, which is a constant. Inside the causal cone, one would expect the correlations between the two $\psi_i$ fields and the two $\psi_j$ fields to be large, and the OTOC $\mathcal{F}_4(t,x)$ should strongly deviate from the decoupled value outside the causal cone.

\section{The chiral SYK model in the large $N$ limit}\label{Sec-SYK}
Now we proceed to study the $1+1$ dimensional chiral SYK model in the large $N$ limit, and examine its analogy and difference compared with the $0+1$ dimensional SYK model. We show that the two point function of chiral Majorana fermions is exactly solvable in the large $N$ limit, and the interaction strength $J$ is exactly marginal. We will also show the model with a nonzero interaction strength within the physical parameter range $0<J<2\pi$ is chaotic in the large $N$ limit, and its maximal velocity dependent Lyapunov exponent in the OTOC approaches the maximal chaos bound $2\pi/\beta$ when $J\rightarrow 2\pi$. The butterfly cone of the OTOC is asymmetric.  For any $0<J<2\pi$, the time direction Lyapunov exponent at the right butterfly edge always saturates the chaos bound $2\pi/\beta$, while at the left butterfly edge the time direction Lyapunov exponent is always $-2\pi/\beta$.


\subsection{Schwinger-Dyson equation and two point function}
In this subsection, we compute the two point function of the $1+1$ dimensional chiral SYK model in the large $N$ limit by exactly solving the Schwinger-Dyson equation. For convenience, we will work in Euclidean spacetime where the time $\tau=it$ is imaginary. Since the model is translationally invariant in both space and time, its two point function $G(-i\tau_1,x_1;-i\tau_2,x_2)=G[-i(\tau_1-\tau_2),(x_1-x_2)]$ will only depend on the spacetime coordinate difference of the two points. Here we define $G(-i\tau,x)$ to be the imaginary time ordered two point Green's function averaged over all Majorana fermion flavors. For zero temperature, it is defined by
\begin{equation}
\begin{split}
G(-i\tau,x)&\equiv\frac{1}{N}\sum_{i=1}^N\langle T \psi_i(\tau,x)\psi_i(0,0)\rangle = \frac{1}{N}\sum_{i=1}^N \langle\psi_i(\tau,x)\psi_i(0,0)\Theta(\tau)-\psi_i(0,0) \psi_i(\tau,x)\Theta(-\tau)\rangle ,
\end{split}
\end{equation}
where $T$ stands for time ordering, $\tau=it$ is the imaginary time, $\psi_i(\tau,x)=e^{H\tau}\psi_i(x)e^{-H\tau}$ is the Majorana field at imaginary time $\tau$, and $\Theta(\tau)$ is the Heaviside step function. For finite temperature $\beta^{-1}$, the definition becomes
\begin{equation}
G(-i\tau,x)\equiv\frac{1}{N}\sum_{i=1}^NZ^{-1}\text{Tr}\left[e^{-\beta H}T\psi_i(\tau,x)\psi_i(0,0)\right] ,
\end{equation}
where $Z=\text{Tr} e^{-\beta H}$ is the partition function, and $T$ still represents time ordering. Due to the anticommutation of fermions, the two point function satisfies the Kubo-Martin-Schwinger condition $G(-i\tau,x)=-G(-i\tau-i\beta,x)$.

The analytical continuation of $G(-i\tau,x)$ allows us to derive different kinds of real time Green's functions. The real time two point function without time ordering $G(t,x)$ can be obtained by substituting $\tau\rightarrow it+0^+$, namely, $G(t,x)\equiv G(t-i0^+,x)$. In Sec. \ref{Sec-chaos}, we will need to use the retarded Green's function $G_{R}(t,x)=N^{-1}\sum_i\langle\{\psi_i(t,x),\psi_i(0,0)\}\rangle =\Theta(t)[G(t-i0^+,x)-G(t+i0^+,x)]$,
and the Wightman function with half thermal circle separation defined by $G_{lr}(t,x)=G(t-i\frac{\beta}{2},x)$. Besides, the real-time ordered Green's function (i.e., the Feynman propagator) is given by $G_{T}(t,x)=G(t-i0^+\text{sgn}(t),x)$.

\subsubsection{The zero temperature solution}\label{Sec-zero-solution}

We first solve the Schwinger-Dyson equation for the two point function $G(-i\tau,x)$ at zero temperature. As we will see, the scaling invariance allows us a short cut to obtain the zero temperature solution exactly.

The imaginary-time ordered Green's function of free chiral Majorana fermions at zero temperature can be easily derived to be
\begin{equation}
G^f(-i\tau,x)=\frac{1}{N}\sum_{i=1}^N\langle T \psi_i(\tau,x)\psi_i(0,0)\rangle_f=\frac{1}{2\pi} \frac{1}{\tau-ix}\ ,
\end{equation}
which has a Fourier transform
\begin{equation}
G^f(i\omega_\tau,k)=\int d\tau dx e^{-i\omega_\tau\tau-ikx} G^f(-i\tau,x)=\frac{-i}{\omega_\tau-ik}\ .
\end{equation}
Here $\omega_\tau$ is the frequency along the imaginary time $\tau$, and $k$ is the momentum along the spatial direction.

\begin{figure}[htbp]
\begin{center}
\includegraphics[width=6in]{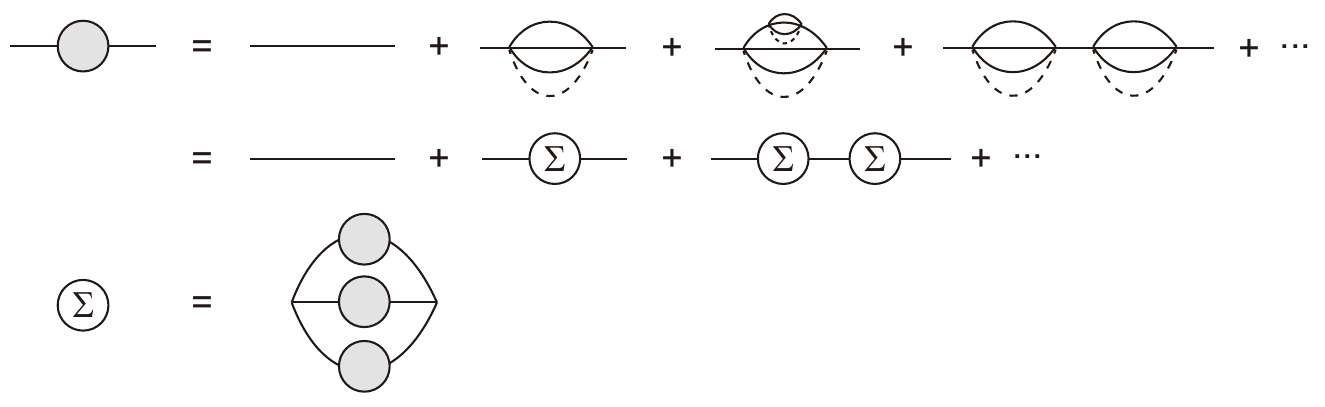}
\end{center}
\caption{The melon diagrams which contribute to the two point function to the leading order of $1/N$, which can be rewritten into the Schwinger-Dyson equation (\ref{Eq-SD}).}
\label{melon}
\end{figure}

When the interaction strength $J>0$, similar to the $0+1$ dimensional SYK model, the two point function $G$ to the leading order of $1/N$ is contributed by the melon diagrams as shown in Fig. \ref{melon}. The solid lines and dashed lines in Fig. \ref{melon} denote the free two point function $G^{f}$ and the average over $J_{ijkl}$, respectively. By defining a translationally invariant self energy $\Sigma(-i\tau,x)$, we can summarize the Feynman diagams into the Schwinger-Dyson equation for the imaginary time ordered two point function $G$ and the self energy $\Sigma$:
\begin{equation}\label{Eq-SD}
\frac{1}{G(i\omega_\tau,k)}=i\omega_\tau+k-\Sigma(i\omega_\tau,k)\ ,\qquad\qquad \Sigma(-i\tau,x)=J^2[G(-i\tau,x)]^3\ .
\end{equation}
The scaling invariance of the action (\ref{Eq-action}) tells us that $G(-i\omega_\tau,k)$ and $\Sigma(-i\omega_\tau,k)$ at zero temperature have scaling dimensions $-1$ and $+1$, respectively.
Therefore, one cannot ignore the $i\omega_\tau+k$ term in the first equation of (\ref{Eq-SD}) in any cases, which has the same scaling dimension as $\Sigma(i\omega_\tau,k)$. As a consequence, the Schwinger-Dyson equation of the $1+1$ dimensional chiral SYK model here does not have a reparametrization symmetry, which is present at low energies in the $0+1$ dimensional SYK model Schwinger-Dyson equation because of the irrelevance of the $i\omega_\tau$ term therein.

The exact scaling invariance, however, helps us constrain the form of $G$ and $\Sigma$.
By introducing the polar coordinates $(\kappa,\theta_k)$ in momentum space
\begin{equation}
\omega_\tau+ik=\kappa e^{i\theta_k}\ ,
\end{equation}
we can write down the following ansatz for $G$ and $\Sigma$ based on their scaling dimensions:
\begin{equation}\label{Eq-ansatz}
\Sigma(i\omega_\tau,k)=\kappa f(e^{i\theta_k})\ ,\qquad \qquad G(i\omega_\tau,k)=\frac{-i}{\kappa} \frac{1}{e^{-i\theta_k}+if(e^{i\theta_k})}\ ,
\end{equation}
where $f(z)=-f(-z)$ is an odd function (corresponding to fermionic statistics) defined on the unit complex circle $|z|=1$. The unknown part of the two point function $G$ and self energy $\Sigma$ is only its angular dependence on $\theta_k$, which comes form the Lorentz symmetry breaking by the interaction $J$. Note that the form of this ansatz already satisfies the first equation of Eq. (\ref{Eq-SD}).

\begin{figure}[htbp]
\begin{center}
\includegraphics[width=2in]{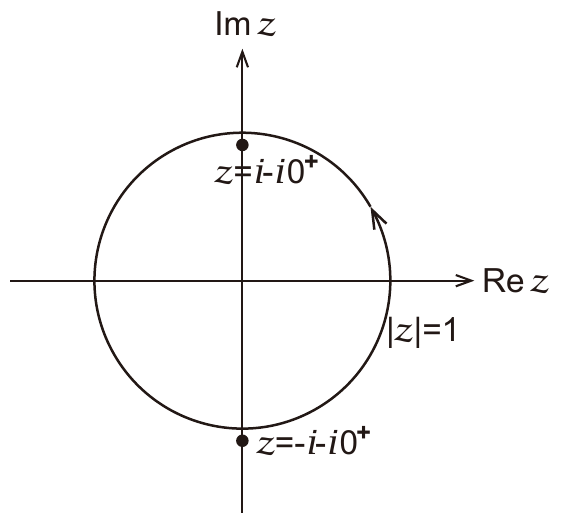}
\end{center}
\caption{The integration contour $|z|=1$ used in Eqs. (\ref{Eq-Gz}) and (\ref{Eq-Sz0}).}
\label{contour}
\end{figure}

The remaining task is to solve for the function $f(z)$ from the second equation of Eq. (\ref{Eq-SD}). This is an equation in real space, so we need to Fourier transform our ansatz (\ref{Eq-ansatz}). For convenience, we also use polar coordinates $(r,\theta)$ for real space defined by
\begin{equation}
\tau+ix=re^{i\theta}\ .
\end{equation}
By integrating along the radial coordinate $\kappa$ first, we can write the Green's function $G$ in real space as
\begin{equation}
\begin{split}
G(-i\tau,x)=&\int\frac{d\omega_\tau dk}{(2\pi)^2}e^{i\omega_\tau\tau+ikx}G(i\omega_\tau,k) =\int_0^{2\pi}\frac{d\theta_k}{(2\pi)^2}\int_0^\infty d\kappa \frac{-i e^{i\kappa r \cos(\theta-\theta_k)}}{e^{-i\theta_k}+if(e^{i\theta_k})}\\
=& \frac{1}{r}\int_0^{2\pi}\frac{d\theta_k}{(2\pi)^2} \frac{1}{[e^{-i\theta_k}+if(e^{i\theta_k})][\cos(\theta_k-\theta)+i0^+]}\ .
\end{split}
\end{equation}
By defining $w=e^{-i\theta}$ and denoting $z=we^{i\theta_k}$, we can view the above integral as a contour integral along the circle $|z|=1$ in the complex $z$ plane. By assuming $f(z)$ can be analytically continued to the complex $z$ plane as a holomorphic function, and has no branch points or intrinsic singularities for $|z|\le1$, we can perform the contour integral to obtain
\begin{equation}\label{Eq-Gz}
\begin{split}
&G(-i\tau,x)=-\frac{i}{rw}\oint_{|z|=1}\frac{dz}{(2\pi)^2}\frac{2}{(z+z^{-1}+i0^+)[1+izw^{-1}f(zw^{-1})]}\\
=&\frac{1}{2\pi r}\left\{\frac{1}{w-f(iw^{-1})}-\sum_{|\xi|<1,\xi f(\xi)=i} \frac{2iw\xi}{(1+w^2\xi^2)[f(\xi)+\xi f'(\xi)]}\right\}\ ,
\end{split}
\end{equation}
where $f'(z)$ denotes the first derivative of $f(z)$. The integral picks up the residues at the pole $z=i-i0^+$ as shown in Fig. \ref{contour} and other possible poles $\xi$ satisfying $\xi f(\xi)=i$ inside the unit circle. Whether or not such additional poles $\xi$ exist depends on the unknown function $f(z)$. The pole at $z=-i-i0^+$ is outside the contour and thus has no contribution.

The Fourier transform of the self energy $\Sigma$ to the real space can be calculated in a similar way. First, we can integrate over the radial momentum $\kappa$ to rewrite the Fourier transform of $\Sigma$ as a contour integral:
\begin{equation}\label{Eq-Sz0}
\begin{split}
&\Sigma(-i\tau,x)=\int\frac{d\omega_\tau dk}{(2\pi)^2}e^{i\omega_\tau\tau+ikx}\Sigma(i\omega_\tau,k) =\int_0^{2\pi}\frac{d\theta_k}{(2\pi)^2}\int_0^\infty d\kappa \kappa^2 f(e^{i\theta_k}) e^{i\kappa r \cos(\theta-\theta_k)}\\
=&-\frac{2i}{r^3}\int_0^{2\pi}\frac{d\theta_k}{(2\pi)^2} \frac{f(e^{i\theta_k})}{[\cos(\theta_k-\theta)+i0^+]^3} =-\frac{4}{\pi^2r^3}\oint_{|z|=1} \frac{f(zw^{-1})dz}{z(z+z^{-1}+i0^+)^3}\ ,\\
\end{split}
\end{equation}
where we again defined $w=e^{-i\theta}$ and $z=we^{i\theta_k}$, and the integral contour is the unit circle $|z|=1$ as shown in Fig. \ref{contour}. In this case, the integrand has two poles at $z=\pm i-i0^+$, and other possible poles from the unknown function $f(z)$. By adding up the residues of poles inside the contour, we find the contour integration leads to a real space self energy function
\begin{equation}\label{Eq-Sz}
\Sigma(-i\tau,x)=\frac{1}{2\pi r^3}\left[-f(iw^{-1})+iw^{-1}f'(iw^{-1})-w^{-2}f''(iw^{-1})-\sum_{|\xi|<1} \frac{16iw^2\xi^2 \text{Res}f(\xi)}{(w^2\xi^2+1)^3}\right]\ ,
\end{equation}
where $f''$ and $f'$ are the second and first derivatives of $f$, respectively, and $\xi$ runs over all poles of the function $f(z)$ in the unit disk $|z|<1$.

For now we assume the function $f(z)$ in the unit disk $|z|<1$ satisfies $|zf(z)|<1$ and has no intrinsic singularities or branch points. We will come back to verify this after we obtain the solution of $f(z)$. This assumption ensures that $f(z)$ has at most one pole at $z=0$, and the last residue terms contributed by $|\xi|<1$ in Eq. (\ref{Eq-Gz}) and in Eq. (\ref{Eq-Sz}) all vanish. Therefore, the expressions of $G(-i\tau,x)$ in Eq. (\ref{Eq-Gz}) and $\Sigma(-i\tau,x)$ in Eq. (\ref{Eq-Sz}) are greatly simplified, and the second Schwinger-Dyson equation in Eq. (\ref{Eq-SD}) becomes a second order ordinary differential equation (ODE) for $f(z)$:
\begin{equation}
f(z)-zf'(z)-z^2f''(z)=\frac{J^2}{4\pi^2}\frac{-iz^3}{[1+izf(z)]^3}\ .
\end{equation}
By changing the variable $z$ to $s=z^2$, and defining a new function $g(s)=-izf(z)$, we can further simplify the above differential equation into
\begin{equation}\label{Eq-dgds}
\frac{d^2g}{ds^2}=\frac{J^2}{16\pi^2}\frac{1}{(1-g)^3}\ .
\end{equation}
Such a second order ODE can be solved by two steps of integration. First, we note that the second derivative of $g$ can be rewritten as $\frac{d^2g}{ds^2}=\frac{dg}{ds} \frac{d}{dg} \Big(\frac{dg}{ds}\Big)=\frac{1}{2}\frac{d}{dg}\Big[\Big(\frac{dg}{ds}\Big)^2\Big]$. Therefore, we can integrate Eq. (\ref{Eq-dgds}) over $g$ to obtain
\begin{equation}
\Big(\frac{dg}{ds}\Big)^2=\frac{J^2}{16\pi^2}\Big[\frac{1}{(1-g)^2}-c_0\Big]\ ,
\end{equation}
where $c_0$ is the constant of integration. The square root of the above equation is a simple first order ODE. The second step is then to solve this first order ODE by direct integration, after which we arrive at a general solution
\begin{equation}
g(s)=1-\sqrt{\frac{1}{c_0}-\frac{c_0J^2}{16\pi^2}(s-s_0)^2}\ ,
\end{equation}
where $s_0$ is the constant of integration in the second step. The function $f(z)$ is then given by
\begin{equation}\label{eqn-f}
	f(z)=iz^{-1}g(z^2)=iz^{-1}\left [1-\sqrt{{1\over c_0}-{c_0 J^2\over 16\pi ^2}(z^2-s_0)^2}\right ]\ .
\end{equation}

Next, we need to determine the constants $c_0$ and $s_0$. When $J=0$, the system is free, so $f(z)$ should vanish identically, which is proportional to the self energy $\Sigma(-i\omega_\tau,k)$. This fixes the constant $c_0=1$. The other constant $s_0$ is determined by the choice of point splitting. As we mentioned in Sec. \ref{Sec-PS}, we have chosen the point splitting in the spatial $x$ direction when regularizing the OPE of $\psi_i$, which leads to the commutation relation (\ref{Eq-commu2}). The commutation relation then requires the Green's function $G(-i\tau,x)$ on the constant time slice $\tau=0$ to be the same as the free Green's function, namely, $G(0,x)=G^f(0,x)=\frac{i}{2\pi x}$. From Eq. (\ref{Eq-Gz}), this requires $f(z)=0$ when $z=\pm1$. Therefore, we find the constant of integral $s_0=1$. This fixes the form of the solution to
\begin{equation}\label{Eq-fz}
f(z)=iz^{-1}g(z^2)=f(z)=iz^{-1}\left[1-\sqrt{1-\frac{J^2}{16\pi^2}(z^2-1)^2}\right]\ .
\end{equation}
In particular, our assumption in solving the Schwinger-Dyson equation, that $|zf(z)|<1$ and $f(z)$ has no intrinsic singularities or branch points when $|z|<1$, is satisfied for $0\le J<2\pi$. Therefore, $2\pi$ is the upper bound of $J$ for the above solution (\ref{Eq-fz}) to be self-consistent. We will discuss more on the bound of $J$ at the end of this subsection.

From Eqs. (\ref{Eq-Gz}) and (\ref{Eq-Sz}), we can then obtain the real space two point Green's function in the large $N$ limit as
\begin{equation}\label{Eq-Gxt}
G(-i\tau,x)= \frac{1}{2\pi r}\frac{1}{e^{-i\theta}-f(ie^{i\theta})} =\frac{1}{2\pi}\frac{1}{\sqrt{(u_+\tau-ix)(u_-\tau-ix)}}\ ,
\end{equation}
and the real space self energy
\begin{equation}\label{Eq-Sxt}
\Sigma(-i\tau,x)=\frac{J^2}{8\pi^3}\frac{1}{[(u_+\tau-ix)(u_-\tau-ix)]^{3/2}}\ ,
\end{equation}
where the two velocities are $u_\pm=1\pm J/2\pi$. The above expressions of $G$ and $\Sigma$ have branch cuts so the function is not uniquely defined unless specified. Here we specify the branch of their definitions by requiring $G(-i\tau,x)\rightarrow {i\over 2\pi x}$ and $\Sigma(-i\tau,x)\rightarrow -{i J^2\over 8\pi^3 x^3}$ when $|x|\gg |\tau|$, and arranging the branch cut to be the straight line segment from $-iu_-\tau$ to $-iu_+\tau$. This procedure uniquely determines the functions $G$ and $\Sigma$ (except for the singular point $x=\tau=0$).

\begin{figure}[htbp]
\begin{center}
\includegraphics[width=3in]{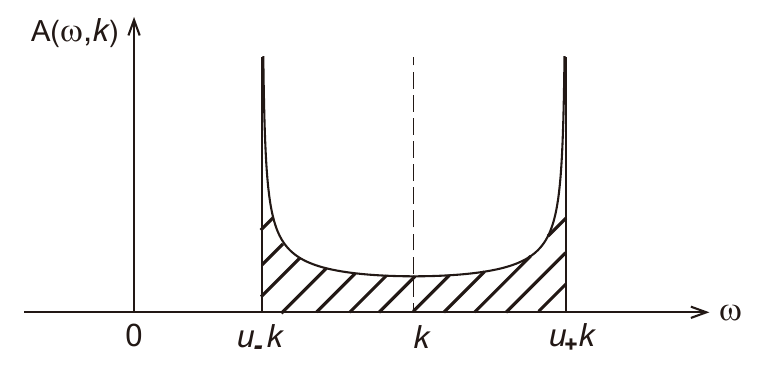}
\end{center}
\caption{The zero temperature spectral weight of the $1+1$ dimensional chiral SYK model in the large $N$ limit at a given momentum $k>0$.}
\label{spectral}
\end{figure}

Remarkably, the two point function in the large $N$ limit in Eq. (\ref{Eq-Gxt}) has exactly the same form as that of $N=4$ given in Eq. (\ref{Eq-G4}) at zero temperature, except that the meanings of $J$ are different: here $J$ is the Gaussian average amplitude of the random couplings $J_{ijkl}$, while for $N=4$ $J$ is defined as the only coupling $J_{1234}$. The form of this solution also indicates that the coupling strength $J$ is exactly marginal, which determines the two velocities $u_\pm$ irrespective of the energy scale.

Besides, similar to the $N=4$ case, the large $N$ solution here also requires a physical bound of the interaction strength $0\le J<2\pi$, which ensures the two velocities $u_\pm$ in Eq. (\ref{Eq-Gxt}) are positive. The physical reason for this bound is analogous to that we explained for $N=4$ in Fig. \ref{largeK}, namely, the total chirality of the system would no longer be preserved for $J>2\pi$ unless certain UV nonlinearities are taken into account. More explicitly, this can be seen from the zero temperature spectral weight obtained by
\begin{equation}
A(\omega,k)=2\text{Im}G(\omega+i0^+,k) =\frac{2\Theta(u_+k-\omega)\Theta(\omega-u_-k)}{\sqrt{(u_+k-\omega)(\omega-u_-k)}}\ ,
\end{equation}
where $G(\omega+i0^+,k)$ is the momentum space retarded Green's function obtained by doing an analytical continuation $i\omega_\tau\rightarrow \omega+i0^+$ of the imaginary time ordered Green's function $G(i\omega_\tau,k)$. Fig. \ref{spectral} shows the spectral weight $A(\omega,k)$ at a fixed momentum $k>0$, which indicates the eigenstate energies of the model at momentum $k$ are distributed within the energy range $[u_-k,u_+k]$. Therefore, if $J>2\pi$, one would have $u_-<0$, and part of the eigenstates will have a negative velocity and reversed chirality. This would lead to a severe change of ground state similar to that illustrated by Fig. \ref{largeK}(c), which cannot be resolved unless certain UV nonlinearities are considered. Here we shall only restrict ourselves in the parameter range $0\le J<2\pi$, so that we do not need to consider any UV nonlinearities.

\subsubsection{Discussion on the real space UV regularization}\label{Sec-Forder}
in Sec. \ref{Sec-zero-solution}, we derived the zero temperature solution (\ref{Eq-Gxt},\ref{Eq-Sxt}) by performing Fourier transforms of $G$ and $\Sigma$ from the momentum space to the real space. In particular, we determined the constant $s_0$ in Eq. (\ref{eqn-f}) by considering the $x$ direction point splitting regularization we chose. This indicates that the real space UV regularization of this model has important physical consequences on correlation functions, and thus has to be handled correctly. An immediate example is to transform the real space $G$ and $\Sigma$ back to the momentum space: since the self energy $\Sigma$ diverges as $1/r^3$ in the real space, its Fourier transform requires a UV regularization, and different UV regularization schemes will give different results as we will see below. In this subsection, we will show that the correct UV regularization scheme for doing real space integrations in consistency with point splitting in $x$ direction is to take a UV cutoff of time $|\tau|\ge\epsilon$. The readers who are solely interested in the main results of the paper can skip this subsection and go directly to Sec. \ref{sec:SDfiniteTemp}.

One may also have noticed that the solution (\ref{Eq-Gxt}) factorizes into the product of two conformal propagators of dimension ${1\over 4}$ and velocities $u_{\pm}$. Naively, this looks like two decoupled $0+1$ dimensional SYK systems moving at different velocities $u_\pm$. However, this cannot be true, since otherwise they will satisfy the low-energy approximate Schwinger-Dyson equation of the $0+1$ dimensional SYK model, rather than the exact $1+1$ dimensional Schwinger-Dyson equation (\ref{Eq-SD}) here. In fact, it is exactly the UV regularization we will discuss below that invalidates such a factorization into two $0+1$ dimensional SYK systems.

\begin{figure}[htbp]
\begin{center}
\includegraphics[width=2in]{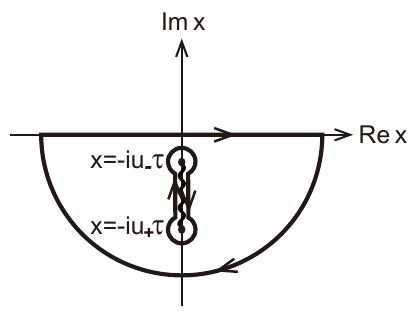}
\end{center}
\caption{Deformation of the integration contour done in Eq. (\ref{Eq-FourierTrans1}) from real $x$ axis closing in lower half plane to the branch cut on the imaginary axis.}
\label{contour2}
\end{figure}

Let us start by considering the Fourier transformations of the real space $G$ and $\Sigma$ in Eqs. (\ref{Eq-Gxt},\ref{Eq-Sxt}) back into the momentum space. As we have said, the Fourier transformation of $\Sigma$ requires a UV regularization. If the UV regularization is correct, we should be able to find that the Fourier transformed $G$ and $\Sigma$ satisfy the first Schwinger-Dyson equation in Eq. (\ref{Eq-SD}).

We first examine the regularization scheme of taking a UV cutoff $|\tau|\ge\epsilon$ in the time direction, where $\epsilon>0$ is a small number. In this case, the Fourier transformations of both $G$ and $\Sigma$ are proportional to the generic form
\begin{equation}
W_\sigma(-i\omega_\tau,k)=\int_{-\infty}^\infty d\tau e^{-i\omega_\tau\tau}\Theta(|\tau|-\epsilon) \int_{-\infty}^\infty dx e^{-ikx} (u_+\tau-ix)^\sigma(u_-\tau-ix)^\sigma\ ,
\end{equation}
where $\sigma$ is a constant. For $\sigma=-1/2$ and $\sigma=-3/2$, one obtains the momentum space two point function $G(-i\omega_\tau,k)=\frac{1}{2\pi}W_{-\frac{1}{2}}(-i\omega_\tau,k)$ and self energy $\Sigma(-i\omega_\tau,k)=\frac{J^2}{8\pi^3}W_{-\frac{3}{2}}(-i\omega_\tau,k)$, respectively.
Without loss of generality, here we assume $k>0$. The integration of $x$ is then along the contour of the real $x$ axis closing in the lower half plane as shown in Fig. \ref{contour2}. When $\tau<0$, the contour integral is zero. When $\tau>0$, the contour encloses a branch cut between $-iu_+\tau$ and $-iu_-\tau$, thus can be deformed into a contour closely surrounding to the branch cut in Fig. \ref{contour2}. This enables us to rewrite the Fourier transformation as
\begin{equation}\label{Eq-FourierTrans1}
W_\sigma(-i\omega_\tau,k)=-2\sin(\pi\sigma)\int_{\epsilon}^\infty d\tau e^{-i\omega_\tau\tau} \int_{u_-\tau}^{u_+\tau} dy e^{-ky} (u_+\tau-y)^\sigma(y-u_-\tau)^\sigma\ ,
\end{equation}
where we have defined $y=ix$. The integration can be further simplified by changing the variables of integration to $y_+=\frac{u_+\tau-y}{u_+-u_-}$ and $y_-=\frac{y-u_-\tau}{u_+-u_-}$, after which the region of integration is bound by $y_+\ge0$, $y_-\ge0$ and $y_++y_-\ge\epsilon$. The integral then takes the form
\begin{equation}\label{Eq-Wpm}
W_\sigma(-i\omega_\tau,k)=-2\sin(\pi\sigma)\left(\frac{J}{\pi}\right)^{1+2\sigma}\int_{0}^\infty dy_+ y_+^\sigma e^{-k_-y_+} \int_{\max(0,\epsilon-y_+)}^{\infty} dy_- y_-^\sigma e^{-k_+y_-}\ ,
\end{equation}
where we have defined $k_\pm=u_\pm k+i\omega_\tau$.  The real parts of $k_{\pm}$ are always positive, so the integral is well convergent in the IR.  We have two possible UV divergences, one is from the $\epsilon$ cutoff, and the other is from the end point $0$ of the integral.  For the latter, we can remind ourselves that the integral is actually a contour integral, and therefore we should take the principle value (or equivalently by analytic continuing $\sigma$).
While for the former, if we neglect the $\epsilon$ piece, the two integrals of $y_+$ and $y_-$ would decouple, and we would get an answer factorized into the product of two decoupled $0+1$ dimensional SYK Green's functions or self energies in the conformal limit along $y_+$ and $y_-$ direction. However, as we shall see, the $\epsilon$ piece is important to produce an additional contribution to the self energy $\Sigma$, which enables $\Sigma$ and $G$ to satisfy the $1+1$ dimensional Schwinger-Dyson equation exactly.

When $y_+<\epsilon$, the integral of $y_-$ yields a result $k_+^{-1-\sigma}\Gamma(1+\sigma,k_+(\epsilon-y_+))$, where $\Gamma(\sigma,\epsilon)=\int_\epsilon^\infty y^{\sigma-1}e^{-y}dy$ is the incomplete Gamma function. While when $y_+>\epsilon$, the integral of $y_-$ simply gives the usual Gamma function $k_+^{-1-\sigma}\Gamma(1+\sigma)$.
By expanding in $\epsilon-y_+$, we have
\begin{equation}\label{Eq-Wa}
\begin{split}
W_\sigma(-i\omega_\tau,k)=&-2\sin(\pi\sigma) \left(\frac{J}{\pi}\right)^{1+2\sigma}\Big\{\Gamma(1+\sigma)^2(k_+k_-)^{-1-\sigma} \\
&+ \int_{0}^\epsilon dy_+ y_+^\sigma e^{-k_-y_+}\Big[-\frac{(\epsilon-y_+)^{1+\sigma}}{1+\sigma} +\frac{k_+(\epsilon-y_+)^{2+\sigma}}{2+\sigma}+\cdots\Big]\Big\}\\
=& -2\sin(\pi\sigma) \left(\frac{J}{\pi}\right)^{1+2\sigma}\Big[\Gamma(1+\sigma)^2(k_+k_-)^{-1-\sigma}\\ & -\frac{\epsilon^{2+2\sigma}\Gamma(1+\sigma)^2}{\Gamma(3+2\sigma)}+ \frac{\epsilon^{3+2\sigma}\Gamma(1+\sigma)\Gamma(2+\sigma)}{\Gamma(4+2\sigma)}(k_++k_-)+\cdots\Big]\ ,
\end{split}
\end{equation}
where we have used the definition of beta function $B(a,b)=\frac{\Gamma(a)\Gamma(b)}{\Gamma(a+b)} =\int_0^1 y^{a-1}(1-y)^{b-1}dy$. For $\sigma=-1/2$, the integral (\ref{Eq-Wa}) is regular as $\epsilon\rightarrow0$, so the $\epsilon$ dependent terms can be ignored, and we find the Fourier transform of the two point function to be
\begin{equation}\label{Eq-Gx1t2}
G(-i\omega_\tau,k)=\frac{1}{2\pi}W_{-\frac{1}{2}}(-i\omega_\tau,k) =\frac{-i}{\sqrt{(\omega_\tau-iu_+k)(\omega_\tau-iu_-k)}}\ ,
\end{equation}
as expected from Eq. (\ref{Eq-ansatz}). This also indicates that the Fourier transform of $G$ does not require a UV regularization. For $\sigma=-3/2$, the leading $1/\epsilon$ divergent term of Eq. (\ref{Eq-Wa}) vanishes, and we arrive at the Fourier transform of the self energy
\begin{equation}\label{Eq-Sx1t2}
\begin{split}
\Sigma(-i\omega_\tau,k)=\frac{J^2}{8\pi^3}W_{-\frac{3}{2}}(-i\omega_\tau,k)
=i\omega_\tau+k -i\sqrt{(\omega_\tau-iu_+k)(\omega_\tau-iu_-k)}\ ,
\end{split}
\end{equation}
again in agreement with Eq. (\ref{Eq-ansatz}). We see that the UV cutoff $\epsilon$ gives rise to a regular piece $i\omega_{\tau}+k$, which is important for the answers to exactly satisfy the first Schwinger-Dyson equation in Eq. (\ref{Eq-SD}). This shows that UV cutoff $|\tau|\ge\epsilon$ is the correct regularization scheme, which we will give an understanding later in this subsection.

We could also try to use a different regularization scheme, for example, take a UV cutoff $|x|\ge\epsilon$ in the spatial direction. Following a similar calculation, one will arrive at an expression analogous to Eq. (\ref{Eq-Wpm}), except that the region of integration is bounded by $u_- y_++u_+ y_-\ge\epsilon$ and $y_\pm\ge0$. This does not affect the two point function $G$ and the $(k_+k_-)^{1/2}$ piece of the self energy $\Sigma$, but does change the UV contributed bare piece. More explicitly, it is easy to check that the resulting Fourier transforms in this case are
\begin{equation}\label{Eq-Gt1x2}
G'(-i\omega_\tau,k)=\frac{-i}{\sqrt{(\omega_\tau-iu_+k)(\omega_\tau-iu_-k)}}\ ,
\end{equation}
\begin{equation}\label{Eq-St1x2}
\begin{split}
&\Sigma'(-i\omega_\tau,k)=i\frac{\omega_\tau}{\sqrt{u_+u_-}}+\sqrt{u_+u_-}k -i\sqrt{(\omega_\tau-iu_+k)(\omega_\tau-iu_-k)}\ .
\end{split}
\end{equation}
which obviously do not satisfy the Schwinger-Dyson equation (\ref{Eq-SD}). In fact, if we instead set $s_0=-1$ in the solution (\ref{eqn-f}), which corresponds to a time $\tau$ direction point splitting, one will find that the spatial UV cutoff $|x|\ge\epsilon$ becomes the correct UV regularization.

The above discussion shows that the real space UV regularization scheme depends on the choice of point splitting direction, and incorrect UV regularization will lead to answers inconsistent with the Schwinger-Dyson equation. Here we give a physical understanding why UV cutoff $|\tau|\ge\epsilon$ is the correct regularization scheme for point splitting in the $x$ direction. A heuristic understanding is the following: taking the cutoff $|\tau|\ge\epsilon$ for the self energy $\Sigma(-i\tau,x)$ is equivalent to setting $\Sigma(0,x)=0$ on the constant time slice $\tau=0$. Recall that point splitting in the $x$ direction leads to the commutation relation (\ref{Eq-commu2}), and requires the two point function $G(-i\tau,x)$ to reduce to the free two point function $G(0,x)=G_f(0,x)=\frac{i}{2\pi x}$ on the constant time slice $\tau=0$ (which is how $s_0=1$ in Eq. (\ref{eqn-f}) is determined). This free nature of $G(0,x)$ therefore intuitively agrees with the vanishing of the self energy $\Sigma(0,x)=0$ on the time slice $\tau=0$.

\begin{figure}[htbp]
\begin{center}
\includegraphics[width=5in]{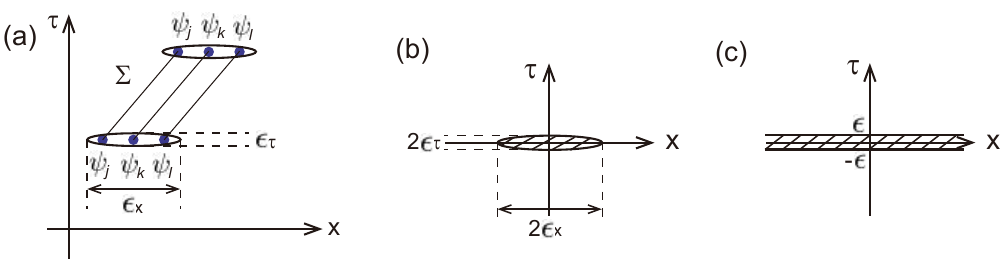}
\end{center}
\caption{Illustration of how point splitting in the $x$ direction leads to a UV cutoff $(x/\epsilon_x)^2+(\tau/\epsilon_\tau)^2\ge1$ with $\epsilon_\tau/\epsilon_x\rightarrow0$ for the self energy $\Sigma$, which is equivalent to the UV cutoff $|\tau|\ge\epsilon$ due to scaling symmetry.}
\label{fig:PointSplitting}
\end{figure}

A more intuitive explanation is illustrated by Fig. \ref{fig:PointSplitting}. In real space, the self energy $\Sigma(-i\tau,x)\sim J^2\langle (\psi_j\psi_k\psi_l)(0,0)(\psi_j\psi_k\psi_l)(\tau,x)\rangle$ is the correlation of the operator product $\psi_j\psi_k\psi_l$, the definition of which needs a point splitting regularization. For point splitting along the $x$ direction we have chosen, the three Majorana fields in $\psi_j\psi_k\psi_l$ should be splitted by a spatial spacing $\epsilon_x$ as shown in Fig. \ref{fig:PointSplitting}(a). Since each $\psi_i$ is a point-like operator, one could view the regularized $\psi_j\psi_k\psi_l$ as a ``flat-shape" operator with size $\epsilon_x$ in the $x$ direction and $\epsilon_\tau$ in the $\tau$ direction, where $\epsilon_\tau/\epsilon_x\rightarrow0$. With such a point splitting, the self energy correlation $\Sigma(-i\tau,x)$ is only meaningful when the two operators $(\psi_j\psi_k\psi_l)(0,0)$ and $(\psi_j\psi_k\psi_l)(\tau,x)$ do not overlap with each other, namely, when $(x/\epsilon_x)^2+(\tau/\epsilon_\tau)^2\ge1$. Therefore, when calculating the Fourier transformation of $\Sigma(-i\tau,x)$, one should take a UV cutoff $(x/\epsilon_x)^2+(\tau/\epsilon_\tau)^2\ge1$ with $\epsilon_\tau/\epsilon_x\rightarrow0$, i.e., perform the Fourier integration outside the shaded area as shown in Fig. \ref{fig:PointSplitting}(b).
This UV cutoff can be further simplified by noting that the zero temperature solution in Eqs. (\ref{Eq-Gxt}) and (\ref{Eq-Sxt}) is scaling invariant. Via a scaling transformation, we can scale $\epsilon_x\rightarrow\infty$ while keeping $\epsilon_t=\epsilon$ small, which still satisfies $\epsilon_t/\epsilon_x\rightarrow0$. The UV cutoff is then simplified to $|\tau|\ge\epsilon$, i.e., outside the shaded area of Fig. \ref{contour2}(c). Therefore, this is the correct UV cutoff corresponding to point splitting in the $x$ direction.

It is interesting that in our case, even though the magnitude of the UV cutoff $\epsilon$ does not affect the final result, the spacetime direction along which we put the cutoff $\epsilon$ does have a physical consequence.
This is not very common in high energy theory, since there we usually study interactions preserving the Lorentz symmetry, which guarantees that the direction of UV cutoff does not matter.
In our particular example, the interaction breaks Lorentz invariance, and such a UV cutoff direction dependence arises.

\subsubsection{The finite temperature solution}\label{sec:SDfiniteTemp}

We now proceed to solve the Schwinger-Dyson equation of our $1+1$ dimensional chiral SYK model at finite temperature $\beta^{-1}$, which is given by Eq. (\ref{Eq-SD}) with $i\omega_\tau$ replaced by the Matsubara frequency $i\omega_n=\frac{(2n+1)\pi}{\beta}$ ($n\in\mathbb{Z}$). Unlike the zero temperature case, the temperature $\beta^{-1}$ sets an energy scale for the problem, and the Schwinger-Dyson equation is no longer scaling invariant. Therefore, the scaling invariant ansatz method in Sec. \ref{Sec-zero-solution} no longer applies. Besides, the $1+1$ dimensional Schwinger-Dyson equation (\ref{Eq-SD}) has no reparameterization symmetry,  so we cannot do conformal transformation from zero temperature to finite temperature. This makes the finite temperature problem much more difficult.

However, there are two hints from the zero temperature solution (\ref{Eq-Gxt},\ref{Eq-Sxt}) which help us guess the finite temperature solution. The first hint is that the large $N$ zero temperature two point function $G$ has exactly the same form as that of $N=4$ (see Eq. (\ref{Eq-G4}) with $\beta\rightarrow\infty$). The second hint is that the large $N$ zero temperature $G(-i\tau,x)$ and $\Sigma(-i\tau,x)$ factorize into the product of two zero temperature two point functions and self energies of the $0+1$ dimensional SYK model in the conformal limit along $\tau-iu_+^{-1}x$ and $\tau-iu_-^{-1}x$ directions, respectively. Although in Sec. \ref{Sec-Forder} we showed such a factorization is invalidated in the momentum space by an indispensable UV cutoff regularization contribution, we could expect it is valid in the IR (since the model is scaling invariant, any finite length scale is IR if no UV cutoff is put in by hand). This motivates us to guess that the above two facts still hold at finite temperature $\beta^{-1}$, so that the finite temperature two point function in the large $N$ limit is given by
\begin{equation}\label{Eq-Ginf}
G_\beta(-i\tau,x)=\frac{1}{2\beta\sqrt{u_+u_-}}\frac{1}{\sqrt{\sin\left[ \frac{\pi}{\beta}(\tau-iu_+^{-1}x)\right] \sin\left[ \frac{\pi}{\beta}(\tau-iu_-^{-1}x)\right]}}\ ,
\end{equation}
and the finite temperature self energy is $\Sigma_\beta(-i\tau,x)=J^2G_\beta(-i\tau,x)^3$ by the second Schwinger Dyson equation in Eq. (\ref{Eq-SD}). Namely, we guess the large $N$ finite temperature two point function $G_\beta$ is equal to the $N=4$ finite temperature two point function as well, and also factorizes into the product of two finite temperature $0+1$ dimensional SYK two point functions in the conformal limit along $\tau-iu_+^{-1}x$ and $\tau-iu_-^{-1}x$ directions. In the remainder of this subsection, we shall verify that Eq. (\ref{Eq-Ginf}) is indeed the large $N$ finite temperature solution.

We shall verify the validity of the two point function $G_\beta$ in Eq. (\ref{Eq-Ginf}) and the corresponding self energy $\Sigma_\beta$ by Fourier transforming them into the momentum space, and check whether they satisfy the first Schwinger-Dyson equation in Eq. (\ref{Eq-SD}). For convenience, we set $\beta=2\pi$ in the Fourier calculations below, while the answers for generic temperature $\beta^{-1}$ can be restored by a rescaling $(i\omega_n,k)\rightarrow \frac{\beta}{2\pi}(i\omega_n,k)$ in the end.
As we have shown in Sec. \ref{Sec-Forder}, the correct UV regularization for the Fourier transform is to take a UV cutoff $|\tau|\ge\epsilon$. Therefore, we calculate the following general function
\begin{equation}\label{Eq-Wba}
W_\sigma(i\omega_n,k)=\int_{\epsilon}^{2\pi-\epsilon} {d\tau\over 2\pi} e^{-i\omega_n\tau} \int_{-\infty}^\infty dx e^{-ikx} \left[\sin\left({\tau-iu_+^{-1}x\over 2}\right) \sin \left(\frac{\tau-iu_-^{-1}x}{2}\right)\right]^\sigma
\end{equation}
with a cutoff $\epsilon$ for small $\tau$, where $\sigma$ is a constant. The Fourier transformations of the two point function and the self energy are by definition given by $G(i\omega_{n},k)={1\over 2\sqrt{u_+u_-}}W_{-1/ 2}(i\omega_{n},k)$ and $\Sigma(i\omega_n,k)={J^2\over 32\pi^2 (u_+u_-)^{3/2}}W_{-3/2}(i\omega_{n},k)$, respectively.

\begin{figure}[htbp]
\begin{center}
\includegraphics[width=2in]{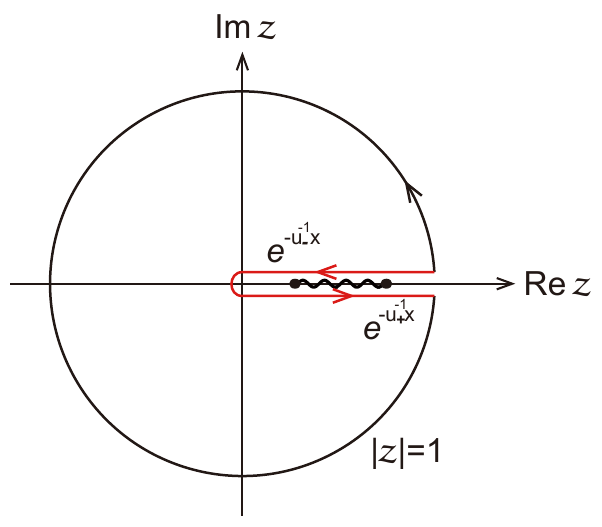}
\end{center}
\caption{Illustration of the contour deformation in Eq. (\ref{Eq-Wdf}).}
\label{fig:Contour3}
\end{figure}

We first change $\tau$ to the complex variable $z=e^{i \tau }$, so that the integration over $\tau$ in Eq. (\ref{Eq-Wba}) becomes an integration of $z$ along the unit circle contour $|z|=1$ counterclockwise from $1+i\epsilon$ to $1-i\epsilon$ as shown in Fig. \ref{fig:Contour3}. At a given $x$, the integrand has three branch points $0$, $e^{-u_-^{-1}x} $, and $e^{-u_+^{-1}x}$ on the nonnegative real axis of $z$. For half-integral $\sigma$ which we are interested in, $z=0$ is not a branch point. Therefore, we can deform the contour along $|z|=1$ into a contour closely bypassing the nonnegative real axis of $z$ from $1+i\epsilon$ to $0$ and then to $1-i\epsilon$, as shown in Fig. \ref{fig:Contour3}. Eq. (\ref{Eq-Wba}) can then be rewritten as
\begin{equation}\label{Eq-Wdf}
	W_{\sigma}={1\over 2\pi i (2i)^{2\sigma}}\int_{1+i\epsilon}^{1-i\epsilon} dz \int_{-\infty}^{\infty} dx z^{-1-\omega_n-\sigma}e^{-ikx-{\sigma \over 2}(u_+^{-1}+u_{-}^{-1})x}(ze^{u_+^{-1}x }-1)^{\sigma}(ze^{u_-^{-1} x }-1)^{\sigma}
\end{equation}
where the integration over $z$ is from $1+i\epsilon$ to $0$ above the real axis, and then from $0$ to $1-i\epsilon$ below the real axis. For convenience, we denote $z=e^{-t}$, where $t=-i\tau$ is the Wick rotation of the imaginary time $\tau$. The deformed integral contour of $z$ is then equivalent to a contour of $t$ going from $-i\epsilon$ to $+\infty$ and then back to $+i\epsilon$ in straight lines.

For a given $z=e^{-t}$ (and $\sigma$ being a half integer), the integrand has branch cuts for $x$ along the line segments from $u_-(t+i{2\pi m \over \beta})$ to $u_+(t+i{2\pi m \over \beta})$ ($m\in\mathbb{Z}$). Without loss of generality, we assume the momentum $k>0$, so that the integration over $x$ is along a contour of the real $x$ axis and closed in the lower half plane at infinity. Therefore, the integration of $x$ picks up the contributions of branch cuts from $u_-(t+i{2\pi m \over \beta})$ to $u_+(t+i{2\pi m \over \beta})$ with $m\le0$ when the imaginary part of $t$ is infinitesimally negative (i.e., when $t$ is integrated from $-i\epsilon$ to $+\infty$), and picks up the contributions of branch cuts from $u_-(t+i{2\pi m \over \beta})$ to $u_+(t+i{2\pi m \over \beta})$ with $m<0$ when the imaginary part of $t$ is infinitesimally positive (i.e., when $t$ is integrated from $+\infty$ to $i\epsilon$). After $t$ is integrated, the contributions of all $m<0$ branch cuts will cancel, and only the contribution of the $m=0$ branch cut from $u_-t$ to $u_+t$ remains. Considering these facts, we can deform the $x$ contour into a contour closely surrounding the branch cut, and write Eq. (\ref{Eq-Wdf}) as
\begin{equation}
	W_{\sigma}={-\sin(\pi\sigma)\over \pi (2i)^{2\sigma}}\int_{-i\epsilon}^{+\infty} dt \int_{u_- t}^{u_+ t} dx e^{(\omega_n+\sigma) t}e^{-ikx-{\sigma \over 2}(u_+^{-1}+u_{-}^{-1})x}({1-e^{u_+^{-1}x-t}})^{\sigma}({e^{u_-^{-1}x-t}-1})^{\sigma}\ .
\end{equation}
Since the physical result should not depend on whether the cutoff $\epsilon$ is real or not, by adiabatical continuation we can replace the lower bound $-i\epsilon$ of the $t$ integral by $\epsilon>0$. To further simplify the expression, we define two new variables $y_+=t- u_+^{-1}x$ and $y_-= u_-^{-1}x-t$. The integral then becomes
\begin{equation}
\begin{split}
W_{\sigma}={-\sin(\pi \sigma) u_+ u_-\over  (2i)^{2\sigma}J }\int_{0}^{+\infty}d y_+ e^{(k_++{\sigma\over 2})y_+} (1-e^{-y_+})^{\sigma}\int_{\Delta(\epsilon,y_+)}^{+\infty}dy_- e^{(k_-+{\sigma\over 2})y_-}(1-e^{-y_-})^{\sigma},
\end{split}
\end{equation}
where we have defined $k_{\pm}={ u_{\pm}\over u_+-u_-}\omega_n-i{u_+u_-\over u_+-u_-}k$, and $\Delta(\epsilon,y_+)=\max[(\frac{u_+}{u_-}-1)\epsilon-\frac{u_+}{u_-}y_+,0]$. Here the integration of $y_\pm$ are all done along the real axis. If the $\epsilon$ piece is ignored, by formula $\int_0^\infty dy e^{ky}(1-e^{-y})=B(1+\sigma,-k)=\frac{\Gamma(1+\sigma)\Gamma(-k)}{\Gamma(1+\sigma-k)}$, we see that $W_\sigma$ factorizes into the product of two Beta functions. With a finite $\epsilon>0$, we can expand in $\Delta(\epsilon,y_+)$, which follows exactly what we did from Eq. (\ref{Eq-Wpm}) to Eq. (\ref{Eq-Wa}) in Sec. \ref{Sec-Forder}. This gives a result
\begin{equation}
\begin{split}
W_{\sigma}&={-\sin(\pi \sigma) u_+ u_-\over (2i)^{2\sigma}J }\Big[ \frac{\Gamma(1+\sigma)^2\Gamma(-k_+-\frac{\sigma}{2})\Gamma(-k_--\frac{\sigma}{2})} {\Gamma(1+\frac{\sigma}{2}-k_+)\Gamma(1+\frac{\sigma}{2}-k_-)} \\
&-\frac{\tilde{\epsilon}^{2+2\sigma}\Gamma(1+\sigma)^2}{\Gamma(3+2\sigma)}- \frac{\tilde{\epsilon}^{3+2\sigma}\Gamma(1+\sigma)\Gamma(2+\sigma)}{\Gamma(4+2\sigma)} \Big(\sqrt{\frac{u_-}{u_+}} k_++ \sqrt{\frac{u_+}{u_-}}k_-\Big)+\cdots \Big]\ ,
\end{split}
\end{equation}
where we have defined $\tilde{\epsilon}=(\sqrt{\frac{u_+}{u_-}}-\sqrt{\frac{u_-}{u_+}})\epsilon$. For $\sigma=-1/2$, the $\epsilon$ pieces vanishes as $\epsilon\rightarrow0$. For $\sigma=-3/2$, the leading $1/\epsilon$ term vanishes, while there is a constant piece contributed by the $\epsilon$ expansion. Therefore, we find the Fourier transformation $G(i\omega_{n},k)={1\over 2\sqrt{u_+u_-}}W_{-1/ 2}$ and $\Sigma(i\omega_n,k)={J^2\over 32\pi^2 (u_+u_-)^{3/2}}W_{-3/2}$ to be given by
\begin{equation}\label{Eq-Gbeta}
G_\beta(-i\omega_n,k)=\frac{i\beta \sqrt{u_+u_-}}{2J} \frac{\Gamma\left[\frac{\beta u_+u_-}{2J}(-\omega_n u_+^{-1}+ik)+\frac{1}{4}\right]\Gamma\left[\frac{\beta u_+u_-}{2J}(-\omega_n u_-^{-1}+ik)+\frac{1}{4}\right]}{ \Gamma\left[\frac{\beta u_+u_-}{2J}(-\omega_n u_+^{-1}+ik)+\frac{3}{4}\right] \Gamma\left[\frac{\beta u_+u_-}{2J}(-\omega_n u_-^{-1}+ik)+\frac{3}{4}\right]}\ ,
\end{equation}
and
\begin{equation}\label{Eq-Sbeta}
\Sigma_\beta(-i\omega_n,k)=i\omega_n+k+\frac{2iJ}{\beta \sqrt{u_+u_-}} \frac{\Gamma\left[\frac{\beta u_+u_-}{2J}(-\omega_n u_+^{-1}+ik)+\frac{3}{4}\right]\Gamma\left[\frac{\beta u_+u_-}{2J}(-\omega_n u_-^{-1}+ik)+\frac{3}{4}\right]}{ \Gamma\left[\frac{\beta u_+u_-}{2J}(-\omega_n u_+^{-1}+ik)+\frac{1}{4}\right] \Gamma\left[\frac{\beta u_+u_-}{2J}(-\omega_n u_-^{-1}+ik)+\frac{1}{4}\right]},
\end{equation}
where we have recovered the answer to arbitrary temperature $\beta^{-1}$. One can then easily see that they satisfy the Schwinger-Dyson equation (\ref{Eq-SD}) at finite temperature (with $i\omega_\tau$ replaced by $i\omega_n$ therein).
Therefore, we have proved that Eq. (\ref{Eq-Ginf}) is the correct finite temperature two point function in the large $N$ limit. One can also check that in the limit $\beta\rightarrow\infty$, Eqs. (\ref{Eq-Gbeta}, \ref{Eq-Sbeta}) reduce to the zero temperature solution (\ref{Eq-Gx1t2}, \ref{Eq-Sx1t2}) in momentum space.

\subsection{Thermal quantities and the ground state entropy density}

The thermal quantities of the large $N$ chiral SYK model can be derived from the exact finite temperature solution we obtained in Sec. \ref{sec:SDfiniteTemp}. In fact, since the large $N$ two point function $G_\beta$ in Eq. (\ref{Eq-Ginf}) is of exactly the same form as that of $N=4$ in Eq. (\ref{Eq-G4}), the energy density $\mathcal{E}$ and the energy current $j_\mathcal{E}$ in the large $N$ limit can be calculated in exactly the same way as we did for $N=4$ in Eqs. (\ref{Eq-E4}) and (\ref{Eq-jE4}). This gives an energy density
\begin{equation}
\mathcal{E}=\langle T^0_{\ 0}\rangle_\beta=-i\frac{N}{4}(\partial_t-\partial_x) G_\beta(t,x)\Big|_{t\rightarrow 0, x\rightarrow\epsilon}=\frac{N\pi}{48\beta^2}(u_+^{-1}+u_-^{-1})\ ,
\end{equation}
and an energy current
\begin{equation}\label{Eq-jEN}
j_\mathcal{E}=\langle T^x_{\ 0}\rangle_\beta=-i\frac{N}{2}\left(\partial_t-\frac{J^2}{4\pi^2}\partial_x\right) G_\beta(t,x)\Big|_{t\rightarrow 0, x\rightarrow\epsilon}=\frac{N\pi}{24\beta^2}\ ,
\end{equation}
where we have eliminated the unphysical $1/\epsilon^2$ terms. This gives a thermal Hall conductance $\kappa_{xy}=\partial j_\mathcal{E}/\partial_\beta=N\pi/12\beta$ if the system is an edge of a $2+1$ dimensional bulk gapped topological state. In particular, $\kappa_{xy}$ does not depend on $J$ and is quantized at order $\mathcal{O}(N)$, which agrees with the general belief that $\kappa_{xy}$ is a topologically invariant quantity. The entropy density $S$ at temperature $\beta$ can be calculated from $\beta^{-1}[\partial \mathcal{S}/\partial(\beta^{-1})]=\partial \mathcal{E}/\partial(\beta^{-1})$, which gives $\mathcal{S}=\mathcal{S}_0+(N\pi/24\beta)(u_+^{-1}+u_-^{-1})$, where $\mathcal{S}_0$ denotes the ground state entropy density. In the below, we discuss the ground state entropy density $\mathcal{S}_0$ of our $1+1$ dimensional chiral SYK model.

A remarkable feature of the $0+1$ dimensional SYK model is that it has a large ground state entropy of order $N$, which indicates an unusual ground state degeneracy of order $e^{N}$. This implies a duality between a $0+1$ dimensional SYK model on the boundary of a $1+1$ dimensional AdS spacetime and a black hole in the bulk, which has a large entropy proportional to the event horizon area. As a comparison, in the $1+1$ dimensional chiral SYK model here we can examine the ground state entropy density $\mathcal{S}_0$. However, unlike the $0+1$ dimensional SYK model where the Hilbert space dimension $2^{N/2}$ is finite, the model here as a field theory has an infinite dimensional Hilbert space per unit length, which therefore makes the ground state entropy density possibly divergent and ill-defined. Therefore, we need to impose a physical spatial UV cutoff $\epsilon_x$ to make the Hilbert space dimension per unit length finite.

The entropy density at temperature $\beta^{-1}$ is defined by $\mathcal{S}=L^{-1}(1-\beta\partial_{\beta})\log Z$, where $L\rightarrow\infty$ is the spatial size of the system, and $Z$ is the partition function. By taking the limit $\beta\rightarrow\infty$, we obtain the ground state entropy density $\mathcal{S}_0$. In the large $N$ limit, we can take a Gaussian average over the random interactions $J_{ijkl}$, and write the partition as \footnote{Here we write the disorder average over $J_{ijkl}$ as the annealed disorder $Z=\langle Z\rangle_J$, however, for large $N$ it is equal to the quenched disorder $Z=e^{\langle \log Z\rangle_J}$ to the leading order \cite{Kitaev:2017awl}. This is because in the replica treatment $\langle Z^n\rangle_J$ is equal to $\langle Z\rangle_J^n$ up to order $1/N^3$, which leads to $\langle \log Z\rangle_J=\lim_{n\rightarrow 0}\frac{\langle Z^n\rangle_J-1}{n}=\lim_{n\rightarrow 0}\frac{\langle Z\rangle^n_J-1}{n}= \log\langle Z\rangle_J$.}
\begin{equation}
Z=\int \prod_{1\le i<j<k<l\le N}dJ_{ijkl}e^{- N^3J^2_{ijkl}/(2\cdot3!J^2)}\int\mathcal{D}\psi_i e^{iS[\psi_i,J_{ijkl}]}\ .
\end{equation}
One can first integrate out $J_{ijkl}$. Then, by inserting $1=\int\mathcal{D}\bar{G}\mathcal{D}\bar{\Sigma} \exp[-\frac{1}{2}\int d\tau dx\int d\tau'dx'\bar{\Sigma}(N\bar{G}-\sum_i\psi_i\psi_i)]$, and integrating out the Majorana fermion fields $\psi_i$, one can rewrite the action as a functional of two functions $\bar{G}(-i\tau,x;-i\tau',x')$ and $\bar{\Sigma}(-i\tau,x;-i\tau',x')$ in the below:
\begin{equation}
Z=\int \mathcal{D}\bar{G}\mathcal{D}\bar{\Sigma}\exp\left[\frac{N}{2}\left\{ \log\det(\partial_\tau -i\partial_x-\bar{\Sigma})- \int d\tau dx\int d\tau'dx'\left[\bar{\Sigma} \bar{G}-\frac{J^2}{4}\bar{G}^4\right] \right\}\right]\ .
\end{equation}
One can easily check that the Euler-Lagrange equation for $\bar{G}$ and $\bar{\Sigma}$ is exactly the Schwinger-Dyson equation satisfied by the two point function $G$ and the self energy $\Sigma$. Therefore, the on-shell values of the functions $\bar{G}$ and $\bar{\Sigma}$ are simply equal to the two point function $G$ and the self energy $\Sigma$ we solved in Sec. \ref{sec:SDfiniteTemp}. To the leading order of $1/N$, the partition function $Z$ is just given by fixing $\bar{G}$ and $\bar{\Sigma}$ to their the on-shell values $G$ and $\Sigma$, namely,
\begin{equation}\label{Eq-FN}
\frac{\log Z}{L}=\frac{N}{2 L}\text{tr} \log (\partial_\tau -i\partial_x-\Sigma) -\frac{N\beta}{2}\int_{-\beta/2}^{\beta/2} d\tau \int_{-\infty}^{\infty} dx \Big[G(-i\tau,x)\Sigma(-i\tau,x)-\frac{J^2}{4}G(-i\tau,x)^4\Big].
\end{equation}
Instead of calculating the above complicated expression directly, we can compare its value with the free case $J=0$. This can be done by evaluating the derivative $\partial_J(\log Z/L)$ with respect to $J$. Making use of the Schwinger-Dyson equations $\partial_\tau -i\partial_x-\Sigma=G^{-1}$ (which should be understood as matrices in the Hilbert space) and $\Sigma(-i\tau,x)=J^2G(-i\tau,x)^3$, one would find that all the terms proportional to $\partial_JG$ and $\partial_J\Sigma$ vanish, and one arrives at
\begin{equation}\label{Eq-plogZ}
\begin{split}
\frac{\partial_J\log Z}{L}&= \frac{N\beta J}{4}\int_{-\beta/2}^{\beta/2} d\tau \int_{-\infty}^{\infty} dx G(-i\tau,x)^4\\
&=\frac{NJ}{64\beta^3u_+^2u_-^2}\int_{-\beta/2}^{\beta/2} d\tau \int_{-\infty}^{\infty} dx \frac{1}{\sin^2[{\pi\over \beta}(\tau-iu_+^{-1}x)]\sin^2[{\pi\over \beta}(\tau-iu_-^{-1}x)]}\ .
\end{split}
\end{equation}
In order for the above integral to be convergent, we need to impose a UV cutoff $(x/\epsilon_x)^2+(\tau/\epsilon_\tau)^2\ge1$ with $\epsilon_\tau/\epsilon_x\rightarrow0$ as shown in Fig. \ref{fig:PointSplitting}(b), where $\epsilon_x>0$ is now a physical spatial cutoff for the Hilbert space dimension per unit length not to diverge. Importantly, note that $\sin^2[{\pi\over \beta}(\tau-iu_+^{-1}x)]\sin^2[{\pi\over \beta}(\tau-iu_-^{-1}x)]$ is invariant under $\beta\rightarrow-\beta$, therefore the above expression is an odd function of $\beta$. As a result, we would find the following to be also an odd function of $\beta$:
\begin{equation}\label{Eq-logZ}
\frac{\log Z}{L}\Big|_{J}-\frac{\log Z}{L}\Big|_{0}=\int_0^{J}\frac{\partial_J\log Z}{L}=a_0(J,\epsilon_x,\epsilon_\tau)\beta+a_1(J)\beta^{-1} +a_2(J,\epsilon_x,\epsilon_\tau)\beta^{-3}+\cdots\ ,
\end{equation}
where $a_j$ ($j\ge0$) are constants depending on $\epsilon_x,\epsilon_\tau$ and $J$. The coefficient $a_j$ is of order $\mathcal{O}(\epsilon_{\tau,x}^{2j-2})$, and in particular, $a_1(J)$ does not depend on $\epsilon_\tau$ and $\epsilon_x$. In the free case $J=0$, the entropy density is easily known to be given by $\mathcal{S}_f=\mathcal{S}_{f0}+N\pi/(12\beta)$, where $\mathcal{S}_{f0}$ denotes the ground state entropy density of $N$ flavors of free chiral Majorana fermions. Therefore, we find the entropy density at interaction $J$ to be given by
\begin{equation}
\mathcal{S}=(1-\beta\partial_\beta)\frac{\log Z}{L}=\mathcal{S}_{f0}+\left[\frac{N\pi}{12}+2a_1(J)\right]\beta^{-1}+\mathcal{O}(\epsilon_x^2\beta^{-3})\ .
\end{equation}
It can be directly calculated from Eq. (\ref{Eq-plogZ}) that $\partial_Ja_1(J)=\frac{N}{96}(u_-^{-2}-u_+^{-2})$ \footnote{since $\partial_Ja_1(J)$ is $\epsilon_x$ and $\epsilon_\tau$ independent, it can be calculated using UV cutoff $|\tau|\ge\epsilon$ as illustrated by Fig. \ref{fig:PointSplitting}(b)-(c), instead of the more complicated cutoff $(x/\epsilon_x)^2+(\tau/\epsilon_\tau)^2\ge1$ with $\epsilon_\tau/\epsilon_x\rightarrow0$.}. Therefore, the coefficient of the thermal part of entropy density is $\frac{N\pi}{12}+2a_1(J)=\frac{N\pi}{24}(u_+^{-1}+u_-^{-1})$, which is in agreement with our calculations below Eq. (\ref{Eq-jEN}).
The ground state entropy density $\mathcal{S}_0=\lim_{\beta\rightarrow\infty}\mathcal{S}$ for any $0\le J<2\pi$ is therefore equal to that in the free fermion case, namely, $\mathcal{S}_0=\mathcal{S}_{f0}$. It is known that $N$ flavors of free chiral Majorana fermions has a unique ground state for anti-periodic spatial boundary condition, and has $2^N$ degenerate ground states for periodic spatial boundary condition due to $N$ zero modes. This gives a total ground state entropy either $\mathcal{S}_{f0}L=0$ or $\mathcal{S}_{f0}L=N\log2$. Therefore, in the infinite spatial size limit $L\rightarrow\infty$, the ground state entropy density of the free fermions is $\mathcal{S}_{f0}\rightarrow0$. We then conclude that the ground state entropy density for any interaction strength $0\le J<2\pi$ is $\mathcal{S}_0=\mathcal{S}_{f0}=0$. Basically, in order to have $\mathcal{S}_0$ different from $\mathcal{S}_{f0}$, one has to have a constant $1/\epsilon_x$ piece independent of $\beta$ in $L^{-1}\log Z$, which is absent in Eq. (\ref{Eq-logZ}) here.

The fact that the ground state entropy density $\mathcal{S}_0=\mathcal{S}_{f0}=0$ for any $0\le J<2\pi$ can also be physically understood from the spectral weight $A(\omega,k)$ in the large $N$ limit plotted in Fig. \ref{spectral}. Since the spectral weight $A(\omega,k)$ is nonzero only when $u_-\le\omega/ k\le u_+$, this suggests that any many-body eigenstate $|n,k\rangle$ with momentum $k$ (we use $n$ to denote different states) has an energy bounded by $E_{n,k}\in [u_-k,u_+k]$, and all the eigenstates have $k\ge0$ (with both the energy and momentum of the ground state energy defined as zero). This means all the many-body state energy eigenvalues $E_{n,k}$ at momentum $k$ of the chiral SYK system are no smaller than that of $N$ flavors of free chiral Majorana fermions with velocity $u_-$, and no larger than that of $N$ flavors of free chiral Majorana fermions with velocity $u_+$. Accordingly, the partition function $Z$ of the chiral SYK model at any temperature $\beta^{-1}$ will also be bounded by $Z_+\le Z\le Z_-$, where $Z_\pm$ is the partition function of $N$ flavors of free chiral Majorana fermions with velocity $u_\pm$, respectively. Since the ground state entropy density of free chiral Majorana fermions $\mathcal{S}_{f0}=0$ is independent of their velocity, the above bound of partition function $Z$ indicates that the ground state entropy density of the chiral SYK model has to be $\mathcal{S}_{0}= \mathcal{S}_{f0}=0$ for any $0\le J<2\pi$. From this argument, we can see that the chiral nature of the model (and translational invariance) protects its ground state entropy density to be at zero.

\subsection{The OTOC, chaos exponent and butterfly cone}\label{Sec-chaos}

In this section, we derive the chaos exponent and butterfly velocity in the OTOC of the model. Following \cite{Maldacena:2016hyu}, we define the regularized OTOC four point function in real time as
\begin{equation}\label{Eq-4pfInf}
\mathcal{F}(t_1,x_1;t_2,x_2)=\frac{1}{N^2}\sum_{i,j}\text{Tr}\left[y\psi_j(t_1,x_1)y \psi_i(0,0)y \psi_j(t_2,x_2)y \psi_i(0,0)\right]\ ,
\end{equation}
where $y=e^{-\beta H/4}=\rho(\beta)^{1/4}$ separates evenly the four fermion fields by a quarter of the thermal circle. The leading contribution in the early time OTOC comes from the contractions between two $\psi_i$ and between two $\psi_j$, which gives an order $1$ piece $-G(-i\frac{\beta}{2},0)G(t_1-t_2-i\frac{\beta}{2},x_1-x_2)$. The next order contribution comes from contraction of $\psi_i$ with $\psi_j$, and is of order $1/N$ at the early time. Therefore, we can separate the order $1$ and order $1/N$ pieces of the OTOC as
\begin{equation}
\mathcal{F}(t_1,x_1;t_2,x_2)=-G(-i\frac{\beta}{2},0)G(t_1-t_2-i\frac{\beta}{2},x_1-x_2)-\frac{1}{N}\delta \mathcal{F}(t_1,x_1;t_2,x_2)\ ,
\end{equation}
where the function $\delta \mathcal{F}$ is then of order $1$ at the early time by definition.

\begin{figure}[htbp]
\begin{center}
\includegraphics[width=6in]{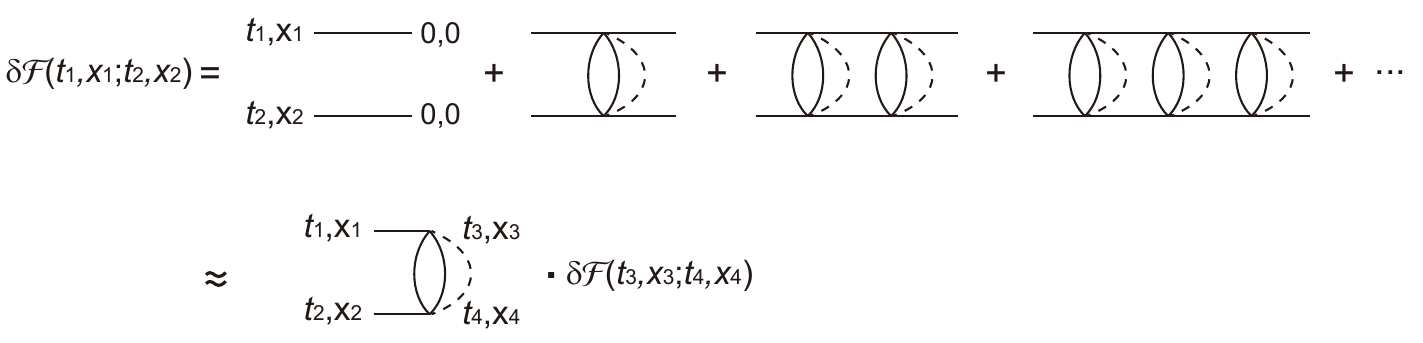}
\end{center}
\caption{The ladder diagrams contributing to the $1/N$ piece $\delta\mathcal{F}$ of the OTOC, which are the leading order contributions in the $1/N$ expansion. The solid lines are the two point function, and the dashed line stands for averaging over $J_{ijkl}$. This implies $\delta\mathcal{F}$ is approximately an eigenfunction of kernel $K_R$ with eigenvalue $1$.}
\label{ladder}
\end{figure}

To the leading order of $1/N$, the Feynman diagrams contributing to $\delta \mathcal{F}$ are the ladder diagrams as shown in Fig. \ref{ladder} \cite{Maldacena:2016hyu,Kitaev:2017awl}.
For chaotic systems where there is a Lyapunov regime (which usually requires unbounded local Hilbert spaces), when $t_1=t_2=t$, one expects $\delta \mathcal{F}$ to grow exponentially with respect to time $t$ in the Lyapunov regime. For large $N$ models like ours, the Lyapunov regime is in the interval $\beta\lesssim t\lesssim \beta\log N$. In this case, at late time $t$ when $\delta \mathcal{F}$ becomes large, it will approximately satisfy the self-consistent equation
\begin{equation}\label{FKR}
\delta\mathcal{F}(t_1,x_1;t_2,x_2)=\int dt_3dx_3dt_4dx_4 K_R(t_1,x_1;\cdots t_4,x_4)\delta\mathcal{F}(t_3,x_3;t_4,x_4)\ ,
\end{equation}
where $K_R$ is the retarded kernel defined as
\begin{equation}
K_R(t_1,x_1;\cdots t_4,x_4)=3J^2 G_R(t_{13},x_{13})G_R(t_{24},x_{24})G_{lr}^2(t_{34},x_{34})
\end{equation}
as illustrated by the second line of Fig. \ref{ladder}. Here we have denoted for short $t_{ij}=t_i-t_j$, and $x_{ij}=x_i-x_j$. The functions $G_R$ and $G_{lr}$ are the retarded Green's function and the Wightman correlator with half thermal circle separation, respectively, which can be explicitly derived via the analytic continuation of Eq. (\ref{Eq-Ginf}) as
\begin{equation}
G_R(t,x)= \frac{1}{\beta\sqrt{u_+u_-}}\frac{\Theta(t-u_+^{-1}x)\Theta(u_-^{-1}x-t)}{\sqrt{\sinh\left[ \frac{\pi}{\beta}(t-u_+^{-1}x)\right] \sinh\left[ \frac{\pi}{\beta}(u_-^{-1}x-t)\right]}}\ ,
\end{equation}
and
\begin{equation}
G_{lr}(t,x)= \frac{1}{2\beta\sqrt{u_+u_-}}\frac{1}{\sqrt{\cosh\left[ \frac{\pi}{\beta}(t-u_+^{-1}x)\right] \cosh\left[ \frac{\pi}{\beta}(u_-^{-1}x-t)\right]}}\ .
\end{equation}
Therefore, the function $\delta\mathcal{F}$ is an eigenfunction of kernel $K_R$ with eigenvalue $1$. To derive the exponential growth of $\delta\mathcal{F}$, we first need to compute all the eigenfunctions of kernel $K_R$ which have eigenvalue $1$.

To simplify the computation of the eigenfunctions, we make a spacetime coordinate transformation from $(t,x)$ to a new basis
\begin{equation}
(t^+,t^-)=(t-u_{+}^{-1}x,\ t-u_{-}^{-1}x)\ ,\qquad dtdx=\frac{u_+u_-}{u_+-u_-}dt^+dt^-= \frac{\pi u_+u_-}{J}dt^+dt^-\ ,
\end{equation}
where we have used the relation $u_\pm=1\pm J/2\pi$. The transformation is nonsingular as long as the interaction strength $J>0$. After this change of coordinates, Eq. (\ref{FKR}) can be rewritten as
\begin{equation}\label{Eq-FKRtpm}
\begin{split}
\delta\mathcal{F}&(t_1^+,t_1^-;t_2^+,t_2^-)=\frac{3\pi^2}{4\beta^4}\int dt_3^+dt_3^-dt_4^+dt_4^-\\ &\times\frac{\Theta(t_{13}^+)\Theta(t_{24}^+)\Theta(-t_{13}^-) \Theta(-t_{24}^-)}{\left(\sinh\frac{\pi t^+_{13}}{\beta} \sinh\frac{-\pi t^-_{13}}{\beta} \sinh\frac{\pi t^+_{24}}{\beta} \sinh\frac{-\pi t^-_{24}}{\beta}\right)^{1/2} \left(\cosh\frac{\pi t^+_{34}}{\beta} \cosh\frac{-\pi t^-_{34}}{\beta}\right)} \delta\mathcal{F}(t_3^+,t_3^-;t_4^+,t_4^-)\ .
\end{split}
\end{equation}
One can see the equation does not have an explicit dependence on the interaction strength $J$ any more. In addition, the kernel $K_R$ factorizes into the product of a function $k_R^+(t_i^+)$ of coordinates $t_i^+$ multiplied by a function $k_R^-(-t_i^-)$ of $-t_i^-$ ($1\le i\le4$), each of which up to a factor is nothing but the kernel of a $1$d $q=4$ SYK model along the $t^+$ or the $-t^-$ time direction in the conformal limit. Therefore, an eigenfunction of the $1+1$ dimensional kernel $K_R$ here can also be factorized into the product of the eigenfunctions of two $0+1$ dimensional SYK model along $t^+$ and $-t^-$ directions, respectively. More explicitly, the eigenfunction of the kernel $K_R$ can be written into the form
\begin{equation}\label{Eq-eigf}
f_{h_+,h_-}(t_1^+,t_1^-;t_2^+,t_2^-)=\frac{e^{-\frac{\pi}{\beta}[h_+(t_1^++t_2^+) -h_-(t_1^-+t_2^-)]}}{\left(\cosh\frac{\pi t_{12}^+}{\beta}\right)^{\frac{1}{2}-h_+} \left(\cosh\frac{\pi t_{12}^-}{\beta}\right)^{\frac{1}{2}-h_-}}\ ,
\end{equation}
where $h_+$ and $h_-$ are constants to be determined. The corresponding eigenvalues of $K_R$ are
\begin{equation}\label{Eq-eigv}
k_R(h_+,h_-)=\frac{3}{(1-2h_+)(1-2h_-)}\ .
\end{equation}
The above eigenfunction (\ref{Eq-eigf}) and eigenvalue (\ref{Eq-eigv}) can be directly verified by substituting them into Eq. (\ref{Eq-FKRtpm}). Since there is no UV divergence in the above calculations, we do not have the point-splitting issue that arises in the calculation of the self energy $\Sigma$. This is consistent with our expectation that the chaos is controlled by the IR physics.

\begin{figure}[tbp]
\begin{center}
\includegraphics[width=6.5in]{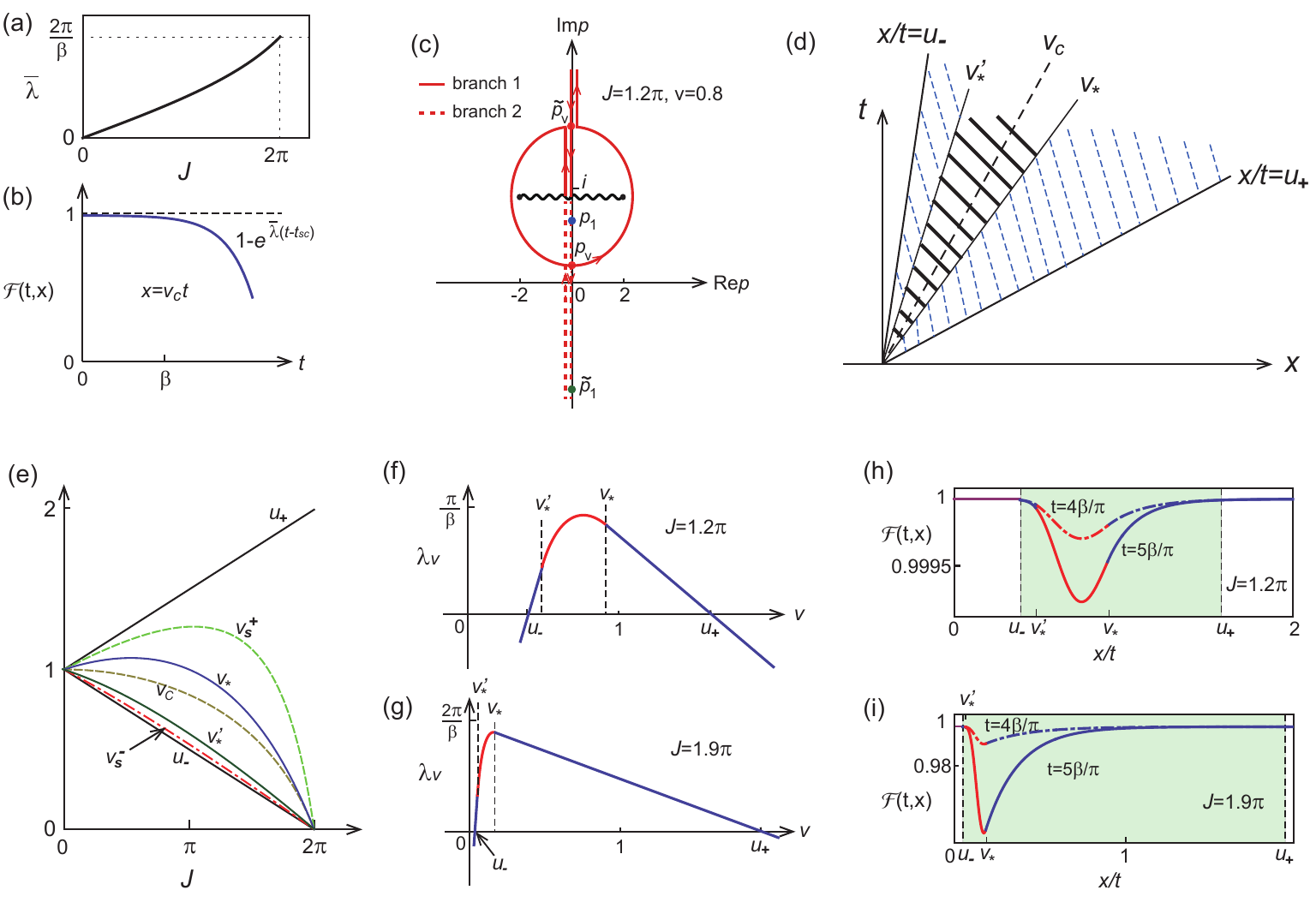}
\end{center}
\caption{(a) The leading Lyapunov exponent $\overline{\lambda}$ of the OTOC for $0\le J<2\pi$, which saturates the maximal chaos bound $2\pi/\beta$ as $J\rightarrow2\pi$. (b) Illustration of the exponential growth behavior of the OTOC $\mathcal{F}(t,x)$ along the maximal growth direction $x=v_ct$, where $t_{sc}\sim\mathcal{O}(\beta \log N)$ is the scrambling time. (c) The deformed contour for $J=1.2\pi$ at velocity $v=0.8$, which are the steepest descent (stationary phase) contours of function $e^{2\pi[\varkappa(p)t+ip x]/\beta}$ passing saddle points $p_v$ and $\widetilde{p}_v$, respectively. The wavy line denotes the branch cut of $\varkappa(p)$. The part of contour in the first (second) branch of the Riemann surface is represented by solid (dashed) lines. In this example, the contour deformation does not cross either pole $p_1$ or pole $\widetilde{p}_1$, and the integral is dominated by saddle point $p_v$. (d) Illustration of the butterfly cone between velocities $u_-$ and $u_+$, where $\delta\mathcal{F}$ grows exponentially along a constant velocity line $x=vt$. The OTOC in the solid shaded region and in the dashed shaded region are saddle dominated given by Eq. (\ref{Eq-butterfly1}) and pole dominated given by Eq. (\ref{Eq-butterfly2}) or (\ref{Eq-butterfly3}), respectively. $v_c$ is the velocity along which the velocity dependent Lyapunov exponent $\lambda_v$ reaches its maximal value $\overline{\lambda}$. (e) The seven velocities (from below to above) $u_-,v_s^-,v_*',v_c,v_*,v_s^+,u_+$ mentioned in our discussions as a function of $J$. (f)-(g) The velocity dependent Lyapunov exponent $\lambda_v$ as a function of velocity $v=x/t$ for (f) $J=1.2\pi$ and (g) $J=1.9\pi$, respectively. (h)-(i) The constant time slices of the OTOC at time $t=4\beta/\pi$ (dashed-dotted lines) and $t=5\beta/\pi$ (solid lines) for (h) $J=1.2\pi$ and (i) $J=1.9\pi$, respectively, where we used $\mathcal{F}(t,x)=1-e^{\lambda_v(t-t_{sc})}$ and took the scrambling time $t_{sc}=6\beta/\pi$ ($t_{sc}\sim\mathcal{O}(\beta \log N))$. The green shaded area denotes the butterfly region.}
\label{exponent}
\end{figure}

When the spacetime coordinates of the two $\psi_j$ fields in Eq. (\ref{Eq-4pfInf}) coincide, namely, $t_1=t_2=t$ and $x_1=x_2=x$, the eigenfunction can be rewritten as
\begin{equation}
f_{h_+,h_-}= e^{-\frac{2\pi}{\beta}(h_+t^+ -h_-t^-)}= \exp\left[\frac{2\pi}{\beta}(\varkappa t+ip x)\right]\ ,~~~h_{\pm}=-{u_{\pm}\varkappa \over u_+-u_-}-ip{u_+ u_-\over u_+-u_-}\ .
\end{equation}
Since the four point function $\delta\mathcal{F}$ at any fixed time $t$ cannot diverge when $x\rightarrow\pm\infty$, we need to constrain $p$ to be real, and $f_{h_+,h_-}$ is then simply a plane wave solution with momentum $2\pi p/\beta$. By solving Eq. (\ref{Eq-eigv}), we can derive $\varkappa$ as a function of $p$ to be
\begin{equation}\label{Eq-varkappa}
	\varkappa(p)={-\mathcal{J}-i p (1-\mathcal{J}^2)+\mathcal{J}\sqrt{3(1-\mathcal{J}^2)+(\mathcal{J}+(1-\mathcal{J}^2)i p)^2}\over 1-\mathcal{J}^2},~~~~\mathcal{J}={J\over 2\pi}\ .
\end{equation}
The function is defined in the branch where $\chi(p)\rightarrow -i(1-\mathcal{J})p-\frac{\mathcal{J}}{1+\mathcal{J}}$ as $p\rightarrow\infty$. Note that $\varkappa(p)$ in general takes complex values, so the Lyapunov exponent of the eigenfunction $f_{h_+(p),h_-(p)}$ is given by $\lambda(p)=\text{Re}\left[{2\pi \over \beta}\varkappa(p)\right]$ (where $p$ is real), which is an even function of $p$ since $\varkappa(p)=\varkappa(-p)^*$.
Its maximum value $\overline{\lambda}$ is always achieved at $p=0$, which has a magnitude
\begin{equation}\label{Eq-chaos-exp}
	\overline{\lambda}=\lambda(0)={2\pi\over \beta}\varkappa(0)={2\pi\over \beta}{\mathcal{J}(\sqrt{3-2\mathcal{J}^2}-1)\over 1-\mathcal{J}^2}\ .
\end{equation}
As $J$ grows from $0$ to $2\pi$ (i.e., $\mathcal{J}$ grows from $0$ to $1$), the maximal Lyapunov exponent $\overline{\lambda}$ grows from $0$ to the maximal chaos bound $2\pi/\beta$, as shown in Fig. \ref{exponent}(a).

We now discuss the chaotic behavior and butterfly cone of the function $\delta\mathcal{F}(t,x)=\delta\mathcal{F}(t,x;t,x)$ with $t_1=t_2=t>0$ and $x_1=x_2=x$ set in Eq. (\ref{Eq-4pfInf}), namely, the $1/N$ piece of the OTOC of our chiral SYK model. For this purpose, we need to know the weight coefficient $\rho(p)$ of the eigenfunction $f_{h_+(p),h_-(p)}$ in $\delta\mathcal{F}$. Due to the translation symmetry, one can check that each total momentum $2\pi p/\beta$ component of the four point function $\delta\mathcal{F}(p,x_{12},t_1,t_2)=\int d x e^{i\frac{2\pi}{\beta}px}\delta\mathcal{F}(t_1,x_1+x;t_2,x_2+x)$ satisfies the ladder diagram of Fig. \ref{ladder} by itself. This allows us to employ the remarkable identity between magnitude and the chaos exponent derived in \cite{Gu:2018jsv} to determine the coefficient $\rho(p)$.
Up to multiplication of some mild function, the weight coefficient is dominated by the factor
\begin{equation}\label{Eq-rho}
	\rho(p)\sim {1\over \cos[{\pi\varkappa(p)/ 2}]}\ .
\end{equation}
Therefore, to a good approximation the OTOC function $\delta\mathcal{F}$ is given by the integral
\begin{equation}\label{Eq-OTOC}
	\delta \mathcal{F}(t,x)\sim
\int_{-\infty}^{\infty} dp {e^{{2\pi\over \beta}[\varkappa(p)t+ip x]}\over \cos [{\pi\varkappa(p)/ 2}]}\ .
	\end{equation}
As discussed in details in \cite{Gu:2018jsv}, the integral at large $x$ and $t$ can be done by using saddle point approximation for $p$. The saddle point $p_v$ of the integrand of Eq. (\ref{Eq-OTOC}) in the complex plane of $p$ is given by
\begin{equation}
	\varkappa'(p_v)+iv=0,~~~~~x=vt\ ,
\end{equation}
which depends on the velocity $v=x/t$ we look at.
Since $\varkappa(p)$ is real on the imaginary $p$ axis, $\varkappa'(p)$ is purely imaginary on the imaginary $p$ axis. Besides, $\varkappa(p)$ has two Riemann surface branches connected by a branch cut.
For velocity $u_-\le v\le u_+$, this yields two saddle points $p_v$ and $\tilde{p}_v$ on the imaginary axes of the two Riemann surface branches, as shown in Fig. \ref{exponent}(c).
The original integration contour along the real axis in Eq. (\ref{Eq-OTOC}) can be deformed into two steepest descent contours passing the two saddle points $p_v$ and $\widetilde{p}_v$, respectively, as shown in Fig. \ref{exponent}(c). If the contour deformation does not cross any poles, the integral of Eq. (\ref{Eq-OTOC}) will be dominated by the contribution of the first saddle point $p_v$ as (Appendix \ref{Sec-SAP})
\begin{equation}\label{Eq-butterfly1}
	\delta \mathcal{F}(t,vt)\sim e^{{2\pi t\over \beta}[\varkappa(p_v)+ip_v v]}=e^{\lambda_vt}\ ,~~~~~(t\gg \beta)
\end{equation}
where the velocity $v$ dependent Lyapunov exponent $\lambda_v=2\pi[\varkappa(p_v)+ip_v v]/\beta$ is real.
The maximal value of $\lambda_v$ in Eq. (\ref{Eq-butterfly1}) is equal to $\overline{\lambda}$ given by Eq. (\ref{Eq-chaos-exp}), which is reached at velocity
\begin{equation}\label{Eq-vc}
v_c=i\varkappa'(0)=1-\frac{\mathcal{J}^2}{\sqrt{3-2\mathcal{J}^2}}\ ,\qquad \lambda_{v_c}=\overline{\lambda}\ ,
\end{equation}
namely, when the saddle point $p_v=0$. For velocity $v$ in the vicinity of $v_c$, the $v$ dependent Lyapunov exponent is approximately given by $\lambda_v\approx \overline{\lambda}-\frac{2\pi\xi}{\beta}(v-v_c)^2$, and thus
\begin{equation}
	\delta \mathcal{F}(t,x)\sim e^{\overline{\lambda}t-\frac{2\pi\xi(x-v_ct)^2}{\beta t}}\ ,~~~~~(t\gg \beta\ ,\quad |x-v_ct|\ll t)
\end{equation}
where $\xi=\frac{(3-2\mathcal{J}^2)^{3/2}}{6\mathcal{J}(1-\mathcal{J}^2)^2}$. Besides, $\lambda_v$ drops to zero at a velocity $v_s$ satisfying $v_s=i\varkappa'(p_s)=i{\varkappa(p_s)/ p_s}$, where $p_s$ is the saddle point position for velocity $v_s$. Solving this equation gives us two velocities $v_s^\pm$ satisfying $\lambda_{v_s^\pm}=0$, which have the expressions
\begin{equation}
	v_s^-=1 - \mathcal{J} {\sqrt{2} + \sqrt{3} \mathcal{J}\over\sqrt{ 2} \mathcal{J} + \sqrt{3}}\ ,~~~~~~~ v_s^+=1 + \mathcal{J} {\sqrt{2} - \sqrt{3} \mathcal{J}\over \sqrt{3}-\sqrt{ 2} \mathcal{J}}\ .
\end{equation}
When the velocity $v<v_s^-$ or $v>v_s^+$, we have $\lambda_v<0$, so the saddle point contribution (\ref{Eq-butterfly1}) no longer gives an exponential growth in $t$, and is not valid for large $t$. In particular, we note that for any $0< J< 2\pi$ we have $v_s^->u_-$ and $v_s^+<u_+$ (Fig. \ref{exponent}(e)), so $v_s^\pm$ are within the causality cone of the system between $u_-$ and $u_+$ (where the retarded Green's function $G_R$ is nonzero).

If the contour deformation for Eq. (\ref{Eq-OTOC}) crosses a pole of the integrand, the integral will pick up both the pole contribution and the saddle point contribution. The weight coefficient $\rho(p)$ has two poles $p_1$ and $p_2$ satisfying $\varkappa(p_1)=\varkappa(p_2)=1$ on the positive imaginary axes of the two Riemann surface branches:
\begin{equation}
p_1={i\over u_+}\ ,~~~~~ p_2=i\left(\frac{2}{u_-}-\frac{1}{u_+}\right)\ ,
\end{equation}
and two poles $\widetilde{p}_1$ and $\widetilde{p}_2$ satisfying $\varkappa(\widetilde{p}_1)=\varkappa(\widetilde{p}_2)=-1$ on the imaginary axes of the two branches
\begin{equation}
\widetilde{p}_1=-\frac{i}{u_-}\ ,~~~~~ \widetilde{p}_2=-i\left(\frac{2}{u_+}-\frac{1}{u_-}\right)\ .
\end{equation}
The poles $p_2$ and $\widetilde{p}_2$ can be shown to never give a dominant contribution (Appendix \ref{Sec-SAP}).

As the velocity $v$ increases to $v>v_*=i\varkappa'(p_1)={2-2\mathcal{J}^2\over 2- \mathcal{J}}$, the saddle point $p_v$ will move upwards and pass the pole $p_1$, and (i.e., $v>v_*$) the integral (\ref{Eq-OTOC}) will pick up and be dominated by the residue of the pole $p_1$.
In particular, one always has $v_c<v_*<v_s^+$ for $0<J<2\pi$ (Fig. \ref{exponent}(f)), so this transition from saddle to pole contribution at $v=v^*$ happens before $\lambda_v$ given by the saddle point contribution decreases to zero. Therefore, for $v=x/t>v_*$, the OTOC function has the form
\begin{equation}\label{Eq-butterfly2}
	\delta \mathcal{F}(t,x)\sim e^{{2\pi \over \beta}(t+ip_1 x)}=e^{{2\pi \over \beta}(t-u_+^{-1}x)}\ ,~~~~~(x>v_* t)
\end{equation}
and the velocity dependent Lyapunov exponent $\lambda_v$ will be given by the pole contribution as $\lambda_v=2\pi(1-vu_+^{-1})/\beta$. In particular, one sees that $\lambda_v$ in this spacetime region decreases to zero exactly at velocity $v=u_+$, i.e., at the right causality edge. For velocity $v>u_+$, the growth behavior of $\delta\mathcal{F}$ disappears.

As the velocity $v$ decreases to $v<v_*'=i\varkappa'(\widetilde{p}_1)={2-2\mathcal{J}^2\over 2+ \mathcal{J}}$, the saddle point $p_v$ will move downwards and pass the pole $\widetilde{p}_1$. Accordingly, the integral (\ref{Eq-OTOC}) will be dominated by the residue of the pole $\widetilde{p}_1$. Similarly, one has $v_s^-<v_*'<v_c$ for $0<J<2\pi$ (Fig. \ref{exponent}(f)), so this transition from saddle to pole contribution at $v=v_*'$ also happens before $\lambda_v$ decreases to zero. So the OTOC for $v=x/t<v_*'$ is given by
\begin{equation}\label{Eq-butterfly3}
	\delta \mathcal{F}(t,x)\sim e^{{2\pi \over \beta}(-t+i\widetilde{p}_1 x)}=e^{{2\pi \over \beta}(-t+u_-^{-1}x)}\ ,~~~~~(x<v_*' t)
\end{equation}
and the velocity dependent Lyapunov exponent is $\lambda_v=2\pi(vu_-^{-1}-1)/\beta$. Therefore, one sees $\lambda_v$ also decreases to zero at velocity $v=u_-$, i.e., the left causality edge. For velocity $v<u_-$, the growth of $\delta\mathcal{F}$ disappears.

The numerical accuracy of the above saddle point approximation is demonstrated in Appendix \ref{Sec-SAP}. Fig. \ref{exponent}(f)-(g) shows the velocity dependent Lyapunov exponent $\lambda_v$ as a function of velocity $v=x/t$ for two cases $J=1.2\pi$ and $J=1.9\pi$, respectively. We can therefore define the butterfly cone of our model as $u_-<x/t<u_+$ as shown by the shaded regions in Fig. \ref{exponent}(d), inside (outside) which $\lambda_v>0$ ($\lambda_v<0$). Roughly speaking, the butterfly cone characterizes the range of spreading of local chiral operators as a function of time \cite{vedika2018}. It coincides exactly with the causality cone. We can summarize the behavior of the OTOC function $\delta\mathcal{F}$ as follows:

1) $v_-<{x/t}<v_*'$, in which case the pole $\widetilde{p}_1$ contribution dominates, and $\delta\mathcal{F}$ grows for $\beta\lesssim t\lesssim \beta\log N$ according to Eq. (\ref{Eq-butterfly3}). This is illustrated by the left dashed shaded region in Fig. \ref{exponent}(d). The velocity dependent Lyapunov exponent in this region is given by $\lambda_v=2\pi(vu_-^{-1}-1)/\beta$. Note that the $t$ direction Lyapunov exponent in this region is $-2\pi/\beta<0$, although the velocity dependent Lyapunov exponent is positive.

2) $v_*'<{x/t}<v_*$, where the saddle point $p_v$ contribution dominates, and $\delta\mathcal{F}$ grows exponentially in $t$ as given by Eq. (\ref{Eq-butterfly1}), which is valid for time $\beta\lesssim t\lesssim \beta\log N$. This butterfly region is illustrated by the solid shaded region in Fig. \ref{exponent}(d). The velocity dependent Lyapunov exponent $\lambda_v$ reaches its maximal value $\overline{\lambda}$ at velocity $v_c$ in this range (Eq. (\ref{Eq-vc})). Fig. \ref{exponent}(b) illustrates the OTOC $\mathcal{F}(t,v_ct)$ along the $x=v_ct$ direction, which is proportional to $1-e^{\overline{\lambda}(t-t_{sc})}$, where $t_{sc}\sim\mathcal{O}(\beta \log N)$ is the scrambling time.

3) $v_*<{x/t}<u_+$, where the pole $p_1$ contribution dominates, and the exponential growth of $\delta\mathcal{F}$ for $\beta\lesssim t\lesssim \beta\log N$ is given by Eq. (\ref{Eq-butterfly2}). This butterfly region is illustrated by the right dashed shaded region in Fig. \ref{exponent}(d). The velocity dependent Lyapunov exponent in this region is given by $\lambda_v=2\pi(1-vu_+^{-1})/\beta$. In addition, it is also worthwhile to note that for fixed $x>0$ in this spacetime region, $\delta\mathcal{F}$ grows in $t$ with an exponent $2\pi/\beta$, saturating the maximal chaos bound.

4) $x/t>u_+$ or $x/t<v_s^{-}$. In this region, $\delta\mathcal{F}$ no longer shows a growing behavior in $t$. Accordingly, the normalized OTOC $\mathcal{F}\sim 1-\frac{1}{N}\delta\mathcal{F}$ will tend to $1$ at large time $t$.

As one can see from the constant time $t$ slices of the (normalized) OTOC $\mathcal{F}(t,x)$ shown in Fig. \ref{exponent}(h) and (i) for $J=1.2\pi$ and $J=1.9\pi$, respectively, the exponential growth is peaked at velocity $v_c$ inside butterfly cone represented by the green shaded area. The peak becomes sharper and tends to $u_-$ as $J$ increases. In some sense, the behavior of the chiral SYK model looks like a nonchiral chaotic model with its edges of butterfly cone tilted to velocities $u_\pm$, and its time direction tilted to velocity $v_c$.

\section{The model at other small finite $N$}\label{Sec-56}
We have seen in Sec. \ref{Sec-N=4} that the $1+1$ dimensional chiral SYK model (\ref{Eq-action}) is integrable for $N=4$, and in Sec. \ref{Sec-SYK} that the model is chaotic in the large $N$ limit. Therefore, we conjecture there exists a critical number of Majorana flavors $N_c$, below which ($N<N_c$) the model is integrable, while equal to or above which ($N\ge N_c$) the model is non-integrable and chaotic. We note that here this is a well-defined statement, since the Hilbert space dimension of the $1+1$ dimensional model (\ref{Eq-action}) is infinite for any $N$, and the energy spectrum cannot be exactly solved (i.e., integrable) unless the model has infinite number of conserved quantities defined in terms of local operators. In contrast, the $0+1$ dimensional SYK model with $N$ Majorana fermions has a finite Hilbert space at any finite $N$, therefore it can always be exactly solved by diagonalizing a finite dimensional Hamiltonian matrix. In this sense, the $0+1$ dimensional SYK model is ``integrable" at any finite $N$, and there is no clear boundary between integrable and non-integrable during the increase of $N$.

In this section, we show that the $1+1$ dimensional chiral SYK model is also integrable by bosonization for $N=5$ and $N=6$. However, the interactions of the model becomes increasingly complicated for $N\ge7$. This leads us to conjecture that the boundary between integrable and chaotic is $N_c=7$.

The difficulty in solving the chiral SYK model (\ref{Eq-action}) for a finite $N>4$ comes from the fact that there are large number of interaction terms $J_{ijkl}\psi_i\psi_j\psi_k\psi_l$ which do not mutually commute. However, we can always rotate the chiral Majorana basis $\psi_i$ to a new basis $\psi_i'=O_{ij}\psi_j$, where $O$ is an SO(N) matrix, under which $J_{ijkl}$ transforms as a rank-4 tensor. In some cases, one can find a proper basis to reduce the number of interaction terms. This is the basic idea how we can solve the model for $N=5$ and $N=6$.

\subsection{Exact solution for $N=5$}

For the chiral model with $N=5$, there are $5$ independent interactions $J_{ijkl}$, which forms a total antisymmetric rank-4 tensor. We can redefine them using their Hodge dual as $I_i=\frac{1}{4!}\epsilon_{ijklm}J_{jklm}$ (repeated indices are summed), where $\epsilon_{ijklm}$ is the Levi-Civita symbol. $I_i$ is nothing but a rewriting of the $5$ interactions $J_{ijkl}$. The Lagrangian density of the model then becomes
\begin{equation}
\mathcal{L}=\frac{i}{2}\sum_{i=1}^5\psi_i(\partial_t+\partial_x)\psi_i +\frac{1}{4!}\epsilon_{ijklm}I_i \psi_j\psi_k\psi_l\psi_m\ .
\end{equation}
Under an SO(5) rotation $\psi_i'=O_{ij}\psi_j$, the interactions $I_i$ transform as an SO(5) vector. Therefore, we can find a proper SO(5) matrix $O_{ij}$ to rotate the vector $I_i$ to $I_i'=O_{ij}I_j=(0,0,0,0,I)$, where $I=\sqrt{\sum_{i=1}^5 I_i^2}$. Under this new basis, the Lagrangian becomes
\begin{equation}
\mathcal{L}=\frac{i}{2}\sum_{i=1}^5\psi_i'(\partial_t+\partial_x)\psi_i' +I \psi_1'\psi_2'\psi_3'\psi_4'\ .
\end{equation}
One easily sees that $\psi_5'$ is a free chiral Majorana fermion decoupled from the other four fermions $\psi_{1\le i\le4}'$. The Lagrangian for $\psi_{1\le i\le4}'$ simply reduces to the $N=4$ model (with interaction strength $I$) we discussed in Sec. \ref{Sec-N=4}, which is integrable by bosonization. Therefore, we see the $N=5$ model is integrable, which is equivalent to two free bosons with velocities $u_\pm$ and a free chiral Majorana fermion with velocity $1$. We can then easily obtain the averaged two point function for $N=5$ at temperature $\beta^{-1}$ as
\begin{equation}
\begin{split}
&G(-i\tau,x)\equiv\frac{1}{5}\sum_{i=1}^5\langle T \psi_i(\tau,x)\psi_i(0,0)\rangle= \frac{1}{5}\sum_{i=1}^5\langle T \psi_i'(\tau,x)\psi_i'(0,0)\rangle\\
&=\frac{2}{5\beta\sqrt{u_+u_-}}\frac{1}{\sqrt{\sin\left[ \frac{\pi}{\beta}(\tau-iu_+^{-1}x)\right] \sin\left[ \frac{\pi}{\beta}(\tau-iu_-^{-1}x)\right]}}+\frac{1}{10\beta}\frac{1}{\sin\left[ \frac{\pi}{\beta}(\tau-ix)\right]}\ ,
\end{split}
\end{equation}
where $u_\pm=1\pm \frac{I}{2\pi}$. Any higher $n$ point function can be calculated using the Wick's theorem. From the two point function $G(-i\tau,x)$ or from the bosonic picture, we can derive the finite temperature energy density to be $\mathcal{E}=\frac{\pi}{12\beta^2}(u_+^{-1}+u_-^{-1}+\frac{1}{2})$, and the energy current $j_\mathcal{E}=5\pi/24\beta^2$. This yields a quantized thermal Hall conductance $\kappa_{xy}=5\pi/12\beta$, and a thermal entropy density $\mathcal{S}=\frac{\pi}{6\beta}(u_+^{-1}+u_-^{-1}+\frac{1}{2})$.

\subsection{Exact solution for $N=6$}

Now we turn to the $N=6$ case. Following a similar idea, we can redefine the interaction rank-4 tensor $J_{ijkl}$ using its Hodge dual $I_{ij}=\frac{1}{4!}\epsilon_{ijklmn}J_{klmn}$, which is antisymmetric in $i$ and $j$. Under an SO(6) rotation $\psi_i'=O_{ij}\psi_j$, the interactions $I_{ij}$ transform as an antisymmetric rank-2 tensor. Therefore, there exists an SO(6) transformation $O_{ij}$ which brings $I_{ij}$ into the standard $2\times2$ block diagonal form
\begin{equation}
I_{ij}'=(O^TIO)_{ij}=
\left(\begin{array}{cccccc}
0&I_a&&&&\\
-I_a&0&&&&\\
&&0&I_b&&\\
&&-I_b&0&&\\
&&&&0&I_c\\
&&&&-I_c&0\\
\end{array}\right)\ .
\end{equation}
After changing into this new basis of $\psi_i'$, the Lagrangian of the model at $N=6$ becomes
\begin{equation}
\mathcal{L}=\frac{i}{2}\sum_{i=1}^6\psi_i'(\partial_t+\partial_x)\psi_i' +I_a\psi_3'\psi_4'\psi_5'\psi_6' +I_b\psi_1'\psi_2'\psi_5'\psi_6' +I_c\psi_1'\psi_2'\psi_3'\psi_4'\ .
\end{equation}
We can then bosonize the chiral Majorana fermions as $\psi_1'+i\psi_2'=e^{i\phi_1}$, $\psi_3'+i\psi_4'=e^{i\phi_2}$ and $\psi_5'+i\psi_6'=e^{i\phi_3}$, respectively, after which the Lagrangian becomes
\begin{equation}
\mathcal{L}=\frac{1}{4\pi}\sum_{i,j=1}^3\left(\delta_{ij}\partial_x\phi_i\partial_t\phi_j +V_{ij}\partial_x\phi_i\partial_x\phi_j\right)\ ,
\end{equation}
where the velocity matrix is a real symmetric matrix given by
\begin{equation}
V_{ij}=
\left(\begin{array}{ccc}
1&I_c/2\pi& I_b/2\pi\\
I_c/2\pi&1&I_a/2\pi\\
I_b/2\pi&I_a/2\pi&1\\
\end{array}\right)\ .
\end{equation}
Note that we have only quadratic terms of $\phi_i$ in the above Lagrangian. The velocity matrix can then be diagonalized into $V_{ij}'=\text{diag}(u_1,u_2,u_3)$ by an SO(3) transformation $\phi_i'=Q_{ij}\phi_j$, where $Q_{ij}$ is a $3\times3$ special orthogonal matrix. After the transformation, one gets $3$ decoupled free boson fields $\phi_i'$ with velocities $u_i$ ($i=1,2,3$), respectively, and the Lagrangian becomes
\begin{equation}
\mathcal{L}=\frac{1}{4\pi}\sum_{i=1}^3\partial_x\phi_i'\left(\partial_t +u_i\partial_x\right) \phi_i'\ .
\end{equation}
Therefore, the model is also integrable via bosonization for $N=6$. One can then show the averaged fermion two point function for $N=6$ at temperature $\beta^{-1}$ to be given by
\begin{equation}
G(-i\tau,x)= \frac{1}{6}\sum_{i=1}^6\langle T \psi_i'(\tau,x)\psi_i'(0,0)\rangle =\frac{1}{6\beta}\sum_{j=1}^3\prod_{i=1}^3\left\{u_i\sin\left[ \frac{\pi}{\beta}(\tau-iu_i^{-1}x)\right]\right\}^{-Q_{ij}^2} \ .
\end{equation}
Any higher $n$ point function can be calculated using the Wick's theorem.
It is also straightforward to derive the finite temperature energy density $\mathcal{E}=\frac{\pi}{12\beta^2}(u_1^{-1}+u_2^{-1}+u_3^{-1})$, and the energy current $j_\mathcal{E}=\pi/4\beta^2$. This yields a quantized thermal Hall conductance $\kappa_{xy}=\pi/2\beta$, and a thermal entropy density $\mathcal{S}=\frac{\pi}{6\beta}(u_1^{-1}+u_2^{-1}+u_3^{-1})$.

\subsection{$N\ge7$ and thermalization of chiral Majorana fermions}

The interaction terms $J_{ijkl}$ of the chiral SYK model become increasingly complicated when $N\ge7$. For instance, for $N=7$, the rank-4 total antisymmetric tensor $J_{ijkl}$ has $35$ parameters in total. One can try to make some of them zero by an SO($7$) rotation $\psi_i'=O_{ij}\psi_j$. However, since the SO(7) group has only $21$ generators, the SO(7) rotation matrix $O_{ij}$ can at most reduce $21$ or less of the interaction parameters $J_{ijkl}$ to zero, while there are still $14$ or more nonzero parameters $J_{ijkl}$ remaining. After bosonization, these remaining $14$ or more interaction terms cannot all be bilinears of boson fields $\phi_i$ of the form $\partial_x\phi_i\partial_x\phi_j$ but necessarily contain many nonlinear terms of the form $\cos(\phi_i\pm\phi_j\pm\phi_k\pm\phi_l)$. Therefore, the model is no longer free bosons for $N=7$, which makes the model extremely complicated. This is also the case for all $N>7$. Therefore, we conjecture the model is no longer integrable for $N\ge N_c$ and starts to show chaotic behaviors in some aspects, where the critical number $N_c\ge7$, and is probably equal to $7$.

\begin{figure}[htbp]
\begin{center}
\includegraphics[width=2.5in]{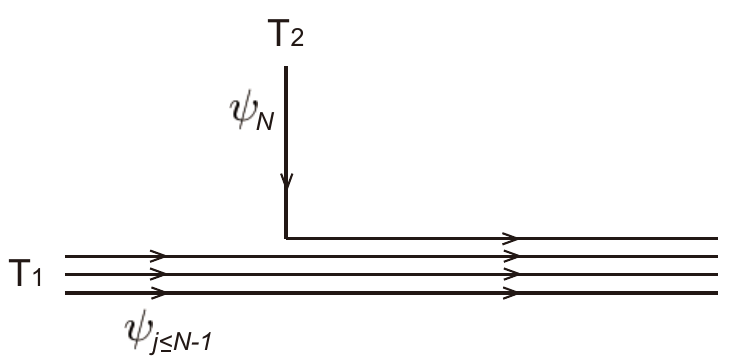}
\end{center}
\caption{A configuration that different flavors of chiral Majorana fermions have different initial temperatures, and may or may not thermalize with each other later.}
\label{thermal}
\end{figure}

One of the physical differences between $N<N_c$ and $N\ge N_c$ will be the thermalization of the model, which is closely related to integrability. For instance, consider the spatial configuration of a set of chiral Majorana edge states as shown in Fig. \ref{thermal}, where the $N$-th mode $\psi_N$ is initially spatially separated from the other $N-1$ modes $\psi_{j\le N-1}$, and merges with them after propagating a distance. Such a configuration can be realized by using the edges of $N$ copies of the $p+ip$ chiral topological superconductor. Assume $\psi_{j\le N-1}$ and $\psi_N$ are connected with thermal baths of temperature $T_1$ and $T_2$, respectively. For $N<N_c$, the model is integrable, so one expects the $N$ chiral Majorana fermion modes do not thermalize with each other during propagation. Instead, the model with $N\ge N_c$ is non-integrable, and will have all the chiral Majorana fermion modes thermalize among themselves after they merge together.

The thermalization of the chiral edge states of a topological condensed matter system may affect the experimental measurement of its thermal Hall conductance. For instance, the recent experiment observed a thermal Hall conductance of the filling $\nu=5/2$ fractional quantum Hall state disagreeing with theoretical predictions \cite{banerjee2018,mross2018,wangc2018,lian2018}, and one of the possible explanations is suggested to be the lack of thermal equilibrium among the chiral edge modes \cite{simon2018,feldman2018,ma2019}. Our model here might be too idealized to describe a realistic edge of condensed matter system, which always have spatial disorders and coupling to phonons, etc. However, our results still suggest that thermalization is more difficult for fewer number of chiral edge modes.

\section{In closing}\label{Sec-con}

We have introduced a new chiral model in $1+1$ dimensions which is solvable in the large $N$ limit like its $0+1$ dimensional namesake and tensor models. Indeed, the exact scaling symmetry of our $1+1$ dimensional chiral SYK model makes it exactly solvable at all energy scales, thus makes it a ``simpler" chaotic model than the the $0+1$ dimensional SYK model which is only exactly solvable at low energies. We have calculated the two point function of the model, and verified that the thermal Hall conductance of the chiral model at nonzero interaction strength $J>0$ in the large $N$ limit is quantized at $\kappa_{xy}=N\pi/12\beta$, which is generically believed to be a topological invariant. We also find the ground state entropy at nonzero interaction strength $J$ is no different from that of $N$ free chiral Majorana fermions, which leads to a vanishing ground state entropy density.

The OTOC in the large $N$ limit generically has a Lyapunov regime and an asymmetric butterfly cone. The velocity dependent Lyapunov exponent $\lambda_v$ along $x=vt$ has its maximum reaching the chaos bound $2\pi/\beta$ when the interaction strength $J$ approaches the upper bound $2\pi$. Therefore, the model can approach the maximally chaotic limit without having a nonzero ground state entropy density, which is not expected from a black hole picture. Besides, for any interaction strength $J>0$, the $t$ direction Lyapunov exponent $\lambda$ for a fixed $x$ near the right butterfly edge always saturates the chaos bound $2\pi/\beta$, while near the left butterfly edge the $t$ direction Lyapunov exponent $\lambda$ is always $-2\pi/\beta$ (although the velocity dependent Lyapunov exponent is positive). This feature near the left butterfly edge is different from that in nonchiral models. Furthermore, the model is integrable for small $N\le6$ as can be seen by bosonization, but becomes increasingly complicated for $N\ge7$. Therefore, we conjecture there is a transition from integrablity to chaos as $N$ increases to some number $N_c$, and the transition is probably at $N_c=7$.

Somewhat surprisingly we find that the two point function of the chiral SYK model in the large $N$ limit is exactly the same as that for $N=4$, although the four point functions of the two cases are completely different. We conjecture such a coincidence of two-point functions is because of the presence of the SO($N$) symmetry in both the large $N$ case (effectively) and the $N=4$ case.

It will also be interesting to investigate whether this model has Schwarzian like soft modes and an effective Schwarzian action as in \cite{Turiaci:2017zwd}. If so, we could ask whether there is a holographic interpretation.
Besides, in this paper we only calculate the OTOC in the late time chaos regime. To get more information about the operators in the model, one can study the OPE expansion of the four point function. Furthermore, we have been restricting ourselves to interaction strengths within the range $0\le J<2\pi$ in this paper. One may ask what happens for large $N$ if $J>2\pi$, with proper UV nonlinearities included to preserve the total chirality (as shown in Fig. \ref{largeK}(c)). In this case, the model is no longer a purely chiral model. We shall leave these questions to future studies.

We briefly discuss how spatial disorder may affect the model, which always exists in practical condensed matter systems. The most relevant disordered term one can add to the Hamiltonian (\ref{Eq-H}) is a quadratic fermion term
\begin{equation}
\Delta\mathcal{H}_1(x)=2i\sum_{1=i<j}^N A_{ij}(x)\psi_i\psi_j\ ,
\end{equation}
where $A_{ij}(x)$ is an $x$ dependent real antisymmetric matrix. However, such a term can be eliminated by a spatially dependent SO(N) rotation $\psi_i\rightarrow (e^{-\int^x_{-\infty} A(x')dx'})_{ij}\psi_j$. In addition, the 4-fermion coupling $J_{ijkl}$ may also has an $x$ dependence due to disorder or the above spatially dependent SO(N) rotation. However, as long as the couplings $J_{ijkl}$ still have long range spatial correlations, namely, $\langle J_{ijkl}(x)J_{ijkl}(x')\rangle\rightarrow 3!J^2/N^3$ as $|x-x'|\rightarrow\infty$, we expect the physics in this paper remains unchanged in the IR.


Lastly, we comment that interacting chiral models with a large number of degrees of freedom (either Majorana fermions or complex fermions) arise naturally in two settings. The first is, naturally enough, that of ``infinite layer''
quantum Hall systems \cite{chalker1995,balents1996,naud2000,naud2001} or $p+ip$ chiral superconductors. The second is the low energy theory of a piece of Fermi surface (of complex fermions with a conserved U(1) charge) in spatial dimension $d_s\ge2$ with forward scatterings \cite{Patel:2017,Patel:2017b}. In this case, the fermion dispersion $\omega=v_Fk_\perp$ is linear in the momentum $k_\perp$ normal to the Fermi surface, where $v_F$ is the Fermi velocity. Therefore, one can treat the fermion at each fixed transverse momentum $\mathbf{k}_\parallel$ tangent to the Fermi surface as a chiral fermion in the $k_\perp$ direction, and regard $\mathbf{k}_\parallel$ as a flavor index which runs over large $N$ number of values. We expect our chiral SYK model to further shed lights on the general understanding of such chiral systems.


\section{Acknowledgments}

We thank Y. Gu, F.D.M. Haldane, I. Klebanov, J. Maldacena, X.L. Qi, N. Seiberg, D. Stanford, H. Verlinde, J.Y. Chen and X.Q. Sun for helpful discussions. We also thank G. S\'arosi and M. Mezei for valuable comments on our OTOC calculations \cite{mezei2019}. BL is supported by Princeton Center for Theoretical Sciences at Princeton University. SLS acknowledges support from the United States Department of Energy via grant No. DE-SC0016244. ZY is supported by Charlotte Elizabeth Procter Fellowship from Princeton University.

\appendix
\section{Bosonization and point splitting}\label{App-OPE}
In this appendix, we show how to deal with the $x$ direction point splitting regularization of chiral Majorana fermion fields $\psi_i$ and $\psi_j$ via bosonization, and derive the commutation relation (\ref{Eq-commu2}) and the energy momentum tensor $T^\mu_{\ \nu}$.

\subsection{Bosonization and normal ordering}
We shall follow the procedure of bosonization to regularize the point splitting. Given two chiral Majorana fermion fields $\psi_i$ and $\psi_j$ ($i\neq j$), we can define a complex fermion $c$ and its bosonization
\begin{equation}
\frac{\psi_i(t,x)+i\psi_j(t,x)}{\sqrt{2}}=c(t,x)=\normord{e^{i\phi(t,x)}}\ ,\qquad \frac{\psi_i(t,x)-i\psi_j(t,x)}{\sqrt{2}}=c^\dag(t,x)=\normord{e^{-i\phi(t,x)}}
\end{equation}
where $\phi$ is a bosonic field, and $\normord{\mathcal{O}}$ stands for the normal ordering of operator $\mathcal{O}$. Since all the point splitting we are considering is along the $x$ direction, we assume all the fields in the below are on the same time slice, i.e., at the same $t$. For simplicity, we shall therefore omit the time variable $t$ of all fields.

On the constant time slice, the chiral Majorana fermion fields satisfy the anticommutation relation $\{c(x),c^\dag(x')\}=\{c(x),c(x')\}=\{c^\dag(x),c^\dag(x')\}=0$ for $x\neq x'$, which requires the boson commutation relation
\begin{equation}\label{Eq-bc1}
[\phi(x),\phi(x')]=i\pi \text{sgn}(x-x')\ ,
\end{equation}
and thus
\begin{equation}\label{Eq-bc2}
[\partial_x\phi(x),\phi(x')]=2\pi i \delta(x-x')\ .
\end{equation}
To correctly deal with normal ordering, we can separate the boson field $\phi$ into
\begin{equation}
\phi(x)=\varphi(x)+\varphi^\dag(x)\ ,
\end{equation}
where $\varphi(x)$ and $\varphi^\dag(x)$ are the collection of annihilation and creation operators in the mode expansion, respectively. The normal ordering rule is then that $\varphi^\dag(x)$ should always be on the left of $\varphi(x)$. They satisfy $[\varphi(x),\varphi(x')]=[\varphi^\dag(x),\varphi^\dag(x')]=0$. Using Eq. (\ref{Eq-bc2}) and the fact that
\begin{equation}
2\pi i\delta (x-x')=\frac{1}{x-x'-i0^+}+\frac{1}{x'-x-i0^+}
\end{equation}
where $0^+$ represents positive infinitesimal, one finds that
\begin{equation}
\begin{split}
[\partial_x\varphi(x),\varphi^\dag(x')]=\frac{1}{x'-x-i0^+}\ ,\qquad [\partial_x\varphi^\dag(x),\varphi(x')]=\frac{1}{x-x'-i0^+}\ .
\end{split}
\end{equation}
Integrating the two equations yields
\begin{equation}\label{Eq-bc3}
[\varphi(x),\varphi^\dag(x')]=-\log [-2\pi i(x-x'+i0^+)]\ ,
\end{equation}
where the constant of integration is chosen for normalization reason. In particular, from Eq. (\ref{Eq-bc3}) one can verify that $[\phi(x),\phi(x')] =-\log\left(\frac{x-x'+i0^+}{x'-x+i0^+}\right)=i\pi\text{sgn}(x-x')$, in agreement with Eq. (\ref{Eq-bc1}).

\subsection{Operator product expansion and commutation relation}
We can now calculate the operator product expansion of chiral Majorana fermion fields, with a point splitting in the $x$ direction. First, we consider the bilinear $c^\dag(x)c(x')$ with small $x$ direction separation $x-x'$:
\begin{equation}\label{Eq-ope-cdc}
\begin{split}
&c^\dag(x)c(x')=\normord{e^{-i\phi(x)}}\normord{e^{i\phi(x')}}
=e^{-i\varphi^\dag(x)}e^{-i\varphi(x)} e^{i\varphi^\dag(x')}e^{i\varphi(x')}\\
=&e^{-i\varphi^\dag(x)}\left(e^{[\varphi(x),\varphi^\dag(x')]} e^{i\varphi^\dag(x')} e^{-i\varphi(x)}\right)e^{i\varphi(x')}
=\frac{e^{-i\left(\varphi^\dag(x)-\varphi^\dag(x')\right)} e^{-i\left(\varphi(x)-\varphi(x')\right)}}{-2\pi i(x-x'+i0^+)}\\
=&\frac{i}{2\pi(x-x'+i0^+)}\Big\{1-i(x-x')\left(\partial_x \varphi^\dag +\partial_x \varphi\right)\\
&-\frac{(x-x')^2}{2}\Big[(\partial_x\varphi^\dag)^2+ 2\partial_x\varphi^\dag\partial_x\varphi +(\partial_x\varphi)^2 +i\partial_x^2\varphi^\dag+i\partial_x^2\varphi
\Big]\Big\}\\
=&\frac{i}{2\pi(x-x'+i0^+)}\Big\{1-i(x-x')\partial_x \phi
-\frac{(x-x')^2}{2}\Big[\normord{(\partial_x\phi)^2} +i\partial_x^2\phi
\Big]+\mathcal{O}\Big((x-x')^3\Big)\Big\}\\
=&\frac{i}{2\pi(x-x'+i0^+)}+\frac{\partial_x \phi}{2\pi}-\frac{i(x-x')}{4\pi} \Big[\normord{(\partial_x\phi)^2} +i\partial_x^2\phi
\Big]+\mathcal{O}\Big((x-x')^2\Big)\ .
\end{split}
\end{equation}
The first constant term $\frac{i}{2\pi(x-x'+i0^+)}$ is the vacuum term. Therefore, by taking the limit $x-x'\rightarrow0$, we find the normal ordered density operator is
\begin{equation}\label{Eq-ope-n}
\normord{i\psi_i(x)\psi_j(x)}=\normord{c^\dag(x)c(x)}=\frac{\partial_x \phi(x)}{2\pi}\ .
\end{equation}
In particular, the density operator is proportional to $\partial_x\phi$ because the point splitting is in the $x$ direction. Then using Eq. (\ref{Eq-bc2}), we can also derive the following expected commutation relation:
\begin{equation}
\begin{split}
&[\normord{\psi_i(x)\psi_j(x)},\psi_i(x')]=\left[-i\frac{\partial_x \phi(x)}{2\pi}, \frac{\normord{e^{i\phi(x')}}+\normord{e^{-i\phi(x')}}}{\sqrt{2}}\right] \\ =&\delta(x-x')\frac{i(\normord{e^{i\phi(x')}}-\normord{e^{-i\phi(x')}})}{\sqrt{2}} =-\delta(x-x')\psi_j(x)\ .\\
\end{split}
\end{equation}

Next, we consider the kinetic term by point splitting:
\begin{equation}\label{Eq-ope-k}
\begin{split}
&-\frac{i}{2}\left[\psi_i(x)\partial_x\psi_i(x)+\psi_j(x)\partial_x\psi_j(x)\right] =-ic^\dag(x)\partial_xc(x)\\
\approx &i\left(\frac{c^\dag(x)+c^\dag(x')}{2}\right) \left(\frac{c(x)-c(x')}{x-x'}\right)\\
=&-\frac{i}{2}\left[\frac{c^\dag(x)c(x)-c^\dag(x')c(x')}{x-x'} +\frac{c^\dag(x')c(x)-c^\dag(x)c(x')}{x-x'}\right]\\
=&-\frac{i}{2}\left\{\frac{\partial_x\phi(x)-\partial_x\phi(x')}{2\pi(x-x')}-\frac{i}{2\pi (x-x')^2}\left[2-(x-x')^2\left(\normord{(\partial_x\phi)^2} +i\partial_x^2\phi \right)\right]\right\}\\
=&-\frac{1}{2\pi(x-x')^2}+\frac{1}{4\pi}\normord{(\partial_x\phi)^2}\ ,
\end{split}
\end{equation}
where we have split the two fermion fields at points $x$ and $x'$, respectively, and used the OPE in Eq. (\ref{Eq-ope-cdc}). By taking the limit $x-x'\rightarrow0$, we find the normal ordered kinetic term is
\begin{equation}
-\frac{i}{2}\normord{\left[\psi_i(x)\partial_x\psi_i(x)+\psi_j(x)\partial_x\psi_j(x)\right]} =-i\normord{c^\dag(x)\partial_xc(x)} =\frac{1}{4\pi}\normord{[\partial_x\phi(x)]^2}\ .
\end{equation}

When there are more than $2$ chiral Majorana fermion fields $\psi_i$, we can define multiple boson fields, for instance
\begin{equation}
\frac{\psi_{2i-1}(x)+i\psi_{2i}(x)}{\sqrt{2}}=c_i(x)=\eta_i\normord{e^{i\phi_i(x)}}\ ,
\end{equation}
where $1\le i\le[N/2]$, and $\eta_i$ are the Klein factors satisfying $\{\eta_i,\eta_j\} =\{\eta_i^\dag,\eta_j\} =\{\eta_i^\dag,\eta_j^\dag\}=0$ for $i\neq j$, and $\eta_i^\dag\eta_i =\eta_i\eta_i^\dag=1$ \cite{delft1998}. Different flavors of boson $\phi_i$ commute with each other. The use of the Klein factors is only to ensure the anticommutation relation between different fermion fields, and can usually be neglected in calculations where only bosonic terms are involved.

We can now derive the commutation relation (\ref{Eq-commu2}) of fermion bilinears, $[(\psi_i\psi_j)(x),(\psi_k\psi_l)(x')]$. First, the commutation is trivially zero when $i=j$ or $k=l$. Secondly, when $i,j,k,l$ are all distinct, the commutation is also obviously zero. Thirdly, when $i=k\neq j=l$, we can define $(\psi_i+i\psi_j)/\sqrt{2}=\normord{e^{i\phi}}$, and use Eqs. (\ref{Eq-bc2}) and (\ref{Eq-ope-n}) to find
\begin{equation}
[(\psi_i\psi_j)(x),(\psi_i\psi_j)(x')]=-\left[\frac{\partial_x \phi(x)}{2\pi},\frac{\partial_{x'} \phi(x')}{2\pi}\right] =\frac{i}{2\pi}\partial_x\delta(x-x')\ .
\end{equation}
Lastly, when $i=k\neq j$ and $j\neq l\neq i$, we can define two boson fields $\phi$ and $\phi'$ by $(\psi_i+i\psi_j)/\sqrt{2}=\eta\normord{e^{i\phi}}$ and $(\psi_l+i\psi_m)/\sqrt{2}=\eta'\normord{e^{i\phi'}}$, where $\psi_m$ is another arbitrary fermion field, while $\eta$ and $\eta'$ are the Klein factors. The commutator in this case is then
\begin{equation}
\begin{split}
&[(\psi_i\psi_j)(x),(\psi_i\psi_l)(x')]=-i\left[\frac{\partial_x \phi(x)}{2\pi},\frac{(\normord{e^{i\phi(x')}}+\normord{e^{-i\phi(x')}}) (\normord{e^{i\phi'(x')}}+\normord{e^{-i\phi'(x')}})}{2}\right] \\
=&\delta(x-x') \frac{i(\normord{e^{i\phi(x')}}-\normord{e^{-i\phi(x')}}) (\normord{e^{i\phi'(x')}}+\normord{e^{-i\phi'(x')}})}{2} =-\delta(x-x')\psi_j\psi_l\ .
\end{split}
\end{equation}
We can then summarize these cases into the commutation relation (\ref{Eq-commu2}). Note that since all $i\psi_i\psi_j$ ($1\le i<j\le N$) form the generators of SO$(N)$ group, so Eq. (\ref{Eq-commu2}) is nothing but the SO$(N)_1$ Kac-Moody algebra.

Besides, for each pair of fermion fields $\psi_i$ and $\psi_j$ ($i\neq j$), from Eqs. (\ref{Eq-ope-n}) and (\ref{Eq-ope-k}) we can formally rewrite their kinetic term as
\begin{equation}
-\frac{i}{2}\left[\psi_i(x)\partial_x\psi_i(x)+\psi_j(x)\partial_x\psi_j(x)\right] =-\pi \normord{\psi_i(x)\psi_j(x)}\normord{\psi_i(x)\psi_j(x)}\ .
\end{equation}
Therefore, we can rewrite the total kinetic term of $N$ flavors of $\psi_i$ as
\begin{equation}
-\frac{i}{2}\sum_{i}\psi_i(x)\partial_x\psi_i(x)= -\frac{\pi}{N-1} \sum_{i<j}\normord{\psi_i(x)\psi_j(x)}\normord{\psi_i(x)\psi_j(x)}\ ,
\end{equation}
which is simply the Sugawara construction of the kinetic term.

\subsection{Energy momentum tensor}
Here we derive the energy momentum tensor of the $1+1$ dimensional chiral SYK model. The energy density can be simply derived from Legendre transformation of the Lagrangian, as shown in Eq. (\ref{Eq-H}):
\begin{equation}
T^{0}_{\ 0}=\mathcal{H}=-\frac{i}{2}\sum_{i=1}^N\psi_i\partial_x\psi_i -\sum_{1\le i<j<k<l\le N}J_{ijkl}\psi_i\psi_j\psi_k\psi_l\ .
\end{equation}

In general, the full energy momentum tensor can be derived from Noether's theorem. However, to avoid complications from the point-splitting regularization, we will not use Noether's theorem, instead we will directly employ the energy momentum conservation law.

To derive the energy current $T^{x}_{\ 0}$, we consider the energy conservation law $\partial_tT^{0}_{\ 0}+\partial_xT^{x}_{\ 0}=i[H,\mathcal{H}]+\partial_xT^{x}_{\ 0}=0$, namely, $\partial_xT^{x}_{\ 0}(x)=-i\int dx'[\mathcal{H}(x'),\mathcal{H}(x)]$. To calculate the right hand side, we first note that from the bosonized expressions above, one can verify that
\begin{equation}
\begin{split}
&\int dx'\left[-\frac{i}{2}\psi_i(x')\partial_{x'}\psi_i(x'), -\frac{i}{2}\psi_j(x)\partial_x\psi_j(x)\right] 
=\frac{\delta_{ij}}{2}\int dx'\left[\frac{\normord{[\partial_{x'}\phi(x')]^2}}{4\pi}, \frac{\normord{[\partial_{x}\phi(x)]^2}}{4\pi}\right] \\ =&-i \delta_{ij}\int dx' \frac{ \normord{\partial_{x'}\phi(x')\partial_{x}\phi(x)}}{4\pi} \partial_{x'}\delta(x-x') =i\delta_{ij}\partial_x \left(\frac{\normord{[\partial_{x}\phi(x)]^2}}{8\pi}\right) =\frac{\delta_{ij}}{2}\partial_x\left[\psi_j(x)\partial_{x}\psi_j(x)\right],
\end{split}
\end{equation}
\begin{equation}
\begin{split}
&\int dx'\left[-\frac{i}{2}\psi_i(x')\partial_{x'}\psi_i(x'), (\psi_i\psi_j\psi_k\psi_l)(x)\right]= \frac{1}{2}\int dx'\left[\frac{\normord{[\partial_{x'}\phi(x')]^2}}{4\pi}, \frac{\normord{\partial_{x}\phi(x) \partial_{x}\phi'(x)}}{4\pi^2}\right]\\
=& -i \int dx' \frac{\normord{\partial_{x'}\phi(x') \partial_{x}\phi'(x)}}{8\pi^2} \partial_{x'}\delta(x-x') = \frac{i}{2} [\partial_x(\psi_i\psi_j)(x)](\psi_k\psi_l)(x)\ ,
\end{split}
\end{equation}
and
\begin{equation}
\begin{split}
&\int dx'\left[(\psi_i\psi_j\psi_k\psi_l)(x'), -\frac{i}{2}\psi_i(x)\partial_{x}\psi_i(x)\right]= \frac{1}{2}\int dx'\left[\frac{\normord{\partial_{x'}\phi(x') \partial_{x'}\phi'(x')}}{4\pi^2}, \frac{\normord{[\partial_{x}\phi(x)]^2}}{4\pi} \right]\\
=& -i \int dx' \frac{\normord{\partial_{x'}\phi'(x') \partial_{x}\phi(x)}}{8\pi^2} \partial_{x'}\delta(x-x') = \frac{i}{2} (\psi_i\psi_j)(x) [\partial_x (\psi_k\psi_l)(x)]\ .
\end{split}
\end{equation}
We also need to know the commutator $[(\psi_{i'}\psi_{j'}\psi_{k'}\psi_{l'})(x'), (\psi_i\psi_j\psi_k\psi_l)(x)]$. The general expression is rather complicated, which we shall not derive here. Instead, we note that the coefficient of the commutator $[(\psi_{i'}\psi_{j'}\psi_{k'}\psi_{l'})(x'), (\psi_i\psi_j\psi_k\psi_l)(x)]$ in $[\mathcal{H}(x'),\mathcal{H}(x)]$ is $J_{i'j'k'l'}J_{ijkl}$. Here we shall only derive the energy current $T^{x}_{\ 0}$ averaged over $J_{ijkl}$ in the large $N$ limit, so one expects this commutator has a nonzero contribution only if its coefficient $J_{i'j'k'l'}J_{ijkl}$ does not average to zero, namely, only if $(i'j'k'l')=(ijkl)$ up to permutations. The commutator in this case can be easily derived to be
\begin{equation}
\begin{split}
&\int dx'\left[(\psi_i\psi_j\psi_k\psi_l)(x'), (\psi_i\psi_j\psi_k\psi_l)(x)\right]= \int dx'\left[\frac{\normord{\partial_{x'}\phi(x') \partial_{x'}\phi'(x')}}{4\pi^2}, \frac{\normord{\partial_{x}\phi(x) \partial_{x}\phi'(x)}}{4\pi^2} \right]\\
=& -i \int dx' \frac{[\normord{\partial_{x'}\phi'(x') \partial_{x}\phi'(x)} +\normord{\partial_{x'}\phi(x') \partial_{x}\phi(x)}]}{8\pi^3} \partial_{x'}\delta(x-x')\\
=& \frac{1}{8\pi^2} \partial_x \left[\psi_i(x)\partial_{x}\psi_i(x) + \psi_j(x)\partial_{x}\psi_j(x) + \psi_k(x)\partial_{x}\psi_k(x) + \psi_l(x)\partial_{x}\psi_l(x) \right]\ .
\end{split}
\end{equation}
The other terms will take the form such as $J_{ijk'l'}J_{ijkl}\partial_x(\psi_{k'}\psi_{l'}\psi_k\psi_l)(x)$, which we expect will have a zero average value, since $J_{ijk'l'}J_{ijkl}$ is random and has no correlation with $\psi_{k'}\psi_{l'}\psi_k\psi_l$.

Therefore, one finds the energy current $T^{x}_{\ 0}$ averaged over $J_{ijkl}$ satisfying
\begin{equation}
\begin{split}
&\partial_xT^{x}_{\ 0}(x)=-i\int dx'[\mathcal{H}(x'),\mathcal{H}(x)]\\
=& \partial_x\left[-\frac{i}{2}\sum_{i=1}^N\psi_i\partial_x\psi_i- 2\sum_{i<j<k<l}J_{ijkl}\psi_i\psi_j\psi_k\psi_l - \sum_{i<j<k<l}\frac{iJ_{ijkl}^2}{8\pi^2} \left(\sum_{a=i,j,k,l}\psi_a\partial_x\psi_a\right) \right]\\
=& \partial_x\left[\frac{1}{2}\sum_{i=1}^N\psi_i\left(-i\partial_x\psi_i- \sum_{j<k<l}J_{ijkl}\psi_j\psi_k\psi_l\right) - \frac{N!}{4!(N-4)!}\frac{4i\langle J_{ijkl}^2\rangle}{8\pi^2 N} \sum_{i=1}^N\psi_i\partial_x\psi_i \right]\\
=& \partial_x\left[\frac{i}{2}\sum_{i=1}^N\psi_i\left(\partial_t- \frac{J^2}{4\pi^2}\partial_x\right)\psi_i\right]\ ,
\end{split}
\end{equation}
where in the last line we have used the equation of motion (\ref{Eq-EOM}). Therefore, we arrive at the average energy current
\begin{equation}
T^{x}_{\ 0}= \frac{i}{2}\sum_{i=1}^N\psi_i\left(\partial_t- \frac{J^2}{4\pi^2}\partial_x\right)\psi_i\ .
\end{equation}
We note that this expression is exact instead of an average when $N=4$, in which case there is only one interaction $J_{1234}=J$.

We now turn to the momentum density $T^{0}_{\ x}$ and momentum current $T^{x}_{\ x}$. The momentum density can be directly written down by its definition as
\begin{equation}
T^{0}_{\ x}= -\frac{i}{2}\sum_{i=1}^N\psi_i\partial_x\psi_i\ .
\end{equation}
The energy current can then be derived from the momentum conservation law:
\begin{equation}
\begin{split}
&\partial_xT^{x}_{\ x}(x)=-i\int dx'[\mathcal{H}(x'),T^{0}_{\ x}(x)]
= \partial_x\left[-\frac{i}{2}\sum_{i=1}^N\psi_i\partial_x\psi_i- \sum_{i<j<k<l}J_{ijkl}\psi_i\psi_j\psi_k\psi_l\right] \\ =&\partial_x\left[\frac{1}{2}\sum_{i=1}^N\psi_i\left(-i\partial_x\psi_i- \frac{1}{2}\sum_{j<k<l}J_{ijkl}\psi_j\psi_k\psi_l\right)\right]= \partial_x\left[ \frac{i}{4}\sum_{i=1}^N\psi_i(\partial_t-\partial_x)\psi_i\right]\ ,
\end{split}
\end{equation}
where again we have used the equation of motion (\ref{Eq-EOM}). Therefore, we find
\begin{equation}
T^{x}_{\ x}= \frac{i}{4}\sum_{i=1}^N\psi_i(\partial_t-\partial_x)\psi_i\ .
\end{equation}
Note that the momentum density $T^{x}_{\ x}$ is in fact the same as the energy density $T^{0}_{\ 0}$.

\section{Saddle Point Approximation for the OTOC}\label{Sec-SAP}

In this section, we give more details on the saddle point approximation for the OTOC integral (\ref{Eq-OTOC}) in the large $N$ limit. In particular, the contour deformation requires a careful consideration of both branches of the Riemann surface of the function $\varkappa(p)$ defined in Eq. (\ref{Eq-varkappa}).

\begin{figure}[tbp]
\begin{center}
\includegraphics[width=5.5in]{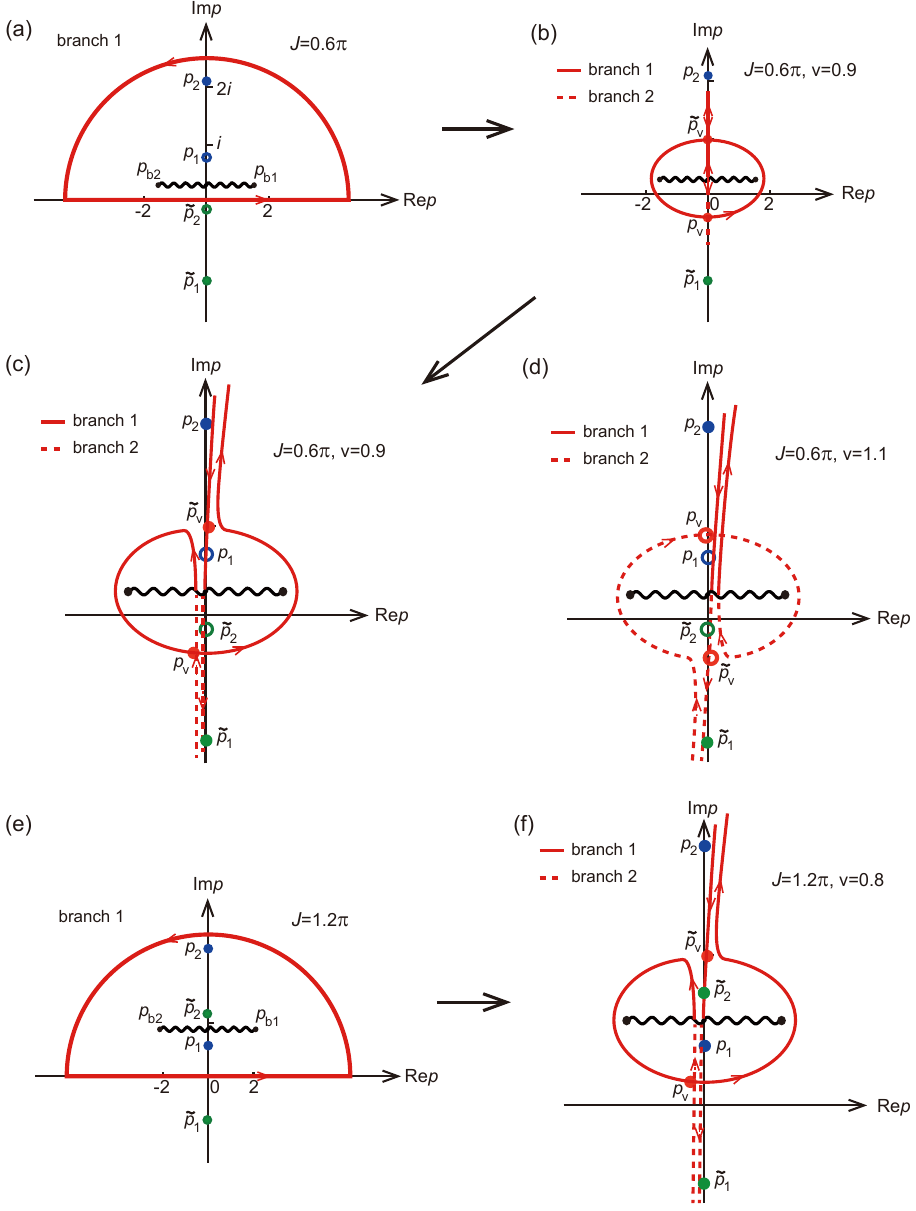}
\end{center}
\caption{Illustration of contour deformation. Contours in branch $1$ (branch $2$) are plotted in solid (dashed) lines, respectively. (a) The original contour of Eq. (\ref{Eq-OTOC}) in branch $1$ for $J=0.6\pi$. (b) The saddle point contours of $p_v$ and $\widetilde{p}_v$, which accidentally hit together. The parameters are $J=0.6\pi$, $v=0.9$. (c) Illustration of the resolved saddle point contours for $u_-<v<1$ ($v=0.9$ as an example) by adding a small imaginary part to velocity $v$. (d) The resolved saddle point contours for $1<v<u_+$ ($v=1.1$ as an example). (e) The original contour in branch $1$ for $J=1.2\pi$. (f) The deformed saddle point contours for $J=1.2\pi$, $v=0.8$.}
\label{saddle}
\end{figure}

The function $\varkappa(p)$ in Eq. (\ref{Eq-varkappa}) has two branch points in the upper half complex plane at
\begin{equation}
p_{b1}=\frac{i\mathcal{J}+\sqrt{3(1-\mathcal{J}^2)}}{1-\mathcal{J}^2}\ ,\qquad  p_{b2}=\frac{i\mathcal{J}-\sqrt{3(1-\mathcal{J}^2)}}{1-\mathcal{J}^2}\ ,
\end{equation}
where $\mathcal{J}=J/2\pi\in[0,1)$ as we have defined previously. As a result, $\varkappa(p)$ lives on a Riemann surface with two branches. To be specific, we define a branch cut along the straight line segment between the two branch points $p_{b1}$ and $p_{b2}$ as shown by the wavy line in Fig. \ref{saddle}(a), which separate the two branches of Riemann surface. We then denote the branch where the original contour of the OTOC integral (\ref{Eq-OTOC}) lies in as branch $1$ (Fig. \ref{saddle}(a)), and the other branch as branch $2$. More specifically, on branch $1$ we have $\varkappa(0)=\frac{\mathcal{J}(\sqrt{3-2\mathcal{J}^2} -1)}{1-\mathcal{J}^2}=\frac{\beta}{2\pi}\overline{\lambda}$ (see Eq. (\ref{Eq-chaos-exp})), while on branch $2$ we have $\varkappa(0)=\frac{\mathcal{J}(-\sqrt{3-2\mathcal{J}^2} -1)}{1-\mathcal{J}^2}$.

On branch $1$, the function $\chi(p)\rightarrow -i(1-\mathcal{J})p-\frac{\mathcal{J}}{1+\mathcal{J}}$ as $p\rightarrow\infty$. Therefore, the numerator $\exp\left[\frac{2\pi t}{\beta}(\varkappa+ipv)\right]$ of the integrand in Eq. (\ref{Eq-OTOC}) at velocity $x/t=v$ tends to $\exp\left[i\frac{2\pi t}{\beta}(v-u_-)p\right]$ at large $p$. Since we are interested in velocities $v$ within the causality cone $v\in[u_-,u_+]$, the original integral contour of Eq. (\ref{Eq-OTOC}) along the real axis should always be closed in the upper half complex plane of branch $1$ (counterclockwise), as shown in Fig. \ref{saddle}(a) and (e).

The weight coefficient $\rho(p)=1/\cos[\pi\varkappa(p)/2]$ in the integrand of Eq. (\ref{Eq-OTOC}) has two poles
\begin{equation}
p_1=\frac{i}{u_+}\ ,\qquad p_2=i\left(\frac{2}{u_-}-\frac{1}{u_+}\right)
\end{equation}
on the imaginary axis satisfying $\varkappa(p_1)=\varkappa(p_2)=1$, and another two poles
\begin{equation}
\widetilde{p}_1=-\frac{i}{u_-}\ ,\qquad \widetilde{p}_2=-i\left(\frac{2}{u_+}-\frac{1}{u_-}\right)
\end{equation}
satisfying $\varkappa(\widetilde{p}_1)=\varkappa(\widetilde{p}_2)=-1$. The two poles $p_1$ and $\widetilde{p}_2$ are symmetric about the branch cut, and so do the two poles $p_2$ and $\widetilde{p}_1$.
For interaction strenth $0<J<\pi$, poles $p_2$ and $\widetilde{p}_1$ are located in branch $1$ above and below the branch cut, respectively, while poles $p_1$ and $\widetilde{p}_2$ are located in branch $2$ above and below the branch cut, respectively. Such an example at $J=0.6\pi$ is given in Fig. \ref{exponent}(a), where we use solid dots for poles lying in branch 1, and hollow dots for poles lying in branch $2$. For interaction strength $\pi<J<2\pi$, all the poles $p_1$, $p_2$, $\widetilde{p}_1$ and $\widetilde{p}_2$ are located in branch $1$.
Fig. \ref{exponent}(e) shows such an example at $J=1.2\pi$.

Meanwhile, for a given velocity $x/t=v$, the numerator $\exp\left[\frac{2\pi t}{\beta}(\varkappa+ipv)\right]$ has two saddle points $p_v$ and $\widetilde{p}_v$ on the imaginary axis satisfying $\varkappa'(p_v)+iv=\varkappa'(\widetilde{p}_v)+iv=0$, which are symmetric about the branch cut:
\begin{equation}\label{Eq-pv}
	p_{v}={i\mathcal{J}\over 1-\mathcal{J}^2}-{i(1-v)\over \sqrt{\mathcal{J}^2-(1-v)^2}}{\sqrt{3}\over \sqrt{1-\mathcal{J}^2}},~~
	\widetilde p_{v}={i\mathcal{J}\over 1-\mathcal{J}^2}+{i(1-v)\over \sqrt{\mathcal{J}^2-(1-v)^2}}{\sqrt{3}\over \sqrt{1-\mathcal{J}^2}}.
\end{equation}
The first (second) saddle point $p_v$ ($\widetilde{p}_v$) moves upwards (downwards) on the imaginary axis as $v$ increases from $v_-$ to $v_+$. When the velocity $u_-<v<1$, both of the two saddle points $p_v$ and $\widetilde{p}_v$ are located in branch $1$ (Fig. \ref{saddle}(b)\&(c)). While when the velocity $1<v<u_+$, the both saddle points $p_v$ and $\widetilde{p}_v$ will move to branch $2$ (Fig. \ref{saddle}(d)).

The saddle point approximation generically works in the following way: if the original integration contour can be deformed into some deepest decent (stationary phase) contour passing a saddle point of the integrand without crossing any poles, the integral can be evaluated by the value of the integrand at the saddle point. If the deformed contour contains multiple deepest decent contours passing several saddle points, the contributions of these saddle points will add, and usually one of the saddle points will dominate the integral. If the contour deformation crosses some pole of the integrand, the integral will be contributed by both the saddle point values and the pole residues.

We first discuss the contour deformation when the velocity $u_-<v<1$, in which case both saddle points $p_v$ and $\widetilde{p}_v$ are in Riemann branch $1$. As shown in Fig. \ref{saddle}(b), the steepest decent contour of saddle point $p_v$ hits the other saddle point $\widetilde{p}_v$, forming an ellipse surrounding the branch cut in branch $1$; then from $\widetilde{p}_v$ the steepest decent contour extends along the imaginary axis to $+i\infty$ of branch $1$ and $-i\infty$ of branch $2$ (across the branch cut). In the figure, we use solid lines to represent contours lying in branch $1$, and dashed lines for contours lying in branch $2$. Hitting the other saddle point $\widetilde{p}_v$ makes the deepest decent contour of saddle point $p_v$ ambiguous, which is accidental. To resolve this ambiguity, we can add a infinitesimal imaginary part to the velocity $v$, so that $p_v$ and $\widetilde{p}_v$ slightly deviates from the imaginary axis. After such a procedure, the steepest decent contours passing $p_v$ and $\widetilde{p}_v$ no longer coincide, which are illustrated by Fig. \ref{saddle}(c): the contour passing $p_v$ extends from $-i\infty$ of branch $2$ to the counterclockwise ellipse surrounding the branch cut in branch $1$, and then extends to $+i\infty$ of branch $1$. While the contour passing $\widetilde{p}_v$ goes directly from $i\infty$ of branch $1$ to $-i\infty$ of branch $2$. One can easily check that the original contour in Fig. \ref{saddle}(a) can be continuously deformed into the sum of the two steepest decent contours passing $p_v$ and $\widetilde{p}_v$. Therefore, both saddle points $p_v$ and $\widetilde{p}_v$ contributes to the integral. If $-i\widetilde{p}_v<-ip_2$ or $-i\widetilde{p}_v<-i\widetilde{p}_2$, the deformation will cross the pole $p_2$ or $\widetilde{p}_2$; while if $-ip_v>-ip_1$ or $-ip_v<-i\widetilde{p}_1$, the contour deformation will cross the pole $p_1$ or $\widetilde{p}_1$. In these cases, the residues of the poles crossed during the deformation will then also contribute.

Second, we discuss the case when the velocity $1<v<u_+$, where both saddle points $p_v$ and $\widetilde{p}_v$ move to Riemann branch $2$. Fig. \ref{saddle}(d) shows such an example, we we use hollow dots for $p_v$ and $\widetilde{p}_v$ to remind the reader that they are in branch $2$. Now the steepest decent contour passing $p_v$ extends from $-i\infty$ of branch $2$ to a clockwise ellipse surrounding the branch cut in branch $2$, and then extends to $i\infty$ of branch $1$ across the branch cut. The other contour passing $\widetilde{p}_v$ still extends from $i\infty$ of branch $1$ to $-i\infty$ of branch $2$. In this case, one may find that deforming the original contour of Fig. \ref{saddle}(a) into the sum of the two steepest decent contours passing $p_v$ and $\widetilde{p}_v$ requires crossing the two branch points, and may wonder whether this is legal. In fact, it is indeed legal to cross the branch points in this configuration. This is because here the Riemann surface with two branches connected by a branch cut is topologically equivalent to a two dimensional sphere, so the branch points are topologically nonsingular and allow a contour to cross through. By drawing the Riemann surface as a sphere, one can show the original contour in Fig. \ref{saddle}(a) can be smoothly deformed into the sum of the two steepest contours of $p_v$ and $\widetilde{p}_v$ in Fig. \ref{saddle}(d). Therefore, the integral is still contributed by the two saddle points $p_v$ and $\widetilde{p}_v$, and residues of any of the poles $p_1$, $p_2$, $\widetilde{p}_1$ and $\widetilde{p}_2$ crossed during the contour deformation.

All together, in the entire velocity range $u_-<v<u_+$, to calculate the integral (\ref{Eq-OTOC}) we only need to add the contributions from saddle points $p_v$ and $\widetilde{p}_v$, and the residues of poles $p_1$, $p_2$, $\widetilde{p}_1$ and $\widetilde{p}_2$ if crossed during the contour deformation. The evaluation at large $t$ can be further simplified by noting the following two facts. First, for velocity $u_-<v<u_+$, the residues of the two poles $p_2$ and $\widetilde{p}_2$ give velocity dependent Lyapunov exponents \begin{equation}
\varkappa(p_2)+ivp_2=1-v\left(\frac{2}{u_-}-\frac{1}{u_+}\right)< 0\ ,\qquad \varkappa(\widetilde{p}_2)+iv\widetilde{p}_2=-1+v\left(\frac{2}{u_+}-\frac{1}{u_-}\right)< 0\ ,
\end{equation}
respectively. Therefore, the residues of poles $p_2$ and $\widetilde{p}_2$ never give an exponential growth in $t$ along velocity $v$, and can always be ignored. Secondly, we can compare the exponent in $t$ contributed by the two saddle points $p_v$ and $\widetilde{p}_v$. Since $p_v$ and $\widetilde{p}_v$ are symmetric about the branch cut, we can denote them as $p_v=\frac{i\mathcal{J}}{1-\mathcal{J}^2}-i\eta$ and $\widetilde{p}_v=\frac{i\mathcal{J}}{1-\mathcal{J}^2}+i\eta$, respectively, where $\eta$ is real (Eq. (\ref{Eq-pv})). Accordingly, the exponents they contribute at velocity $v$ are given by
\begin{equation}
\varkappa(p_v)+ivp_v=-\frac{\mathcal{J} v}{1-\mathcal{J}^2}+\eta(v-1)+\mathcal{J}\sqrt{\frac{3}{1-\mathcal{J}^2} +\eta^2}\ ,
\end{equation}
\begin{equation}
\varkappa(\widetilde{p}_v)+iv\widetilde{p}_v=-\frac{\mathcal{J} v}{1-\mathcal{J}^2}-\eta(v-1)+\mathcal{J}\sqrt{\frac{3}{1-\mathcal{J}^2} +\eta^2}\ ,
\end{equation}
respectively. For velocity $u_-<v<u_+$ which we are interested in, one can easily see that $\varkappa(p_v)+ivp_v > \varkappa(\widetilde{p}_v)+iv\widetilde{p}_v$, namely, the exponent contributed by saddle point $p_v$ is always larger. Therefore, we conclude that the first saddle point $p_v$ always dominates over the second saddle point $\widetilde{p}_v$.

\begin{figure}[tbp]
\begin{center}
\includegraphics[width=4in]{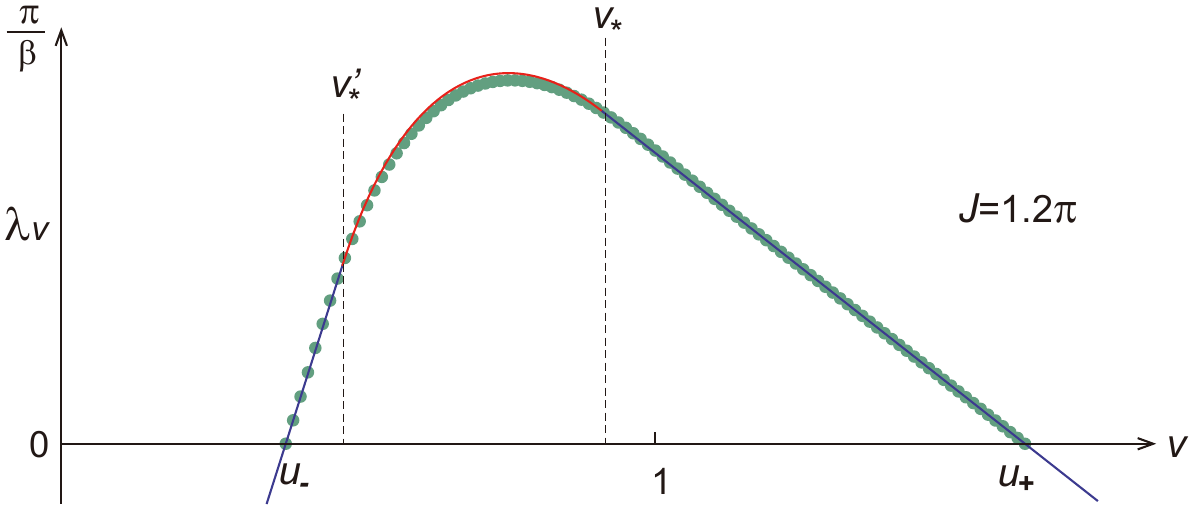}
\end{center}
\caption{The velocity dependent Lyapunov exponent from saddle point approximation (solid line) and from direct numerical integration of Eq. (\ref{Eq-OTOC}) (dotted line) for $J=1.2\pi$.}
\label{Nexp}
\end{figure}

Given the above two facts, we find that we only need to consider the contributions of the saddle point $p_v$ and the two poles $p_1$ and $\widetilde{p}_1$. When $-i\widetilde{p}_1<-ip_v<-ip_1$, the contour deformation does not pass either pole $p_1$ or pole $\widetilde{p}_1$, so the integral is given by the saddle point $p_v$ contribution (Eq. (\ref{Eq-butterfly1})). This is the case in the examples of Fig. \ref{saddle} (c) and (e), where the former (latter) has $0<J<\pi$ ($\pi<J<2\pi$) and thus pole $p_1$ located in branch $2$ (branch $1$). When $-ip_v>-ip_1$, the OTOC integral is contributed by both the saddle point $p_v$ and the residue of pole $p_1$, and the exponent contributed by pole $p_1$ is always larger in this case. Therefore, the integral is dominated by pole $p_1$ (Eq. (\ref{Eq-butterfly2})). An example of this case is shown in Fig. \ref{saddle}(d). Lastly, when $-ip_v<-i\widetilde{p}_1$, the OTOC integral is contributed by both the saddle point $p_v$ and the residue of pole $\widetilde{p}_1$, and is always dominated by pole $\widetilde{p}_1$.

Fig. \ref{Nexp} shows a comparison between the velocity dependent Lyapunov exponent from saddle point approximation and that from the direct numerical integration of Eq. (\ref{Eq-OTOC}) for $J=1.2\pi$. This example demonstrates the accuracy of the above saddle point approximation. \footnote{We thank G. S\'arosi and M. Mezei for pointing out the critical velocity $v_*'$ and the verification via numerical integration \cite{mezei2019}.}

\bibliographystyle{arXiv_new}
\bibliography{CSYK_ref}

\providecommand{\href}[2]{#2}\begingroup\raggedright\begin{thebibliography}{10}

\bibitem{Gu:2018jsv}
Y.~Gu and A.~Kitaev, \emph{{On the relation between the magnitude and exponent
  of OTOCs}}, \href{http://dx.doi.org/10.1007/JHEP02(2019)075}{\emph{JHEP} {\bf
  02} 075 (2019)}, [\href{https://arxiv.org/abs/1812.00120}{{\tt
  arXiv:1812.00120}}].

\bibitem{Sachdev:1992fk}
S.~Sachdev and J.~Ye, \emph{{Gapless spin fluid ground state in a random,
  quantum Heisenberg magnet}},
  \href{http://dx.doi.org/10.1103/PhysRevLett.70.3339}{\emph{Phys. Rev. Lett.}
  {\bf 70} 3339 (1993)}, [\href{https://arxiv.org/abs/cond-mat/9212030}{{\tt
  arXiv:cond-mat/9212030}}].

\bibitem{Polchinski:2016xgd}
J.~Polchinski and V.~Rosenhaus, \emph{{The Spectrum in the Sachdev-Ye-Kitaev
  Model}}, \href{http://dx.doi.org/10.1007/JHEP04(2016)001}{\emph{JHEP} {\bf
  04} 001 (2016)}, [\href{https://arxiv.org/abs/1601.06768}{{\tt
  arXiv:1601.06768}}].

\bibitem{Maldacena:2016hyu}
J.~Maldacena and D.~Stanford, \emph{{Remarks on the Sachdev-Ye-Kitaev model}},
  \href{http://dx.doi.org/10.1103/PhysRevD.94.106002}{\emph{Phys. Rev.} {\bf
  D94} 106002 (2016)}, [\href{https://arxiv.org/abs/1604.07818}{{\tt
  arXiv:1604.07818}}].

\bibitem{Kitaev:2017awl}
A.~Kitaev and S.~J. Suh, \emph{{The soft mode in the Sachdev-Ye-Kitaev model
  and its gravity dual}},
  \href{http://dx.doi.org/10.1007/JHEP05(2018)183}{\emph{JHEP} {\bf 05} 183
  (2018)}, [\href{https://arxiv.org/abs/1711.08467}{{\tt arXiv:1711.08467}}].

\bibitem{Witten:2016iux}
E.~Witten, \emph{{An SYK-Like Model Without Disorder}},
  \href{https://arxiv.org/abs/1610.09758}{{\tt arXiv:1610.09758}}.

\bibitem{gurau}
R.~{Gurau}, \emph{{A review of the large N limit of tensor models}},
  {\emph{arXiv e-prints} arXiv:1209.4295 (2012)},
  [\href{https://arxiv.org/abs/1209.4295}{{\tt arXiv:1209.4295}}].

\bibitem{KPT}
I.~R. {Klebanov}, F.~{Popov} and G.~{Tarnopolsky}, \emph{{TASI Lectures on
  Large $N$ Tensor Models}}, {\emph{arXiv e-prints} arXiv:1808.09434 (2018)},
  [\href{https://arxiv.org/abs/1808.09434}{{\tt arXiv:1808.09434}}].

\bibitem{Gu:2016oyy}
Y.~Gu, X.-L. Qi and D.~Stanford, \emph{{Local criticality, diffusion and chaos
  in generalized Sachdev-Ye-Kitaev models}},
  \href{http://dx.doi.org/10.1007/JHEP05(2017)125}{\emph{JHEP} {\bf 05} 125
  (2017)}, [\href{https://arxiv.org/abs/1609.07832}{{\tt arXiv:1609.07832}}].

\bibitem{Berkooz:2016cvq}
M.~Berkooz, P.~Narayan, M.~Rozali and J.~Simon, \emph{{Higher Dimensional
  Generalizations of the SYK Model}},
  \href{http://dx.doi.org/10.1007/JHEP01(2017)138}{\emph{JHEP} {\bf 01} 138
  (2017)}, [\href{https://arxiv.org/abs/1610.02422}{{\tt arXiv:1610.02422}}].

\bibitem{Blake:2017}
M.~Blake, R.~A. Davison and S.~Sachdev, \emph{Thermal diffusivity and chaos in
  metals without quasiparticles},
  \href{http://dx.doi.org/10.1103/PhysRevD.96.106008}{\emph{Phys. Rev. D} {\bf
  96} 106008 (2017)}.

\bibitem{davison2017}
R.~A. Davison, W.~Fu, A.~Georges, Y.~Gu, K.~Jensen and S.~Sachdev,
  \emph{Thermoelectric transport in disordered metals without quasiparticles:
  The sachdev-ye-kitaev models and holography},
  \href{http://dx.doi.org/10.1103/PhysRevB.95.155131}{\emph{Phys. Rev. B} {\bf
  95} 155131 (2017)}.

\bibitem{Jian:2017unn}
S.-K. Jian and H.~Yao, \emph{{Solvable Sachdev-Ye-Kitaev models in higher
  dimensions: from diffusion to many-body localization}},
  \href{http://dx.doi.org/10.1103/PhysRevLett.119.206602}{\emph{Phys. Rev.
  Lett.} {\bf 119} 206602 (2017)},
  [\href{https://arxiv.org/abs/1703.02051}{{\tt arXiv:1703.02051}}].

\bibitem{chen2017}
Y.~{Chen}, H.~{Zhai} and P.~{Zhang}, \emph{{Tunable quantum chaos in the
  Sachdev-Ye-Kitaev model coupled to a thermal bath}},
  \href{http://dx.doi.org/10.1007/JHEP07(2017)150}{\emph{Journal of High Energy
  Physics} {\bf 2017} 150 (2017)},
  [\href{https://arxiv.org/abs/1705.09818}{{\tt arXiv:1705.09818}}].

\bibitem{cai2018}
W.~Cai, X.-H. Ge and G.-H. Yang, \emph{Diffusion in higher dimensional syk
  model with complex fermions},
  \href{http://dx.doi.org/10.1007/JHEP01(2018)076}{\emph{Journal of High Energy
  Physics} {\bf 2018} 76 (2018)}.

\bibitem{zhangp2018}
P.~{Zhang} and H.~{Zhai}, \emph{{Topological Sachdev-Ye-Kitaev model}},
  \href{http://dx.doi.org/10.1103/PhysRevB.97.201112}{\emph{PRB} {\bf 97}
  201112 (2018)}, [\href{https://arxiv.org/abs/1803.01411}{{\tt
  arXiv:1803.01411}}].

\bibitem{Turiaci:2017zwd}
G.~Turiaci and H.~Verlinde, \emph{{Towards a 2d QFT Analog of the SYK Model}},
  \href{http://dx.doi.org/10.1007/JHEP10(2017)167}{\emph{JHEP} {\bf 10} 167
  (2017)}, [\href{https://arxiv.org/abs/1701.00528}{{\tt arXiv:1701.00528}}].

\bibitem{Murugan:2017eto}
J.~Murugan, D.~Stanford and E.~Witten, \emph{{More on Supersymmetric and 2d
  Analogs of the SYK Model}},
  \href{http://dx.doi.org/10.1007/JHEP08(2017)146}{\emph{JHEP} {\bf 08} 146
  (2017)}, [\href{https://arxiv.org/abs/1706.05362}{{\tt arXiv:1706.05362}}].

\bibitem{Berkooz:2017efq}
M.~Berkooz, P.~Narayan, M.~Rozali and J.~Simon, \emph{{Comments on the Random
  Thirring Model}},
  \href{http://dx.doi.org/10.1007/JHEP09(2017)057}{\emph{JHEP} {\bf 09} 057
  (2017)}, [\href{https://arxiv.org/abs/1702.05105}{{\tt arXiv:1702.05105}}].

\bibitem{narayan2017}
P.~{Narayan} and J.~{Yoon}, \emph{{SYK-like tensor models on the lattice}},
  \href{http://dx.doi.org/10.1007/JHEP08(2017)083}{\emph{Journal of High Energy
  Physics} {\bf 2017} 83 (2017)}, [\href{https://arxiv.org/abs/1705.01554}{{\tt
  arXiv:1705.01554}}].

\bibitem{Klebanov:2016xxf}
I.~R. Klebanov and G.~Tarnopolsky, \emph{{Uncolored random tensors, melon
  diagrams, and the Sachdev-Ye-Kitaev models}},
  \href{http://dx.doi.org/10.1103/PhysRevD.95.046004}{\emph{Phys. Rev.} {\bf
  D95} 046004 (2017)}, [\href{https://arxiv.org/abs/1611.08915}{{\tt
  arXiv:1611.08915}}].

\bibitem{Giombi:2017dtl}
S.~Giombi, I.~R. Klebanov and G.~Tarnopolsky, \emph{{Bosonic tensor models at
  large $N$ and small $\epsilon$}},
  \href{http://dx.doi.org/10.1103/PhysRevD.96.106014}{\emph{Phys. Rev.} {\bf
  D96} 106014 (2017)}, [\href{https://arxiv.org/abs/1707.03866}{{\tt
  arXiv:1707.03866}}].

\bibitem{Shenker:2013pqa}
S.~H. Shenker and D.~Stanford, \emph{{Black holes and the butterfly effect}},
  \href{http://dx.doi.org/10.1007/JHEP03(2014)067}{\emph{JHEP} {\bf 03} 067
  (2014)}, [\href{https://arxiv.org/abs/1306.0622}{{\tt arXiv:1306.0622}}].

\bibitem{Shenker:2014cwa}
S.~H. Shenker and D.~Stanford, \emph{{Stringy effects in scrambling}},
  \href{http://dx.doi.org/10.1007/JHEP05(2015)132}{\emph{JHEP} {\bf 05} 132
  (2015)}, [\href{https://arxiv.org/abs/1412.6087}{{\tt arXiv:1412.6087}}].

\bibitem{Maldacena:2015waa}
J.~Maldacena, S.~H. Shenker and D.~Stanford, \emph{{A bound on chaos}},
  \href{http://dx.doi.org/10.1007/JHEP08(2016)106}{\emph{JHEP} {\bf 08} 106
  (2016)}, [\href{https://arxiv.org/abs/1503.01409}{{\tt arXiv:1503.01409}}].

\bibitem{Sachdev:2010prl}
S.~Sachdev, \emph{Holographic metals and the fractionalized fermi liquid},
  \href{http://dx.doi.org/10.1103/PhysRevLett.105.151602}{\emph{Phys. Rev.
  Lett.} {\bf 105} 151602 (2010)}.

\bibitem{Almheiri:2014cka}
A.~Almheiri and J.~Polchinski, \emph{{Models of AdS$_{2}$ backreaction and
  holography}}, \href{http://dx.doi.org/10.1007/JHEP11(2015)014}{\emph{JHEP}
  {\bf 11} 014 (2015)}, [\href{https://arxiv.org/abs/1402.6334}{{\tt
  arXiv:1402.6334}}].

\bibitem{Jensen:2016pah}
K.~Jensen, \emph{{Chaos in AdS$_2$ Holography}},
  \href{http://dx.doi.org/10.1103/PhysRevLett.117.111601}{\emph{Phys. Rev.
  Lett.} {\bf 117} 111601 (2016)},
  [\href{https://arxiv.org/abs/1605.06098}{{\tt arXiv:1605.06098}}].

\bibitem{Maldacena:2016upp}
J.~Maldacena, D.~Stanford and Z.~Yang, \emph{{Conformal symmetry and its
  breaking in two dimensional Nearly Anti-de-Sitter space}},
  \href{http://dx.doi.org/10.1093/ptep/ptw124}{\emph{PTEP} {\bf 2016} 12C104
  (2016)}, [\href{https://arxiv.org/abs/1606.01857}{{\tt arXiv:1606.01857}}].

\bibitem{Engelsoy:2016xyb}
J.~Engelsoy, T.~G. Mertens and H.~Verlinde, \emph{{An investigation of
  AdS$_{2}$ backreaction and holography}},
  \href{http://dx.doi.org/10.1007/JHEP07(2016)139}{\emph{JHEP} {\bf 07} 139
  (2016)}, [\href{https://arxiv.org/abs/1606.03438}{{\tt arXiv:1606.03438}}].

\bibitem{Maldacena:2017axo}
J.~Maldacena, D.~Stanford and Z.~Yang, \emph{{Diving into traversable
  wormholes}}, \href{http://dx.doi.org/10.1002/prop.201700034}{\emph{Fortsch.
  Phys.} {\bf 65} 1700034 (2017)},
  [\href{https://arxiv.org/abs/1704.05333}{{\tt arXiv:1704.05333}}].

\bibitem{Maldacena:2018lmt}
J.~Maldacena and X.-L. Qi, \emph{{Eternal traversable wormhole}},
  \href{https://arxiv.org/abs/1804.00491}{{\tt arXiv:1804.00491}}.

\bibitem{kim2019}
J.~{Kim}, I.~R. {Klebanov}, G.~{Tarnopolsky} and W.~{Zhao}, \emph{{Symmetry
  Breaking in Coupled SYK or Tensor Models}},
  \href{http://dx.doi.org/10.1103/PhysRevX.9.021043}{\emph{Physical Review X}
  {\bf 9} 021043 (2019)}, [\href{https://arxiv.org/abs/1902.02287}{{\tt
  arXiv:1902.02287}}].

\bibitem{Cotler:2016fpe}
J.~S. Cotler, G.~Gur-Ari, M.~Hanada, J.~Polchinski, P.~Saad, S.~H. Shenker
  et~al., \emph{{Black Holes and Random Matrices}},
  \href{http://dx.doi.org/10.1007/JHEP09(2018)002,
  10.1007/JHEP05(2017)118}{\emph{JHEP} {\bf 05} 118 (2017)},
  [\href{https://arxiv.org/abs/1611.04650}{{\tt arXiv:1611.04650}}].

\bibitem{Saad:2018bqo}
P.~Saad, S.~H. Shenker and D.~Stanford, \emph{{A semiclassical ramp in SYK and
  in gravity}},  \href{https://arxiv.org/abs/1806.06840}{{\tt
  arXiv:1806.06840}}.

\bibitem{Saad:2019lba}
P.~Saad, S.~H. Shenker and D.~Stanford, \emph{{JT gravity as a matrix
  integral}},  \href{https://arxiv.org/abs/1903.11115}{{\tt arXiv:1903.11115}}.

\bibitem{laughlin1981}
R.~B. Laughlin, \emph{Quantized hall conductivity in two dimensions},
  \href{http://dx.doi.org/10.1103/PhysRevB.23.5632}{\emph{Phys. Rev. B} {\bf
  23} 5632--5633 (1981)}.

\bibitem{halperin1982}
B.~I. Halperin, \emph{Quantized hall conductance, current-carrying edge states,
  and the existence of extended states in a two-dimensional disordered
  potential}, \href{http://dx.doi.org/10.1103/PhysRevB.25.2185}{\emph{Phys.
  Rev. B} {\bf 25} 2185--2190 (1982)}.

\bibitem{haldane1988}
F.~D.~M. Haldane, \emph{Model for a quantum hall effect without landau levels:
  Condensed-matter realization of the "parity anomaly"},
  \href{http://dx.doi.org/10.1103/PhysRevLett.61.2015}{\emph{Phys. Rev. Lett.}
  {\bf 61} 2015--2018 (1988)}.

\bibitem{qi2006}
X.-L. Qi, Y.-S. Wu and S.-C. Zhang, \emph{Topological quantization of the spin
  hall effect in two-dimensional paramagnetic semiconductors},
  \href{http://dx.doi.org/10.1103/PhysRevB.74.085308}{\emph{Phys. Rev. B} {\bf
  74} 085308 (2006)}.

\bibitem{moore1991}
G.~Moore and N.~Read, \emph{Nonabelions in the fractional quantum hall effect},
  \href{http://dx.doi.org/http://dx.doi.org/10.1016/0550-3213(91)90407-O}{\emph{Nucl.
  Phys. B} {\bf 360} 362 -- 396 (1991)}.

\bibitem{read2000}
N.~Read and D.~Green, \emph{Paired states of fermions in two dimensions with
  breaking of parity and time-reversal symmetries and the fractional quantum
  hall effect}, \href{http://dx.doi.org/10.1103/PhysRevB.61.10267}{\emph{Phys.
  Rev. B} {\bf 61} 10267--10297 (2000)}.

\bibitem{qi2011}
X.-L. Qi and S.-C. Zhang, \emph{Topological insulators and superconductors},
  \href{http://dx.doi.org/10.1103/RevModPhys.83.1057}{\emph{Rev. Mod. Phys.}
  {\bf 83} 1057--1110 (2011)}.

\bibitem{kitaev2006}
A.~Kitaev, \emph{Anyons in an exactly solved model and beyond},
  \href{http://dx.doi.org/https://doi.org/10.1016/j.aop.2005.10.005}{\emph{Annals
  of Physics} {\bf 321} 2 -- 111 (2006)}.

\bibitem{wen1990}
X.~G. Wen, \emph{Chiral luttinger liquid and the edge excitations in the
  fractional quantum hall states},
  \href{http://dx.doi.org/10.1103/PhysRevB.41.12838}{\emph{Phys. Rev. B} {\bf
  41} 12838--12844 (1990)}.

\bibitem{chang2003}
A.~M. Chang, \emph{Chiral luttinger liquids at the fractional quantum hall
  edge}, \href{http://dx.doi.org/10.1103/RevModPhys.75.1449}{\emph{Rev. Mod.
  Phys.} {\bf 75} 1449--1505 (2003)}.

\bibitem{cano2014}
J.~Cano, M.~Cheng, M.~Mulligan, C.~Nayak, E.~Plamadeala and J.~Yard,
  \emph{Bulk-edge correspondence in (2 + 1)-dimensional abelian topological
  phases}, \href{http://dx.doi.org/10.1103/PhysRevB.89.115116}{\emph{Phys. Rev.
  B} {\bf 89} 115116 (2014)}.

\bibitem{banerjee2018}
M.~{Banerjee}, M.~{Heiblum}, V.~{Umansky}, D.~E. {Feldman}, Y.~{Oreg} and
  A.~{Stern}, \emph{{Observation of half-integer thermal Hall conductance}},
  \href{http://dx.doi.org/10.1038/s41586-018-0184-1}{\emph{Nature} {\bf 559}
  205--210 (2018)}, [\href{https://arxiv.org/abs/1710.00492}{{\tt
  arXiv:1710.00492}}].

\bibitem{simon2018}
S.~H. Simon, \emph{Interpretation of thermal conductance of the
  $\mathbf{\ensuremath{\nu}}=\mathbf{5}/\mathbf{2}$ edge},
  \href{http://dx.doi.org/10.1103/PhysRevB.97.121406}{\emph{Phys. Rev. B} {\bf
  97} 121406 (2018)}.

\bibitem{feldman2018}
D.~E. Feldman, \emph{Comment on ``interpretation of thermal conductance of the
  $\ensuremath{\nu}=5/2$ edge''},
  \href{http://dx.doi.org/10.1103/PhysRevB.98.167401}{\emph{Phys. Rev. B} {\bf
  98} 167401 (2018)}.

\bibitem{ma2019}
K.~K.~W. Ma and D.~E. Feldman, \emph{Partial equilibration of integer and
  fractional edge channels in the thermal quantum hall effect},
  \href{http://dx.doi.org/10.1103/PhysRevB.99.085309}{\emph{Phys. Rev. B} {\bf
  99} 085309 (2019)}.

\bibitem{murugan2018}
J.~{Murugan} and H.~{Nastase}, \emph{{One-dimensional bosonization and the SYK
  model}}, {\emph{arXiv e-prints} arXiv:1812.11929 (2018)},
  [\href{https://arxiv.org/abs/1812.11929}{{\tt arXiv:1812.11929}}].

\bibitem{gross1974}
D.~J. Gross and A.~Neveu, \emph{Dynamical symmetry breaking in asymptotically
  free field theories},
  \href{http://dx.doi.org/10.1103/PhysRevD.10.3235}{\emph{Phys. Rev. D} {\bf
  10} 3235--3253 (1974)}.

\bibitem{kane1997}
C.~L. Kane and M.~P.~A. Fisher, \emph{Quantized thermal transport in the
  fractional quantum hall effect},
  \href{http://dx.doi.org/10.1103/PhysRevB.55.15832}{\emph{Phys. Rev. B} {\bf
  55} 15832--15837 (1997)}.

\bibitem{yang1989}
C.~N. Yang, \emph{\ensuremath{\eta} pairing and off-diagonal long-range order
  in a hubbard model},
  \href{http://dx.doi.org/10.1103/PhysRevLett.63.2144}{\emph{Phys. Rev. Lett.}
  {\bf 63} 2144--2147 (1989)}.

\bibitem{yang1990}
C.~N. Yang and S.~C. Zhang, \emph{So(4) symmetry in a hubbard model},
  \href{http://dx.doi.org/10.1142/S0217984990000933}{\emph{Modern Physics
  Letters B} {\bf 04} 759--766 (1990)},
  [\href{https://arxiv.org/abs/https://doi.org/10.1142/S0217984990000933}{{\tt
  arXiv:https://doi.org/10.1142/S0217984990000933}}].

\bibitem{haldane1981}
F.~D.~M. Haldane, \emph{'luttinger liquid theory' of one-dimensional quantum
  fluids. i. properties of the luttinger model and their extension to the
  general 1d interacting spinless fermi gas}, {\emph{Journal of Physics C:
  Solid State Physics} {\bf 14} 2585 (1981)}.

\bibitem{heidenreich1980}
R.~Heidenreich, R.~Seiler and D.~A. Uhlenbrock, \emph{The luttinger model},
  \href{http://dx.doi.org/10.1007/BF01007986}{\emph{Journal of Statistical
  Physics} {\bf 22} 27--57 (1980)}.

\bibitem{delft1998}
J.~{von Delft} and H.~{Schoeller}, \emph{{Bosonization for beginners -
  refermionization for experts}},
  \href{http://dx.doi.org/10.1002/(SICI)1521-3889(199811)7:4<225::AID-ANDP225>3.0.CO;2-L}{\emph{Annalen
  der Physik} {\bf 7} 225--305 (1998)},
  [\href{https://arxiv.org/abs/cond-mat/9805275}{{\tt
  arXiv:cond-mat/9805275}}].

\bibitem{vedika2018}
V.~Khemani, D.~A. Huse and A.~Nahum, \emph{Velocity-dependent lyapunov
  exponents in many-body quantum, semiclassical, and classical chaos},
  \href{http://dx.doi.org/10.1103/PhysRevB.98.144304}{\emph{Phys. Rev. B} {\bf
  98} 144304 (2018)}.

\bibitem{mross2018}
D.~F. Mross, Y.~Oreg, A.~Stern, G.~Margalit and M.~Heiblum, \emph{Theory of
  disorder-induced half-integer thermal hall conductance},
  \href{http://dx.doi.org/10.1103/PhysRevLett.121.026801}{\emph{Phys. Rev.
  Lett.} {\bf 121} 026801 (2018)}.

\bibitem{wangc2018}
C.~Wang, A.~Vishwanath and B.~I. Halperin, \emph{Topological order from
  disorder and the quantized hall thermal metal: Possible applications to the
  $\ensuremath{\nu}=5/2$ state},
  \href{http://dx.doi.org/10.1103/PhysRevB.98.045112}{\emph{Phys. Rev. B} {\bf
  98} 045112 (2018)}.

\bibitem{lian2018}
B.~Lian and J.~Wang, \emph{Theory of the disordered
  $\ensuremath{\nu}=\frac{5}{2}$ quantum thermal hall state: Emergent symmetry
  and phase diagram},
  \href{http://dx.doi.org/10.1103/PhysRevB.97.165124}{\emph{Phys. Rev. B} {\bf
  97} 165124 (2018)}.

\bibitem{chalker1995}
J.~T. Chalker and A.~Dohmen, \emph{Three-dimensional disordered conductors in a
  strong magnetic field: Surface states and quantum hall plateaus},
  \href{http://dx.doi.org/10.1103/PhysRevLett.75.4496}{\emph{Phys. Rev. Lett.}
  {\bf 75} 4496--4499 (1995)}.

\bibitem{balents1996}
L.~Balents and M.~P.~A. Fisher, \emph{Chiral surface states in the bulk quantum
  hall effect},
  \href{http://dx.doi.org/10.1103/PhysRevLett.76.2782}{\emph{Phys. Rev. Lett.}
  {\bf 76} 2782--2785 (1996)}.

\bibitem{naud2000}
J.~D. Naud, L.~P. Pryadko and S.~L. Sondhi, \emph{Fractional quantum hall
  effect in infinite-layer systems},
  \href{http://dx.doi.org/10.1103/PhysRevLett.85.5408}{\emph{Phys. Rev. Lett.}
  {\bf 85} 5408--5411 (2000)}.

\bibitem{naud2001}
J.~Naud, L.~P. Pryadko and S.~Sondhi, \emph{Notes on infinite layer quantum
  hall systems},
  \href{http://dx.doi.org/https://doi.org/10.1016/S0550-3213(00)00679-9}{\emph{Nuclear
  Physics B} {\bf 594} 713 -- 746 (2001)}.

\bibitem{Patel:2017}
A.~A. Patel and S.~Sachdev, \emph{Quantum chaos on a critical fermi surface},
  \href{http://dx.doi.org/10.1073/pnas.1618185114}{\emph{Proceedings of the
  National Academy of Sciences} {\bf 114} 1844--1849 (2017)},
  [\href{https://arxiv.org/abs/https://www.pnas.org/content/114/8/1844.full.pdf}{{\tt
  arXiv:https://www.pnas.org/content/114/8/1844.full.pdf}}].

\bibitem{Patel:2017b}
A.~A. Patel, D.~Chowdhury, S.~Sachdev and B.~Swingle, \emph{Quantum butterfly
  effect in weakly interacting diffusive metals},
  \href{http://dx.doi.org/10.1103/PhysRevX.7.031047}{\emph{Phys. Rev. X} {\bf
  7} 031047 (2017)}.

\bibitem{mezei2019}
M.~{Mezei} and G.~{S{\'a}rosi}, \emph{{Chaos in the butterfly cone}},
  {\emph{arXiv e-prints} arXiv:1908.03574 (2019)},
  [\href{https://arxiv.org/abs/1908.03574}{{\tt arXiv:1908.03574}}].

\end{thebibliography}\endgroup

\end{document}